%% file: ciesla_cuzzi.tex
\documentclass[12pt]{article}
\usepackage{graphicx}
\usepackage{natbib}
\usepackage{times}
\begin{document}

\begin{center}
\begin{Large}
\textbf{The Evolution of the Water Distribution in a Viscous Protoplanetary
Disk}\\
\end{Large}
\quad \\
\quad
\begin{Large}
Fred J. Ciesla$^{*}$ and Jeffrey N. Cuzzi \\
\end{Large}
$^{*}$Author to whom correspondence should be addressed \\
NASA Ames Research Center; MS 245-3; Moffett Field, CA 94043 \\
ciesla@cosmic.arc.nasa.gov, voice:(650) 604-0328, fax:(650) 604-6779 \\
\end{center}

\quad \\
\quad \\
\begin{flushleft}
pages: 44 \\
tables: 3 \\
figures: 7 \\
\end{flushleft}



\begin{center}
\textbf{Abstract}
\end{center}
\quad

Astronomical observations have shown that protoplanetary disks are dynamic
objects through which mass is transported and accreted by the central star.
This transport causes the disks to decrease in mass and cool over time, and 
such evolution is expected to have occurred in our own solar nebula.
Age dating of meteorite constituents shows that their creation,  evolution,
and accumulation occupied several Myr, and over this time disk properties
would evolve significantly.  Moreover, on this timescale, solid particles 
decouple from the gas in the disk and their evolution follows a different path.
It is in this context that we must understand how our own solar nebula evolved
and what effects this evolution had on the primitive materials contained within
it.  Here we present a model which tracks how the distribution of water 
changes in an evolving disk as the water-bearing species experience 
condensation, accretion, transport, collisional destruction, and vaporization.  
Because solids are transported in a disk at different rates
depending on their sizes, the motions will lead to water being concentrated
in some regions of a disk and depleted in others.  These enhancements and
depletions are consistent with the conditions needed to explain some
aspects of the chemistry of chondritic meteorites and formation of giant 
planets.  The levels of concentration
and depletion, as well as their locations, depend strongly on the combined
effects of the gaseous disk evolution, the formation of rapidly migrating 
rubble, and the 
growth of immobile planetesimals.   Understanding how these processes
operate simultaneously is critical to developing our models for meteorite
parent body formation in the solar system and giant planet formation throughout
the galaxy.  We present examples of evolution under a range of plausible
assumptions and demonstrate how the chemical evolution of the inner region
of a protoplanetary disk is intimately connected to the physical processes
which occur in the outer regions.  \\
\quad \\


\newpage

\section{Introduction}

The solar system contains a wide variety of objects, ranging from small,
rocky bodies at relatively small heliocentric distances to huge, gaseous 
planets at tens of astronomical units from the sun.  The characteristics
of these objects were determined billions of years ago, when they accreted
from primitive material that was present in their respective formation regions.  The
properties of these primitive materials were set by the physical and
chemical environments that they were exposed to prior to accretion.  A long
standing goal in studying primitive objects has been to identify what these
environments were and how they varied with location and time in the solar
nebula.

Observations of protoplanetary disks around young stars have shown that 
unraveling the chemical and physical structure of such disks requires 
understanding how they evolve over time.
Measurements of the excess radiation being emitted from young stars show that
they are accreting mass at rates up to 10$^{-6} M_{\odot}$/yr, decreasing
with time \citep{calvet05}.  
This mass is being fed onto the stars through the disks that surround them, 
implying that these
disks are dynamic objects.  While the driving mechanism for this mass transport
has not been identified, we realize that the disks grow
less massive and cooler over time.  It is in this dynamic, evolving setting that we
must understand what physical and chemical environments existed in the
solar nebula and how they affected the primitive materials
within them.

In order to understand the various conditions that exist within 
a protoplanetary disk, it is necessary to identify the timescales over which 
the environments within it change and the timescales over which material is 
transported from one region to another.  In this paper, we illustrate how this 
mixing and transport take place by investigating how the distribution of water changes
within an evolving disk due to the different processes identified above.  We
focus on water due to its importance in many different aspects of the formation
of our solar system.  In the outer solar nebula, water is expected to have been
a major condensable, making up approximately 50\% of the mass of solids.  In
the hot, inner solar nebula, water would be found in the vapor phase and its
concentration determined
the oxidation state of the gas and controlled the chemistry that
took place and the mineralogy of even high-temperature solids.  The boundary
between these two regions is called the ``snow line,'' and refers to the
location outside of which water freezes out to form a solid.  Where
this transition took place, and how water was incorporated into solids, are also
important to understand so that we can identify how Earth and other
habitable planets acquired their water contents.

The new model of the distribution of water in a protoplanetary 
disk outlined here considers the formation, growth, and destruction
of solid water particles and vapor by collisions, vaporization, and condensation
while simultaneously tracking their transport in an evolving disk.  Our model
also self-consistently tracks the change in opacity of the disk due to the transport
of the solids within it, an effect that is often neglected in disk evolution models.
In developing this model, we build upon a suite
of previous studies that are reviewed in Section 2.  The models for
global nebular evolution and the water distribution are outlined
in Sections 3 and 4 respectively.  Results of some model runs
are presented in Section 5.  The implications of these results for
the chemical evolution of primitive solids, the formation of giant
planets, and protoplanetary disk studies are discussed in Section
6.  In section 7, we
outline our conclusions and discuss future work needed to better
improve our understanding of water transport in the solar nebula.

\section{Previous Models of Water Transport in the Solar Nebula}

Solids and vapor within a protoplanetary disk can be transported in a number of ways.
Diffusion of material along concentration gradients is expected to occur 
regardless of the mechanism responsible for driving disk evolution 
\citep{morfvolk84,stevenson88,hersant01,gail1,taklin02,bitw2, cuzzahn04,boss04,kellgail04}.  
In addition, the  structure of the disk itself will lead to the transport 
of materials. The gas in the disk will
generally be supported in the radial direction by a pressure gradient,  
reducing the effective gravity that the gas feels in its orbit around the sun
and causing it to move at slightly less than Keplerian speed.  Solids do not
feel this gradient and attempt to follow Keplerian orbits, resulting in
the solids experiencing a headwind due to the velocity difference between
them and the gas.  The solids thus lose energy and angular momentum to the
gas, and move inwards over time.  The velocities with which these transport
processes operate are described below \citep[for more details, see][]{cuzzweid04}.
Here we review previous work that
has studied the effects of these transport processes on how water would be
distributed in the solar nebula.

\citet{stevenson88} investigated the diffusion of water vapor from the inner 
solar nebula to the snow line as a way of locally increasing the density of solids 
in order to facilitate the rapid growth of Jupiter's core.  Upon reaching the 
snow line, water vapor was assumed to  condense into ice and be accreted by 
pre-existing planetesimals located there.  It was found that the
diffusional redistribution of the water vapor resulted in an increase in the
surface density of solids immediately outside the snow line by a factor of $\sim$75 in
roughly 10$^{5}$ years.  This enhancement of solids would then allow Jupiter's
massive core to grow much more rapidly than if the core had to grow from the
canonical amount of solids locally present.  \citet{morfvolk84} on the
other hand, did not assume material was trapped on immobile planetesimals, so 
their solutions did not show these substantial effects.

\citet{cyr98} extended this model to examine the diffusion of water vapor
along with the migration of the freshly formed ice particles into the inner 
nebula \emph{via} gas drag.  As the ice particles drifted inwards, they vaporized 
at a finite rate \citep{icevap91} and could drift
significant distances inwards before completely vaporizing.  This drift of
the particles served as a way of replenishing water vapor to the inner nebula,
though \citet{cyr98} found that the eventual accumulation of the condensed
particles onto planetesimals would halt this inward migration and still lead
to the complete dehydration of the inner solar nebula  in $\sim$10$^{5}$ years. 
After this time, the 
concentration of solids beyond the snow line would increase to roughly the same
amount as found by \citet{stevenson88}. 

Both \citet{stevenson88} and \citet{cyr98} focused on the region of the
nebula inside the snow line, ignoring the effects of particles migrating
from the outer nebula inwards, as well as the evolving nature of the
solar nebula.   \citet{stepval97} considered
the distribution of water throughout an extended evolving nebula ($\sim$200 AU).
In this model, the
authors calculated the growth of particles as they collided with one another,
and their subsequent movement due to diffusion and gas drag.  They also 
calculated the simultaneous evolution of the viscous protoplanetary disk, 
accounting for the changes in surface density and temperature.
They reached
similar conclusions to \citet{stevenson88} and \citet{cyr98} in that a large
pile-up of ice would occur just outside of the snow line.  If massive, 
compact (extending to $\sim$20 AU) disks were investigated, the solids rapidly
migrated to the inner region of the disk before planetesimals could form and
were lost.  The
extended ($\sim$ 200 AU) disk ensured that solids would have to migrate over large distances
and were more likely get incorporated into large, immobile planetesimals before
being lost.

Due to the complexity of the different processes involved, \citet{stepval97} 
made a number of simplifying assumptions in developing their model.
In this treatment, the authors allowed for only one size of particles to
exist at a 
given distance from the sun.  Models for the growth of bodies in the solar nebula
\citep[e.g.][]{weidenschilling80,weidenschilling84,weidenschilling97,weidenschilling00,weidcuzz,
dull05} 
have shown that
a range of particle sizes would exist at a given location in the nebula.  
As will be discussed below, different sized objects will have very 
different dynamical behaviors in a viscous disk that must be taken into 
account.  Also, \citet{stepval97} assumed that all collisions between solids 
led to accretion (perfect sticking), and they found that 
more turbulent disks lead to more rapid growth of particles.  This treatment 
contradicts detailed work done by other authors which suggests that
turbulence hinders growth and that even moderate turbulence may cause 
meter-sized bodies to destroy one another in collisions 
\citep[e.g.][]{cuzzweid04}.  Finally, to determine the temperatures of the
nebula at every location, \citet{stepval97} determined the opacity for the
disk assuming a solar abundance of elements at a given location.  As will
be shown below, the rapid transport of material within a disk will result in
all locations in the nebula departing from a solar composition.  This would
lead to variations in the opacity in the disk, which in turn, will affect its
physical evolution.

\citet{cuzzahn04} developed a model to track the evolution of a ``vaporizing''
species in a protoplanetary disk (taken here to be water).  They provided an
analytic solution to an equation describing the simultaneous diffusion of vapor and
dust along with the inward migration of meter-sized bodies throughout
the nebula for steady-state cases.  These authors
found that before the inner nebula would be dehydrated as other studies
had found, it could possibly go
through a period where it was enhanced in water vapor.  This concept was
first explored in the context of silicates by \citet{bitw2}.
\citet{cuzzahn04} argued
that until planetesimals grow outside the snow line, bodies from the outer 
nebula will continuously migrate into the inner nebula and vaporize, enriching
the gas with water and suggest that this could explain the presence of oxidized species in
primitive meteorites.  Once the immobile planetesimals form, they act as a sink by
accreting material rather than allowing it to drift inwards, and the inner
nebula begins to dehydrate as was found by \citet{stevenson88} and
\citet{cyr98}.

In doing their calculations, \citet{cuzzahn04} also made a number of simplifying assumptions.  
The goal was to illustrate the different phases of evolution
that the solar nebula may have experienced due to volatile transport, not to explicitly
model it.  Thus they 
only considered a snapshot of the nebula when the snow line existed at
a fixed location,  assumed that the meter-sized bodies
represented a fixed fraction of the total solids, and did not consider accretion
or destruction of those bodies.  In a real protoplanetary disk, these
disk and particle characteristics will change over time, often on very
different timescales, and will affect how the water distribution evolves.
In addition, \citet{cuzzahn04} assumed that
there was an infinite reservoir of solids in the outer nebula, providing enough
material for the inner nebula to become significantly enhanced (up to several
orders of magnitude, depending on parameters), though the authors state that
this assumption requires detailed numerical testing.

The work reviewed thus far has provided insight into the 
different processes that would affect the distribution of water vapor in 
the solar nebula and has offered explanations of how Jupiter's core formed
so quickly and of the origin of the different oxidation states recorded by 
the various meteorite classes.  The goal here is to develop an internally consistent model
which incorporates the concepts explored in these earlier papers and
is more consistent with detailed studies of particle growth in protoplanetary disks.

In developing the model presented here, we, too, had to make some
simplifying assumptions as will be discussed below.
The variables needed to describe the model are listed
in Table 1.  Despite the attempts to make this model completely 
self-consistent and dependent only on well understood physics, a set of
parameters must be introduced, which, unfortunately, 
are relatively unconstrained by observations (Table 2). They represent estimates 
as to what the structure of a
protoplanetary disk may be and the evolutionary processes that
occur within it.  They are assumed to be constant in both space
and time (unless otherwise stated), though there is no reason to believe
that is the case.  Due to these uncertainties, 
the results presented here do not represent specific, quantitative
predictions of the behavior of water in a protoplanetary disk.  Instead,
they should be interpreted as global trends for disks in general
which provide us with a context within which we can understand how materials
are distributed in protoplanetary disks and, hopefully, help interpret
some observations.  Specific applications will 
be discussed in detail at the end of this paper.

\section{Evolution of the Nebular Gas}

As described above, astronomical observations suggest that protoplanetary
disks are evolving objects, with mass being transported within them 
\citep{calvet00,calvet05}.  The exact driving mechanism for
this evolution is unknown and is the subject of ongoing research.  Proposed
mechanisms include turbulence induced viscosity 
\citep{shak73,pringle81,linpap,rudpoll91,stepinski98} where the turbulence
arises from either magnetohydrodynamical or
shear instabilities \citep{bh91,gammie96,hure01,davis03,richard04} or 
wave-driven gravitational torques 
\citep{laughroz96,boss04}.  It is possible that more than one mechanism played a role
at various times and locations over the lifetime of a disk.

Here we use the $\alpha$-viscosity model to 
describe the evolution of the disk over the course of millions of years.  This
is done without
prejudice as to the source of $\alpha$ and in lieu of the $\beta$-viscosity 
model \citep{richard99,richard04} for simplicity and in order to facilitate 
comparisons to other work.  
This model assumes that the evolution of the disk is caused by a viscosity that
arises as a result of turbulent motions within the gas.  In this prescription, 
the surface 
density of the gas is governed by the equation:
\begin{equation}
\frac{\partial \Sigma}{\partial t} = \frac{3}{r}\frac{\partial}{\partial r}
\left[r^{\frac{1}{2}}\frac{\partial}{\partial r}\left(\Sigma \nu
r^{\frac{1}{2}}\right)\right]
\end{equation}
where $\Sigma$ is the total surface density of the disk, $\nu$ is the kinematic
viscosity, and $r$ is the radius from the central star (the star is 
assumed to grow negligibly in mass over the lifetime of the disk).  

In the $\alpha$-disk model, the kinematic viscosity is assumed to be a function
of nebular parameters and can be described by the equation:
\begin{equation}
\nu = \alpha c_{s} H
\end{equation}
where $c_{s}$ is the local speed of sound and $H$ is the local nebular scale
height.  The turbulence parameter $\alpha$ is assumed to be some value,
generally much less than 1 \citep{linpap,rudpoll91}. For simplicity, and due
to the few constraints placed on it, $\alpha$ is generally assumed to be
both temporally and spatially constant.

The local speed of sound is given by:
\begin{equation}
c_{s} = \left( \frac{\gamma k T_{m}}{\overline{m}} \right)^{\frac{1}{2}}
\end{equation}
where $\gamma$ is the ratio of specific heats for the gas (taken to be 1.4
here), $k$ is Boltzmann's constant, $T_{m}$ is the temperature of the nebula
at the midplane, and $\overline{m}$ is the average molecular weight of the gas
(taken to be 4$\times$10$^{-24}$ g to account for the gas being composed of
mainly H$_{2}$ and He).  The disk scale height is related
to the speed of sound by:
\begin{equation}
H = \frac{c_{s}}{\Omega}
\end{equation}
where $\Omega$ is the angular rotation frequency of a Keplerian orbit at the
radial distance of interest.

The thermal structure of the nebula is determined by balancing the combination
of the local viscous
dissipation and stellar irradiation with the radiative energy loss from 
the surfaces of the disk.
That is:
\begin{equation}
2 \sigma \left(T^{4}_{e}-T^{4}_{irr}-T^{4}_{amb}\right) = \frac{9}{4} \Sigma \nu \Omega^{2}
\end{equation}
where $\sigma$ is the Stefan-Boltzmann constant and $T_{e}$ is the effective
temperature of the surface of the disk.  Here, $T_{irr}$ is the temperature
at the surface of the disk determined by irradiation from the central star and
$T_{amb}$ is the ambient temperature from the local environment in which the
disk is located.  The irradiation temperature is determined by assuming that
the disk is flat:
\begin{equation}
T^{4}_{irr}\left(r\right) = \frac{2}{3\pi}T^{4}_{*}\left(\frac{R_{*}}{r}\right)^{3}
\end{equation}
where $T_{*}$ and $R_{*}$ are the surface temperature and radius of the central
object, taken to be 4000 K and 3$R_{\odot}$ respectively \citep{rudpoll91}.
The ambient temperature, $T_{amb}$ is assumed to be some low constant value,
set to 20 K in the work presented here to represent minimal effects from
outside sources.  By using a low value for the assumed ambient temperature we
are concentrating on how the disk evolves due to the effects of internal 
processes.  It has
recently been suggested that the sun may have formed in close proximity to 
a massive star, which would have led to a hotter ambient radiation field (and thus
higher midplane temperatures) as well as other external perturbations \citep{hester05}.  Such
effects should be considered in future work.

The effective temperature can be
related to the midplane temperature by the equation:
\begin{equation}
T_{m}^{4} = \frac{3}{4} \tau T_{e}^{4}
\end{equation}
where $\tau$ is the optical depth from the midplane of the nebula to the
surface.  This is given by:
\begin{equation}
\tau = \frac{\kappa \Sigma}{2}
\end{equation}
where $\kappa$ is the local opacity of the disk.  

For these initial models we wanted to adopt an opacity prescription 
which was simple, and avoided the huge uncertainties currently facing 
most treatments. The opacity in question is a Rosseland mean opacity; 
that is, wavelength averaged and weighted by local thermal radiation. 
Many models simply adopt the Rosseland mean opacities of 
\citet{pollack94}, which display  explicit temperature dependence 
for two reasons. First, absorption edges appear at
temperatures where different major species (water, organics, silicates)
evaporate and the opacity thereby drops; second, a different kind of
temperature dependence appears at low temperatures because of the 
decreasing absorption efficiency of the grains they modeled as the effective wavelength 
gets much larger than the grain size. Pollack et al (1994) assumed a composition 
of roughly 1/3 water, 1/3 moderately refractory organics, and 1/3 
silicates; this leads to a negligible opacity drop where water evaporates 
but large  ones where the postulated organics and silicates evaporate. 
The  most frequently used results of Pollack et al assume a MRN size distribution
\citep{mrn77}, 
which is a steep powerlaw extending only to 5 micron radius, but in which 
practically all the scattering mass and area is in the smallest particles - 
hence the low-temperature, long-wavelength temperature dependence. Cassen (1994)
simplified the results of Pollack et al., retaining only the silicate
evaporation opacity drop (and the low-temperature dependence) - thus 
still assuming tiny grains ($<<$30 micron radius), which sets the magnitude of 
the opacity.

Clearly, since $\kappa \sim \pi a^2/m \sim 1/a \rho$, opacities drop as 
grains grow. \citet{miyake93} illustrate how particle growth affects 
grain opacity. Once particles become opaque to thermal radiation (in their 
cases, around 10 micron radius), opacity drops by an order of magnitude for 
each order of magnitude change in radius. Moreover, all grain accretion models 
indicate that grain growth from ISM grain size to {\it at least} 100 micron 
radius is quite easy and quite rapid in the nebula, because of the low relative
velocities and high densities (Weidenschilling 1997, Cuzzi and 
Weidenschilling 2005). Thus one can surely overestimate opacities significantly by 
adopting opacities that simply assume micron-sized grains. On the other hand, 
these initial aggregates are also probably quite 'fluffy' so their individual 
{\it mass} is much smaller than one might expect for a solid particle of the 
same size; that is, if they grow as fractals \citep{meakdonn88,beck00}, the 
product $a \rho$ is the same as that for an independent, micron-size monomer
constituent and their opacity is the same. Compaction, of course, 
will change this, so growth to larger sizes has a very large potential effect. 
Particle radii of several cm, still included in our ``dust" population, are 
quite probably compacted and have opacities orders of magnitude smaller than 
those frequently used (Pollack et al 1994, Cassen 1994). Complicating things 
still further is the as-yet unmodeled evolving size distribution itself over 
these long times and large radial ranges - including particle growth, 
collisions, erosion, and destruction, which are here simply modeled by our three 
crude size bins and the total amount of mass in the ``dust" population itself (the 
opacity of the migrators and the planetesimals is negligible). Point models 
such as Weidenschilling (1997) would be prohibitive to run in a model such as 
ours. Yet, this evolution surely does occur. Miyake and Nakagawa (1993, 
figure 10) show that the opacities for size distributions of realistic, porous 
particles ranging in 
size from 0.01$\mu$ to 10cm radius vary by three orders of magnitude, even 
depending on the slope of their powerlaw size distribution (different slopes 
emphasize either the smaller or the larger particles). The several cm$^2$g$^{-1}$ 
(per gram of nebula gas, that is) of Pollack et al (1994) and Cassen (1994) 
are at the high end of this range.

Where, then, does this leave us regarding an opacity model for our 
evolving disk? Fortunately, the midplane temperature depends only weakly on the 
total optical depth ($T_m \propto \tau^{1/4}$); factors of several will only 
change the timing and location of the water evaporation front we want to model, 
not change  the fact that there {\it is} a water evaporation front. Our focus 
is not on its specific location, but the phenomena associated with its presence. 
The likelihood of  significant particle growth leads us to doubt the reality 
of the $T^2$ dependence found by Pollack et al (1994) and assumed by 
Cassen (1994), which is a tiny-particle property, so we will assume a 
temperature-independent opacity per gram of {\it solid material}. 
Results of Weidenschilling (1997, 2000) using detailed particle 
growth models, tend to find the ``dust" regime (microns to tens of
centimeters radius) has a mass distribution $m(a)$ with equal mass per 
bin of width $a$ (equal to the particle radius); that is, then, $n\left(a\right)a^4$ = 
constant, or  $n\left(a\right) \propto a^{-4}$. From figure 10 of 
Miyake and Nakagawa (1993) we see  that, {\it very} roughly, the opacity of such
a distribution, ranging from microns to centimeters,
can be approximated by something like Cassen's adopted 5 cm$^2$g$^{-1}$ 
(because the steep powerlaw emphasizes particles at the small end of the range), 
over the wavelength range 1-100$\mu$m  characterizing nebula blackbody 
radiation.  This opacity 
assumes a full nebula relative abundance of solids (ice plus silicate $\sim$
1\% by mass assumed by Miyake and Nakagawa). We will adopt this value, with 
the advantage of making comparisons with prior work easier. It does indeed 
seem to be consistent with a realistic case. Thus, our local opacity is given 
by
\begin{equation}
\kappa = 5 \left[\frac{\Sigma_d \left( \textrm{H$_{2}$O} \right)}{0.0045 \Sigma} \right] ~{\rm cm}^2{\rm g}^{-1}
\end{equation}
where the opacity is given per gram of nebula gas having local surface 
mass density $\Sigma$.  Because we are only tracking the ice component of
material in the outer disk, the opacity relation is scaled by the ice mass
fraction \citep[0.0045,][]{krot00}, and
the assumption is made that the other solids closely follow the ice distribution.
The dust population can be enhanced or depleted relative to ``cosmic 
abundance." As noted earlier, the opacity of the migrator 
and planetesimal populations is negligible. Inside of the water evaporation 
front, only silicate (and refractory organic) dust remains, which we do not
specifically track in our model. Thus between the ice and silicate evaporation
fronts we assume cosmic abundance of refractory materials, all in the ``dust" 
size bin, with an average mass density (silicate + refractory organics/2) of 
$ \sim 0.005 \Sigma$, giving an opacity of 2.5 cm$^2$g$^{-1}$. More complicated 
models can be imagined, in which the very process which brings migrators inside 
the water evaporation front to evaporate also releases additional 
refractory dust \citep{cuzcatpd}, but we will not model 
that in this paper. To help with numerical stability while keeping the 
models plausible, a small opacity (0.1 cm$^2$g$^{-1}$) is retained inside 
of the silicate evaporation front (1350-1400K). Physically, this is associated 
with (a) the refractory Calcium-Aluminum-rich material, about 5\% of all 
silicates by mass, which survives to approximately 2000K (inside of which 
our model is insensitive to the opacity), and (b) some small amount of 
normal silicate material which will remain solid at high elevations, where 
the temperature decreases below the silicate condensation temperature.

One final refinement is added, because it deals expressly with the 
water evaporation front. The strong enhancement of water vapor just inside 
the water evaporation front, due to migrators evaporating near the 
{\it midplane}, will lead to  condensation of additional ``snow'' 
at high elevations there where temperatures  decrease below the water 
condensation temperature \citep{cassen94,davisfront05}. 
Because of the large amounts of water
involved, this effect can result in noticeable opacity increases and 
outward displacement of the water evaporation front relative to a 
case where it is neglected.  To model this effect we determine what surface
density of ice-rich gas is needed in order to produce a midplane temperature
of 160 K.  The opacity of that gas is assumed to be given by the above
relation, with the water in the vapor phase also contributing to the
opacity (at the higher elevations this vapor will indeed condense out to form
``snow'').  We determine the amount of gas needed by setting $T_{m}$=160 K,
and using equations 5-8 to solve for the needed surface density.  This represents
the column density of material at the very upper layers of the disk which would
contain condensed ice.  The opacity of the rest of the gas in the column would
be provided only by the silicates in this region, as temperatures would be
high enough for water to exist only as a vapor which does not contribute significantly
to the opacity.  The opacity for the column is then
found by a weighted average:
\begin{equation}
\kappa_{mean} = \left(2.5 \Sigma_{T>160} + 5 
\frac{\left(\Sigma_{vap}+\Sigma_{d}\right)}{0.0045 \Sigma} \Sigma_{T<160} \right)
\frac{1}{\Sigma}
\end{equation}
where $\Sigma_{T<160}$ is the surface density of material at temperatures below
160 K, and $\Sigma_{T>160}$ is the remaining surface density at a given location
of the disk.

\section{Evolution of the Water Distribution}

In this model, water exists in two phases: as vapor and as solids.
The solid material considered here is divided into three categories: dust,
migrators, planetesimals.  
Each of these categories represents a range
of sizes of the solids, characterized by some \emph{typical size}, 
and the evolution of these categories is calculated by assuming all the mass
in each category is contained in particles of that single size.  
Here we aim to show the general evolution
of how water vapor and ice could be distributed in protoplanetary disks by
modeling all of the physics that control the evolution of the four
species which have the most distinct dynamical behavior. Thus our model is an
improvement over the single-size model at a given location 
of \citet{stepval97}, but is simpler
to evolve than the full size distribution of \citet{weidenschilling97} (or
other similar studies) and 
accounts for the diversity of physical behaviors involved.

In general, water vapor will be transported by the same processes that 
transport the rest of the nebular gas (diffusion and advection).  The 
transport of solids, on the other hand, will depend on the size
of the solids under consideration.  The dynamical evolution of a solid
is determined by its Stokes number, $St$, which is the ratio of the
particle stopping time to
the turnover time of the largest local turbulent eddy (usually taken to be the
local orbital period). The stopping time, $t_{s}$,
measures the amount of time it takes for a particle to lose its relative
velocity with respect to the gas and is given by 
$a \rho$/$\rho_{g}c_{s}$ for small particles ($a << \lambda$) and 
$a^{2} \rho$/$\rho_{g}c_{s}\lambda$ for larger objects, 
where $a$ is the radius, $\lambda$ is the mean free path in the gas, 
and $\rho$ is the density of the particle.  
Dust particles represent those solids with $St <<$1, which covers solids
with $a <$1 cm.  Because of their short stopping times, these particles 
are very well coupled to the gas.
Planetesimals are those solids with $St >> 1$, corresponding to bodies
larger than 1 km. Due to their large inertia, planetesimal motions
are relatively unaffected by the presence of the gas.  
Migrators are those particles in between with $St \sim$1, corresponding to 
objects roughly 1 meter in size \citep{cuzzweid04}. These bodies are most 
dramatically affected by
gas drag as will be discussed below.

Here we present the equations which describe how water vapor, dust, migrators, 
and planetesimals behave in a viscous accretion disk.  We first present
the equations which describe the transport of the various species, with each 
equation including a term representing the net sources and sinks for
the respective species.  The sources and sinks represent the transfer of
mass from one species to another through accretion, collisional destruction,
vaporization, or condensation, as described below.

\subsection{Water Vapor}

The major transport mechanisms of water vapor will be advection and
diffusion caused by the same viscous interactions which control the
global evolution of the nebula.  The surface density of the water vapor is
thus described by:
\begin{equation}
\frac{\partial \Sigma_{vap}}{\partial t} = \frac{3}{r}\frac{\partial}{\partial r}
\left[r^{\frac{1}{2}}\frac{\partial}{\partial r}\left(\Sigma_{vap} \nu
r^{\frac{1}{2}}\right)\right]
 + S_{vap}\left(r,t\right)
\end{equation}
where $\Sigma_{vap}$ is the surface density of water vapor,
$r$ is the distance from the sun, and
$S_{vap}\left(r,t\right)$ represents the sum of the source and sink terms for the
vapor at a given location and time \citep{stepval96,stepval97}. 
Here, the diffusivity is assumed to be the same as the viscosity, $\nu$, 
for the evolution of the nebula used in equation (1).  That is,
we assume a Prandtl number of 1 \citep{cuzzahn04}.
Note that in some previous studies \citep{stevenson88,cyr98,cuzzahn04}, the evolution of the
\emph{concentration} of water vapor (or solids) was tracked rather than the
surface density.  The concentration is defined as the mass of water present
relative to hydrogen ($\Sigma_{vap}$/$\Sigma$ in the inner disk).  Here we follow
\citet{stepval97} by explicitly tracking how the surface densities of the
various water species evolve.

\subsection{Dust}

\emph{Dust} refers to solid material that is strongly coupled to the
gas ($St << 1$),    and thus
these particles are assumed to move with the gas and their motion is described by a 
similar equation as that which describes the motion of the water vapor:
\begin{equation}
\frac{\partial \Sigma_{d}}{\partial t} = \frac{3}{r}\frac{\partial}{\partial r}
\left[r^{\frac{1}{2}}\frac{\partial}{\partial r}\left(\Sigma_{d} \nu
r^{\frac{1}{2}}\right)\right]
 + S_{d}\left(r,t\right)
\end{equation}
where $\Sigma_{d}$ is the surface density of the dust.
Because the dust particles are so well coupled to the gas, the diffusivity
with which they are transported is again the same as the viscosity of the 
nebula and diffusivity of the water molecules.

\subsection{Migrators}

As bodies grow larger in the disk, they are still affected by the turbulence
in the nebula and diffuse along concentration gradients, though to a lesser 
extent than dust or gas.  The diffusive transport of the migrating population 
is described by:
\begin{equation}
\frac{\partial \Sigma_{m}}{\partial t} = \frac{3}{r}\frac{\partial}{\partial r}
\left[r^{\frac{1}{2}}\frac{\partial}{\partial r}\left(\Sigma_{m} \nu_{m}
r^{\frac{1}{2}}\right)\right]
 + S_{m}\left(r,t\right)
\end{equation}
where $\Sigma_{m}$ is the surface density of migrators 
and $\nu_{m}$ is the effective diffusivity of the migrators, given by
\begin{equation}
\nu_{m}= \frac{\nu}{1+St}
\end{equation}
\citep{cuzzi93,cuzhog03,cuzzweid04}.  Thus for the migrators, $\nu_{m}$ = $\nu$/2.

The lower level of diffusion for these bodies is due to the fact that the 
migrators are not as coupled to the gas as the dust particles are.  \citet{adachi76}
and \citet{weid77mig} showed that as these bodies orbit the sun, they 
would experience headwinds in their orbits due to the fact that the
gas in the disk rotates slower than the Keplerian rate.  These headwinds caused
the particles to drift inward towards the sun.  In a non-uniform
nebula, this same effect may cause objects to migrate outwards if local
pressure maxima exist \citep{haghi03}.
The movement of these 
bodies with respect to the gas is described by the mass flux equation:
\begin{equation}
\frac{\partial \Sigma_{m}}{\partial t} = -\frac{1}{r}
\frac{\partial}{\partial r}\left(r v_{drag}\Sigma_{m}\right)
\end{equation}
where $v_{drag}$ is the inward migration velocity with respect to the gas. 
It is assumed that the
migrating population moves with the maximal radial drift (which equals
the difference between Keplerian velocities and the orbital velocity of the gas):
\begin{equation}
v_{drag}= \frac{\Delta g}{2g} V_{k}
\end{equation}
where $g$ is the acceleration due to the sun's gravity at a given location,
$V_{k}$ is the local Keplerian velocity, and $\Delta g$ is the difference in
gravitational acceleration felt by a solid object and a parcel of gas at
a given location, given by:
\begin{equation}
\Delta g = \frac{1}{\rho_{g}}\frac{dP}{dr}
\end{equation}
where $dP$/$dr$ is the local pressure gradient of the nebular gas.  Typically,
the pressure gradient results in meter-sized bodies migrating inwards at
$\sim$1 AU/century, with only a weak dependence on 
location in the disk \citep{cuzzweid04}.

An important point to stress here is that it is the Stokes number of the
body which determines the category of dynamic behavior that the solid will
belong to, not its size.  The Stokes number is proportional to the product of
the density of the object and its radius.  Thus a consolidated meter-sized
object will behave differently than a fluffy, meter-sized fractal
because the effective densities would be very different.  In the case of
the fractal aggregate, the density would likely be so low that it would fall
into the \emph{dust} category ($St << $1).  

\subsection{Planetesimals}
Planetesimals (assumed to be bodies with radii of 0.5 km in our simulations) 
move on nearly Keplerian orbits in the nebula.  While they 
experience a headwind similar to that of the migrating bodies, 
the resulting drag is not strong enough to cause them to migrate 
significantly due to their large
inertia.  Thus, we assume that the orbits of the planetesimals
do not change over the course of the simulations presented here.
The equation describing the evolution of the planetesimal surface density is
thus simply:
\begin{equation}
\frac{\partial \Sigma_{p}}{\partial t} = S_{p}\left(r,t\right)
\end{equation}
where $\Sigma_{p}$ is the surface mass density of planetesimals and
$S_{p}\left(r,t\right)$ is the source function of the planetesimals.

\subsection{Sources and Sinks}
At the same time that diffusion, migration, and advection are moving the 
different species around the nebula, the vapor, dust,
migrators and planetesimals will be interacting
with one another and their environment to transform from one species to
another.  Below we describe how these transformations take place and
quantify the rates at which mass is exchanged between the different species
that we are tracking.

\subsubsection{Vaporization and Condensation}

As icy solids are transported into the warm regions of the disk
they will begin to vaporize.  To
account for this, after we solve for the temperature profile of the nebula,
we can calculate the equilibrium vapor pressure, $P_{eq}$, for water 
everywhere in the nebula.  If the actual vapor pressure, 
$P_{vap}$, is less than $P_{eq}$, and if solid ice is present, 
then some of the ice mass will be 
converted to vapor.  The rate at which this will occur is given by 
\citet{icevap91}:
\begin{equation}
Z\left(T\right) = \frac{a_{1}}{\sqrt{T}}e^{-\frac{a_2}{T}},
\end{equation}
where $a_{1}$ = 7.08 $\times$ 10$^{31}$ cm$^{-2}$ s$^{-1}$ K$^{\frac{1}{2}}$
and $a_{2}$=6062 K.  The solid mass loss rate is given by:
\begin{equation}
\frac{dm}{dt}= -4 \pi a^{2} \mu m_{p} Z\left(T\right),
\end{equation}
where $a$ is the radius of the particle of interest, $\mu$ is the molecular
weight of the water molecule, $m_{p}$ is the mass of a proton, and $k$ is
Boltzmann's constant.  In this formalism, the equilibrium vapor pressure,
$P_{eq},$ is:
\begin{equation}
P_{eq} = Z\left(T\right) \sqrt{2 \pi \mu m_{p} k T}.
\end{equation}

We can calculate the actual water vapor pressure at the midplane through
the formula:
\begin{equation}
P_{vap}=\frac{\Sigma_{vap}}{2H}\frac{kT}{m_{H_{2}O}}.
\end{equation}

When the criteria for vaporization is met ($P_{vap} < P_{eq}$), we first 
convert the dust needed to vapor in order to increase $P_{vap}$ to $P_{eq}$.
If there is not enough dust present to do this, we then calculate the
evaporation rate for the migrating bodies and convert the mass lost in a
given timestep to vapor.  In a timestep the change in surface density of
the migrators as a result of vaporization is given by:
\begin{equation}
\Delta \Sigma_{m}^{vaporize} = \Sigma_{m} \frac{dm_{m}}{dt} \frac{\Delta t}{m_{m}}
\end{equation}
with the water vapor surface density increasing by an equal but opposite amount.
In this expression, $\Delta t$ is the timestep taken in the model simulations.
We ignore the vaporization rate of the dust particles by treating it as
instantaneous because of their high
surface area-to-volume ratios.

If $P_{vap}$ is greater that $P_{eq}$, then water vapor will condense out 
of the nebula to form solid ice.  In our model, we assume that excess water
vapor is removed from the gas to maintain an equilibrium vapor pressure.
The excess water vapor is assumed to go into the form of dust.
Thus, the excess water vapor is simply added to the dust surface density
at each location in the nebula at the end of each time step.  
The amount of dust added at each timestep is:
\begin{equation}
\Delta \Sigma_{d}^{condense} = (P_{vap}-P_{eq})2H\frac{m_{H_{2}}}{kT}
\end{equation}
and an equal amount is lost from the vapor phase.  

\subsubsection{Growth of Migrators}

In addition to phase transitions, the solid particles will interact with one 
another through collisions.  There are two types of collisions which we 
consider: accretionary and destructive.  Initially, the disk is filled with dust particles
from which the original generation of migrators is formed.  These migrators would
grow through the coagulation of dust particles.  To account for this, we follow
\citet{cassen96} and define a parameter,
$t_{coag}$, which is the coagulation timescale at a given location
in the disk.  As discussed by \citet{cassen96}, to make this model consistent
with the findings of accretion simulations, this timescale is a function of
orbital velocity and thus follows the relation:
\begin{equation}
t_{coag} \left( r\right)= t_{coag} \left( 1 ~\textrm{AU} \right) 
\frac{\Omega\left(1~\textrm{AU}\right)}{\Omega\left(r\right)}
\end{equation}
We assume that in a given time interval, the surface density of new migrators will
grow as:
\begin{equation}
\Delta \Sigma_{m}^{coag} = \Sigma_{d} 
\left(1 - e^{-\frac{\Delta t}{t_{coag}}}\right)
\end{equation}
and the surface density of dust decreases by an equal amount.

The parameter, $t_{coag}$, represents the average timescale for migrators to
grow from the local dust supply.  This timescale is set by the competing 
effects of particle growth from collisions that arise due to brownian motion,
turbulence, or differential drift and particle destruction or erosion
from energetic collisions.  The importance and efficiency of these processes
is uncertain \citep{weidcuzz, weidenschilling97,dull05,cuzzweid04}.  Here we assume that
$t_{coag}$ ranges from 10$^{3}$-10$^{5}$ years at 1 AU, consistent with 
timescales estimated by other authors \citep{cassen96,beck00}.

In addition, if migrators already exist they can grow as they sweep up the dust
particles suspended within the gas.  The rate at which the surface density of the
migrators grows as a result of this can be calculated as:
\begin{equation}
\Delta \Sigma_{m}^{acc} = 
\frac{\pi a_{m}^{2} v_{r} \Sigma_{d} \Sigma_{m}}{H_{d}m_{m}} \Delta t
\end{equation}
where $v_{r}$ is the total relative velocity of the migrators with respect 
to the gas, $m_{d}$ is the average mass of the dust particles, and $H_{d}$ is
the thickness of the dust layer in the nebula.   
We assume $H_{d} \sim  H$ for the purposes of our modeling, meaning that 
little settling has taken place for the dust, due to its low Stokes number.
As the migrators grow, the surface density of dust decreases by an equal amount.

\subsubsection{Destruction of Migrators}

While the above accounts for the collisions between the dust and the migrators,
we must also account for the mutual collisions between migrators once they
have formed.  
We assume that when two migrators collide, they erode one another and 
some fraction of their mass is converted into smaller particles (thus
acting as a sink for migrators and a source for dust). The collision
rate can be calculated such that the surface density lost by 
disruption of the migrators (and gained by the dust) is given by:
\begin{equation}
\Delta \Sigma_{m}^{disrupt} =
-\frac{f \pi a_{m}^{2} \Sigma_{m}^{2} v_{t}}{2H_{m}m_{m}}\Delta t
\end{equation}
where $a_{m}$ is the assumed \emph{average} radius of the migrating particles
(50 cm),
$v_{t}$ is the relative velocity that migrators would have with one another 
due to local turbulence ($v_{t} \sim \sqrt{\alpha}c_{s}$), $m_{m}$ is the
average mass of a migrating particle, $f$ is an efficiency factor, and
$H_{m}$ is the thickness of the 
layer in which the migrators have settled in the nebula.  Because the migrators
have $St \sim 1$, this layer is given roughly by:
\begin{equation}
H_{m} = \sqrt{\frac{\alpha}{2}}H
\end{equation}

The efficiency factor, $f$, represents the fraction of mass that is converted
to dust in a collision.  \citet{benzasp99} showed that bodies of different
sizes have characteristic strengths, $q^{*}$, that determine how they behave
in a collision.  The value of $q^{*}$ represents the energy per unit mass needed
in a collision for a given target to be broken into a number of pieces where
the largest piece is half the mass of the original target ($q^{*} \sim 10^{6}$
erg/g is typical for 1 meter-sized objects).  Thus, particles
in a range of sizes will be produced as a result of the collisions.  To account
for this, we set the efficiency factor, $f$, such that
\begin{equation}
f = \left\{ \begin{array}{ll}
2 & \textrm{if } \frac{1}{2}v_{t}^{2} \ge 2q^{*} \\
\frac{ke}{q^{*}} & \textrm{if } \frac{1}{2}v_{t}^{2} < 2q^{*} 
\end{array} \right.
\end{equation}
where $\frac{1}{2}v_{t}^{2}$ is the kinetic energy per gram of a migrator
in a collision. 
For example, if $\frac{1}{2}v_{t}^{2}=q^{*}$, the energy imparted into one of
the migrators in the collision would be equal to its characteristic strength.
This energy is enough to break up the migrator into pieces such that the largest
remnant would be half the mass of the original body.  Because of the symmetry
of the collision, the bodies would each produce objects that were half the mass
of their parent bodies, summing together to the mass of an original migrator.  
The rest of the mass is assumed to be converted into dust particles.  More
energetic collisions result in more of the mass being converted into dust,
while less energetic collisions are less erosive.
The surface density of the dust particles increases by the same amount
that the surface density of the migrators decreases.

\subsubsection{Growth of Planetesimals}
 How planetesimals formed in the solar nebula is unknown and has been 
 debated for some time.  Some suggested formation mechanisms
include: formation through collisions of particles \citep{weidenschilling84,
weidenschilling97}, turbulent concentration of particles \citep{cuzzi01},
trapping of particles in eddies \citep{klahr97}, buildup of particles
in pressure maxima in a non-uniform nebula \citep{haghi03}
and gravitational instabilities in a dust layer \citep{goldward73,sekiya98,youdin04}.
Each of these mechanisms requires different nebular conditions in order to
work (for example, some require negiligible turbulence to enable settling while
others require higher levels of turbulence to concentrate particles
near eddies), and there is much debate which of those different conditions existed 
within the solar nebula.  Some are formally inconsistent with the
(turbulent) situation modeled here.

The goal of this work is not to prove or disprove any of these
mechanisms for the formation of planetesimals, but rather
to provide insight into the \emph{effects} of planetesimal formation and 
of the timescale over which these objects form.  Thus
here we introduce the parameter $t_{acc}$, which is the timescale for
planetesimals to form from the available solids in the nebula.  We assume
that the planetesimals have some birth function which describes how new 
planetesimals
grow from the migrator population in a similar way that we account
for the birth of migrators by dust coagulation, and that
$t_{acc}$ has the same radial dependence as does $t_{coag}$.  While the birth
functions of the migrators and planetesimals have similar forms, the
timescales are not
necessarily equal as the physical processes responsible for the formation of
each type of object were probably very different.  Thus $t_{coag}$ and $t_{acc}$
are not necessarily equal, but detailed simulations suggest that they may fall
in the same range of values \citep{beck00}.
Like
$t_{coag}$, only estimates of $t_{acc}$ can be made, and thus we have 
investigated a wide range of possible values.
The equation which
describes the increase in surface density of the planetesimal population
is:

\begin{equation}
\Delta \Sigma_{p}^{accretion} = \Sigma_{m} \left(1 - e^{\frac{\Delta t}{t_{acc}}}\right)
\end{equation}
 
Furthermore, the existing planetesimals can grow as they
sweep up solids suspended in the disk, which can be an important effect
once the planetesimal swarm grows in surface density.  Here it is assumed
that the planetesimal surface density grows through the accretion of 
solids suspended in the nebula such that the surface density of 
migrating bodies and dust lost from the nebula (and accreted onto the
planetesimals) is:
\begin{equation}
\Delta \Sigma_{m}^{-,p} = -\frac{\pi a_{p}^{2} \Delta v_{p,m} \Sigma_{m} 
\Sigma_{p}}{H_{m}m_{p}} \Delta t
\end{equation}
\begin{equation}
\Delta \Sigma_{d}^{-,p} = -\frac{\pi a_{p}^{2} \Delta v_{p,d} \Sigma_{d} 
\Sigma_{p}}{H_{d}m_{p}} \Delta t
\end{equation}
and the surface density of the planetesimals increases by an equal, but
opposite, amount.  Here $\Delta v_{p,d}$ and $\Delta v_{p,m}$ represent 
the difference in rotation velocities between a Keplerian orbit and the 
dust and migrators in the nebula at a given location and $a_p$=0.5 km.  
It is assumed the
dust moves with the gas, and thus, the difference in velocity between
the planetesimals and dust, $\Delta v_{p,d}$, is equivalent
to the difference in the orbital velocity of the planetesimals and that of
the gas.  Migrators do not just
move with the gas, but instead move faster in their orbits, and therefore 
the difference in velocity between the
planetesimals and migrators is $v_{drag}$/2 \citep{weid77mig}.  

While more detailed collisional and accretional models have been developed
than those used here,
they have generally focused simply on the formation of planetesimals, not
necessarily the overall distribution of solids in the nebula or their 
movement across condensation fronts.  Furthermore, they are prohibitive to
run over wide ranges of space and time.
Here we focus instead on the general evolution of the 
nebula and the water species inside it by considering the overall motions 
of the particular species of interest.  By selective changes of parameters
we show which are the most important and thus of interest for future study.
More detailed treatments of the 
collisional destruction and accretion of the solids will be the subject 
of future work.

\subsection{Method of Calculation}

In performing the calculations above, we assume that the evolution of the
solids in the disk only affects the evolution of the gas by determining the
opacity of the disk.  That is, there is no angular momentum or energy 
exchanged between the hydrogen-rich gas and the water species.  Thus, we initially solve
for the global evolution of the nebula, first by rewriting the viscosity,
$\nu$, as a function of $\Sigma_{g}$ and $r$.  This allows equation (1) to
be solved by backwards-time finite-differencing.  Once the surface density
profile of the nebula is known for the next time step, the transport equations
for each water-bearing species are solved for the same time interval, again
with backward-time finite differencing.  The
sources and sinks of the species are then determined and the profiles are
updated accordingly.  Once all quantities have been determined, the equations
are solved for the next time interval.

While the backwards-time differencing methods used in solving these equations
are stable for any timestep, numerical problems may arise if too large
a timestep is used when calculating the source and sink terms. 
Thus the timesteps used in these calculations never
exceed half of a year, a time interval that is short compared to the coagulation and
accretion timescales, as well as the timescale for significant transport
of material in the nebula.  The code was developed to run in parallel on 
the SGI Origins clusters at NASA Ames and Goddard.  A single simulation took 
between 1000-3000 computer hours, depending on the extent of the grid needed
to contain the disk.

In doing these calculations, it is critical to set appropriate boundary
conditions.  For the inner boundary of the nebula, we assumed
that the surface density was equal to zero at the center ($r$=0) so that material
would fall directly onto the star.  The outer boundary of the grid 
was chosen such that the expanding outer edge of the disk was always contained
within the grid.  The density at the outer boundary of the grid was held at a
fixed, negligible value (representing interstellar medium).  Initially, 
outside of
$R_{0}$ (the assumed ``edge'' of the disk at $t=0$) 
the surface densities were set such that $\Sigma$=10$^{-20}$ g/cm$^2$,
$\Sigma_{d}$=4.5$\times$10$^{-23}$ g/cm$^2$, $\Sigma_{m}$=0 g/cm$^2$, and 
$\Sigma_{p}$=0 g/cm$^2$.  

\section{Model Results}

We have applied our model to investigate how the distribution of water evolves
in protoplanetary disks exhibiting a variety of structures and evolutionary
parameters.  
In carrying out a model run, the parameters given in Table 2 were assigned 
values and the model equations were solved for a period of
3 million years, which is the observed average lifetime of inner disks around
solar type stars in young clusters \citep{haisch01}.  The initial disk has a
surface density $\Sigma$($r$) = $\Sigma_{0}r^{-p}$, where
$\Sigma_{0}$ is the surface density at 1 AU and the disk is truncated at
$R_{0}$.  In reality, the values of these parameters are determined by 
the mass of the molecular cloud from which the star 
and disk form, as well as the initial angular momentum of the starting
material. There is no way to know the values of these parameters for a
disk like our own solar nebula, thus we consider a wide variety of initial
conditions to investigate how different disk structures affect the evolution
of the water distribution within them.   

Table 3 lists the values of
the parameters used for the cases whose results are
shown in Figures 1-7.  While the number of combinations for the different 
parameters is quite large, we present these seven cases in order to identify
what effects the different parameters have on the evolution of the water
distribution. In Figures 1 through 4, the disk is assumed to have a
mass of 0.2$M_{\odot}$ distributed out to 40 AU.  This type of disk is assumed
to be similar in structure to those disks studied by \citet[][and others by
the same author]{boss01}.  In Figures 5 and 6, the disk is assumed to have the
same mass as in the previous cases, but it is distributed initially out to 
100 AU, leading to a less dense and more extended structure.  In Figure 7, the
disk again has the same mass, but the mass is distributed with a more shallow
power law index, making the disk more massive at larger radii
than in the other cases.

The various panels in Figures 1-7 plot the model results at different stages
during the evolution of the disk
(after 10$^{5}$, 10$^{6}$, 2$\times 10^{6}$, and 3$\times 10^{6}$ years).
In some cases numerical instabilities caused us to end a simulation before
reaching 3 million years, and those cases and the reasons for the instabilities
are identified below.  It should be kept in
mind that these times are model times, where $t$=0 corresponds to an arbitrary
point in the evolution of the protoplanetary disk when we begin to track the
physical processes outlined above.  Our $t$=0 is not necessarily meant to be
taken to be equal
to the time of CAI formation or final collapse of the molecular cloud as it is
uncertain when those events would take place in the formation of a protoplanetary
disk.  Instead it can be thought of as a point in time \emph{very} early in the history
of the disk evolution, $<<$10$^{6}$ years since cloud collapse.
Below we discuss the  evolution
of a subset of cases we have modeled, focusing on identifying
how different aspects of the water distribution are affected by changes in
the model parameters.  In the following section we discuss what implications these
results have for the evolution of meteorite parent bodies, giant planet
formation, and our understanding of protoplanetary disk structures.

\subsection{Disk Mass and Physical Structure}

The surface density evolution of the disks in these models are shown in
panel A of each figure.  In all cases, the mass of the disk decreases over
time as material is transported inward by viscous interactions and 
is eventually accreted by the central star.  In order
to balance the effects of the inward mass movement, the disk expands outwards
as small amounts of mass are pushed away from the star in order to conserve
angular momentum.  This latter effect can be seen as the disk grows larger in its
radial extent.  

The rate at which disk evolution takes place is determined by the viscosity
of the disk.  In the $\alpha$-disk model used here, the viscosity is dependent
on the temperature of the disk ($\nu \propto c_{s}H \propto c_{s}^{2} \propto T_{m}$), and
thus, the change in opacity due to material transport can be important.
An extreme example of this is illustrated in Case 1 whose evolution 
is shown in Figure 1.  In Panel A, 
a noticeable kink in the surface density distribution develops early on, 
indicating the location of the snow line in the disk.  As migrators form in
the outer disk, they are carried inwards where they vaporize as they cross
the snow line.  This results in a local increase in the opacity at the snow
line as vapor diffuses upwards and outwards to condense forming small dust
grains.  This region is then less able to radiate away the energy that is created
by viscous dissipation, and the temperature goes up.  Thus the viscosity locally
increases, leading to more rapid mass transport in this region of the disk.
This local increase in opacity comes at the expense of the outer disk, where
the disk is depleted in its dust.  Thus the viscosity of the outer disk 
decreases at the same time.  This leads to the interesting result that the rate
of mass transport in the outer disk quickly declines in the early
stages of evolution, while the inner disk continues to evolve at a rapid rate.  
Thus the surface density of the inner disk decreases more
rapidly, while the evolution of the outer disk is essentially stalled in 
comparison.   \citet{rudpoll91} also found that the decrease in opacity of the
disk would cause the evolution of the outer disk to halt before the inner disk
and report cases where the surface densities of their model disks increased with
distance from the central star.
In our Case 1, this caused us to halt the evolution after 1.5 million
years as numerical instabilities developed in the water transport equations.

This effect was reduced when dusty material was preserved in the outer disk,
preventing large opacity gradients from developing.  This could
be achieved in two ways.  In Case 2, the same disk structure was used as in
Case 1, but the coagulation timescale is set to be an order of magnitude longer.  This
decreased the rate at which dust was removed from the disk, allowing it to 
provide a source of opacity throughout the lifetime of the disk. 
The second method is shown in Case 4, where we again use the same disk structure
as Case 1, but now assume that $\alpha$ is greater by one order of magnitude.
This higher value of $\alpha$ produces more energetic collisions between the
migrators that form in the outer disk, resulting in more mass being transfered
from migrators to dust than in the less turbulent case.  More importantly,
because the disk has a higher viscosity, the diffusivities of the species within
it are greater.  Thus as dust gets concentrated near the snow line to locally
increase the opacity, the higher diffusivity acts to smooth out this concentration 
gradient by rapidly
carrying dust outward again.  Furthermore, viscous evolution is faster 
for higher values of $\alpha$, leading to more rapid advection of dust which
allows regions to be resupplied on shorter timescales.

As expected, the disks expand as mass is accreted through the disk due to
the outward transport of angular momentum.  The timescale over which a disk will
grow due to its viscosity is given by $t_{vis} \sim R^{2}$/$\nu$, where $R$ is
the radial extent of the disk \citep{hartmann98}.  
Thus, smaller disks and more viscous disks will evolve more rapidly, and this
can be seen in the cases presented here.  Also, the viscosity of the
disks decrease over time as the disk cools due to dust coagulation and transport.
The combination of these effects explains why there are differences in the  
evolution of the disks presented here
versus the analytical models of \citet{bitw2} where in the 
case of a uniform disk viscosity the radial extent of the disk grew by a factor
of 4 and decreased in mass by a factor of 40.  The disks in \citet{bitw2} evolve
to a greater degree than those shown here because the disks used here are initially
larger in radial extent and have their viscosities decrease over their lifetimes.
As will be discussed below, 
while  0.1$M_{\odot}$ remain in some of the disks shown here, the amount
of solids left behind may be a much smaller fraction than was originally
available.  Thus, the solid inventory of a disk is not necessarily representative
of the overall disk structure.

In addition to affecting the evolution of the surface density, the
thermal structure of the disk will change as material is transported within
it. As described above, the localized increase in opacity that occurs near the
snow line will raise temperatures as radiation is more efficiently trapped.
The localized enhancement of water generally reaches a maximum of $\sim$5-10, 
increasing the opacity by roughly the same factor.  As the midplane temperature
of the disk goes as the fourth root of the opacity, this would correspond in
a midplane increase by a factor of $\sim$1.5.  Thus if this opacity increase
were to occur at the snow line, where the temperature was $\sim$160 K, the 
increased opacity would cause the temperature to rise to $\sim$240 K.  The
snow line would then migrate \emph{outwards} to a new position where the
viscous energy dissipated in the disk is balanced by the radiative energy of
the disk to produce conditions where ice would again be stable.  Once the
opacity increase slows, the snow line then moves inward over time as the
disk cools due to mass loss, thinning, and opacity decrease due to coagulation
\citep{cassen94,davisfront05}.

\subsection{Water Vapor Concentration}

In all simulations shown here, the concentration of water vapor in
the inner nebula increases over the canonical solar ratio during the early
stages of evolution, as shown in panel E of each of the
figures. In these panels, the concentration of water vapor relative to hydrogen
is plotted,
normalized to the canonical value 
\citep[$\Sigma_{vap}$/$\Sigma$ $\sim$ 4.5$\times$10$^{-3}$ here,][]{krot00}.  
The snow line in each case would correspond to the 
location where the vapor concentration drops immediately in these panels (that
is the near-vertical line in the distribution).
The early enhancements of water vapor result
from the dust in the outer nebula coagulating to form rapidly drifting migrators.
These migrators move into the hotter regions of the nebula and evaporate.  
The resulting vapor builds up just inside the snow line as illustrated
by the 10$^{5}$ year snapshots.  This is because the mass influx of migrators
is greater than the removal rate of the water vapor by 
diffusion and advection.  The vapor that is produced from the migrators 
is eventually spread throughout the inner nebula.
Over time, the inward mass flux slows for two reasons.  
First, the production rate of the migrators 
in the outer nebula declines due to the decrease in the amount of
source material (dust) present, and second, the probability that the migrators
will be accreted by larger bodies rather than surviving their transport
to the inner disk goes up over time as the planetesimal swarm
grows in the outer disk.  The decrease in the
migrator influx then allows the water vapor to be removed from the inner disk
faster than it can be resupplied, leading to a continual decrease in the 
concentration
of water vapor present there.  Much of the water vapor is advected inwards and
accreted by the central star, while some of it is diffused outwards where it
condenses to form ice particles which are incorporated into the planetesimals
there.

An exception to this rule is Case 1 where removal of the dust
in the outer disk, and its subsequent opacity and viscosity decrease, 
drastically slowed the large scale mass transport in the disk, resulting in the
water vapor concentration constantly increasing in the inner disk.  In these cases 
the hydrogen-rich gas from the inner disk was being removed at a faster rate than
it was being resupplied.  Water, on the other hand, was constantly being
resupplied as migrators from everywhere in the disk continued to move inwards
and dust from just beyond the snow line was carried inward with the gas.  This
led to the water vapor-to-hydrogen ratio continuing to increase, reaching 
very high levels over time.  

The evolution of the water vapor described here is qualitatively similar to
the different evolutionary stages identified by \citet{cuzzahn04}, who also
found that the inner nebula would go through a period of enhanced water vapor
concentration.  The maximum level of enhancement predicted was 
$E_{0}$=$2f_{L}$/$3 \alpha$, where $f_{L}$ was the fraction of solid mass 
contained within migrating bodies, which the authors estimated to be $\sim$0.1.
This would lead to a value of $E_{0} \sim$700 for $\alpha$=10$^{-4}$ and
$\sim$70 for $\alpha$=10$^{-3}$.  In most of our simulations, the maximum enhancement
is $\sim$5-10 for $\alpha$=10$^{-4}$ and $\sim$3 for $\alpha$=10$^{-3}$, 
over an order of magnitude less than was predicted.  This difference is partly 
due to $f_{L}$ being less than 0.1 during the early stages of disk evolution
(starting at levels $<$0.01),
though it also reaches values much greater than 0.1 at different locations
in the disk as the disk evolves.

The major reason for the difference in vapor enhancement shown here and that
predicted by \citet{cuzzahn04} is that here the source of the migrating material
is depleted over time.  \citet{cuzzahn04} assumed a steady state--that is, 
the region of the disk 
outside the snow line always contained its canonical value of water.  In our 
model, as migrators are transported inwards, the outer nebula is depleted
in material, reducing the source of the migrators over time.  This results in the
inward mass flux of migrators decreasing over time, rather than being held
constant.  This effect causes the peak enhancement
experienced by the inner nebula to be substantially lower than would be
estimated in steady-state.   The fact that the finite supply of material was the major
factor in determining the maximum vapor enhancement, rather than the increasing
likelihood of being swept up by a growing planetesimal population, 
was confirmed in runs
where we did not allow planetesimals to form, thus removing the possibility
of losing migrators to immobile objects.  In these
cases it was found that the maximum vapor enhancement did not change 
significantly from those cases presented here.  However, if the Case 1 situation
were to keep evolving, it would develop high water vapor enhancements.

As the influx of material from the outer disk continues to slow, the 
concentration of water vapor eventually decreases as diffusive redistribution
begins to dominate the transport.  Diffusion carries the
water vapor outwards where it condenses to form dust grains.  While these dust
grains can then coagulate together to form migrators which will then move
into the warm inner disk again where the processing is repeated, this recycling
is not perfectly efficient.  This is because planetesimals continue to form in the
outer disk, trapping the dust and migrators, preventing some of the water that
diffuses outward from returning to the inner disk.  Also, the water vapor will
also be advected inwards by the average flow of disk (that generated by the
viscous shear that drives the disk evolution) and some of the vapor is then
accreted onto the central star.  These processes combine to deplete the inner
disk in water vapor, causing the enhancement to decrease over time.

Not only is the excess water vapor removed from the inner disk, but diffusion
works to continue to decrease the water-to-hydrogen ratio below the
canonical solar value.  Examples
of such situations can easily be seen in Cases 3, 4, and 7.  In Case 3, 
planetesimal formation happens on such short timescales that the influx of
migrators to the inner disk is halted early during the disk evolution.  As such,
the inner disk experiences only a minor enhancement in its water vapor content
before diffusion begins to dominate the transport.  At that point the 
concentration of water decreases below the solar value and the inner disk
is depleted by a factor of $\sim$10 after nearly 2 million years of evolution.

Cases 4 and 7 also very quickly reach states where the concentration of water
vapor decreases below the solar value.  In these cases, $\alpha$ was assumed
to be 10$^{-3}$, and thus the diffusivity was 10 times greater than in most
of the other cases shown here.  Thus in higher $\alpha$ cases, diffusion 
transports material much faster, resulting in a larger depletion of material.
While Case 6 also had an $\alpha$ of 10$^{-3}$, it did not reach the
same level of depletions that Cases 4 and 7 did.  The reason for this is that
coagulation proceeded more slowly in Case 6, meaning that the influx of migrators
from the outer nebula did not decrease over such a short timescale.  That is,
the outer disk was able to produce  migrators for a longer
period of time, resupplying the inner disk with vapor over much more of
the lifetime of the disk before allowing diffusion to become the dominant
transport mechanism.

\subsection{Solid Surface Density}

Panels B, C, and D in Figures 1-7 show how the surface densities of the
dust, migrators, and planetesimals of the disks change over time.  In each 
disk the solids initially are distributed uniformly through the outer nebula
in the form of dust.  As discussed above, the dust begins to coagulate to
form migrators which then are accreted to form planetesimals.  Because of
the rapid drift that migrators experience, a particular body may travel
a large distance through the disk before it is accreted--and in many cases,
particularly during the early stages of disk evolution,
it will migrate in to the hot inner regions of the disk and vaporize without
being incorporated into planetesimals at all.  

In all cases shown, the surface density of the dust peaks immediately outside
the snow line at almost all times (the exception being the 3 million year
snapshot in Case 3).  This is similar to the results found by 
\citet{stepval97} and \citet{cuzzahn04}.
The reason for this is that despite having the shortest
coagulation time of anywhere in the outer disk (and therefore the most efficient
dust sink), dust is resupplied to this region in a variety of ways.  First,
the net flow of material in a disk is inward, so dusty material
from the outer nebula will be entrained in the gas and carried towards the
central star.  Secondly, migrators move inward due to gas drag and constantly
resupply the region outside of the snow line with additional material.  There,
the turbulent velocities between the migrators are highest ($v_{t}=\sqrt{\alpha}c_{s}$)
and the migrators have a larger volume density compared to anywhere else
in the disk.  This results in more frequent and destructive collisions between
the migrators, generating more dust there than at any other location.  Finally,
water vapor that diffuses outward from the inner disk will condense to form
water ice, and its concentration will be highest immediately outside the 
snow line.  

Local minima, or dips, in the dust surface density develop in most of 
the cases shown. The outer edges of these dips, where there is a localized
increase with dust surface density, correlate with the local
maxima in the overall disk surface density.  It is at these regions that
the dust began to become depleted as it rapidly coagulates to form migrators, 
causing the disk viscosity to decrease, and as discussed previously, slows the
local rate of mass transfer.  
As a result, the inward flow of material by advection decreases, preventing
material from the outer disk from being carried inwards.  In addition, outward
diffusion of dust does not operate rapidly enough to resupply this region with
dust from smaller heliocentric distances, which itself is advected further
inwards by the large scale flow of the disk. Thus, dust was removed more 
rapidly than it was resupplied,
resulting in a dip in the dust surface density.  These effects are most readily
observed in Cases 1 and 5.

While migrators are most efficiently produced where dust densities are 
high, their distribution in the disks shown do not mirror that of the dust.  
Because of gas drag, migrators are constantly flowing inwards, leading
to a surface density that decreases with distance, despite the gaps that
develop in the dust distribution.  In some cases (most notably Case 1, but also 
2, 5, and 7), local maxima or sharp changes in the slope of the
migrator distribution are noticeable and correspond to local minima in the 
dust distribution.  This 
is not due to the production rate of the migrators, but instead to the effect
that the dust surface density has on the thermal evolution of the disk.  Because
the dust concentration determines the opacity of the disk, regions where dust
is depleted will achieve lower temperatures and lower viscosities.  This results
in nebular gas ``piling up'' there due to the slower evolution.  In addition, the 
midplane temperature gradient will locally decrease, leading to a localized
decrease in the outward pressure gradient of the gas.  It is this pressure
gradient that determines the rate at which the migrators move inwards, and
thus the velocity of the migrators will slow.  Therefore, as migrators move
inwards from the outer disk, they may be traveling at rates on the order
of 1 AU/century as quoted above.  However, as they enter these regions of
shallower pressure gradients, their velocities decrease rapidly, resulting
in a pile up of the migrators as the outer disk continues to rapidly feed
that region with more rubble.

Another important result to notice is that the inner edge of the migrator
surface density does not extend beyond the inner edge of the dust surface density
(the snow line)
by more than a few tenths of an astronomical unit.  This counters the finding
by \citet{cyr98} who reported that icy boulders could exist as far inwards as
2 AU from the snow line due to the rapid rate at which they are transported.
Instead, our results are in agreement with those of \citet{supulver00} who
found that icy bodies would vaporize shortly after crossing the snow line.

In looking at planetesimals, their growth occurs most rapidly immediately 
outside the snow line because the accretion
timescale is shortest there (the local $t_{acc}$ scales with orbital period),
and because it is there, generally, that the largest
amount of icy material passes (both inward from migrators and dust and outward
by dust condensed from vapor).  However, as the disk evolves, the planetesimal
surface density is shaped by the evolution of the migrator distribution and
the migration of the snow line.

As migrators move through the disk, they can be incorporated into planetesimals
by interacting with one another to form new ones (through $t_{acc}$) or by
being swept up by pre-existing ones.  New planetesimals are most easily created close
in to the central star where accretion timescales are shorter or where the surface density
of migrators are highest.  When a significant amount of planetesimals has already
formed, the probability that migrators will pass through this swarm without being
swept up decreases, meaning that the planetesimals at the outer edge of this 
large swarm will grow more rapidly far away from the snow line as migrators
from the outer disk attempt to drift inwards.  Rather than surviving all the
way to the snow line, they instead continuously feed the large swarm further 
away.

As the snow line migrates inwards, planetesimals will still be able to form at smaller
heliocentric distances.  However, the surface density of planetesimals that forms there
will be determined by how much solid material is available at these smaller 
distances.  In the early stages of disk evolution, there is nothing to prevent
material from constantly being carried inward to the snow line, as described
above, and thus the largest planetesimal surface density corresponds to that
location immediately beyond the snow line.  As the disk evolves and
the snow line moves inward, the flow of solids from the outer disk to
the snow line is inhibitted by the existing planetesimals at larger heliocentric
distances.  Thus in the later stages of disk evolution, the planetesimal 
surface density may peak at a location significantly outside the current location
of the snow line.  A similar result was seen by \citet{kornet04}.
This effect is most noticeable when planetesimal formation 
is rapid compared to the rate at which the snow line migrates.  Thus in the
slower evolving disks shown ($\alpha$=10$^{-4}$), the planetesimal surface
density evolves as described.  In those cases where $\alpha$=10$^{-3}$, the
snow line migrates rapidly enough (and dust is transported more readily) so
that the prefered location for planetesimal location is generally always
immediately outside the snow line.

Planetesial growth will determine how much water is retained in the disk at the
end of its evolution.
Below the parameters used in each case, Table 3 lists the initial mass of the
disk, the mass at the end of the simulation, the initial mass of water contained
in the disk, the mass of water contained in the planetesimals at the end of
the simulation as well as the total mass of water in solid form.  
While it is clear that the enhancement of the inner disk
and snow line region comes at the expense of the outer disk, it is also
true that decoupling of solids depletes water from the disk as a whole.
In all cases presented the water-to-hydrogen mass ratio remaining at the
end of a simulation is lower than at the 
beginning.   Because migrators are transported inward by gas drag faster
than the advective velocity of the nebula, these objects speed ahead of the gas
and move towards the sun faster than the rest of the material in the disk.  This
results in solid-forming species being lost from the disks more rapidly than
other species.  This was also found to be the case in the massive disk models
of \citet{stepval97}.  As a result, the amount of mass that would be available to
be incorporated into planetesimals far outside the snow line 
decreases over time, and the amount of
material locked up in such bodies will only be a small fraction of the material
that was available when the disk first formed.  Higher fractions will 
remain if planetesimal formation is very efficient (operates on short timescales)
as will be discussed below.  

Finally, not only does the inward transport of migrators play a significant role
in determining the distribution of the
planetesimals near the snow line, but also at large heliocentric distances as well.
In the outer disk, we find that the planetesimal surface density drops off 
rapidly at large distances, reaching negligible values at smaller 
heliocentric distances than the gas surface density.  These results are similar
to those found by \citet{weidenschilling03}, who found that the inward migration
of bodies as they grow larger in size could produce a planetesimal swarm 
significantly smaller than the gaseous disk that orbited the sun.  In the cases
presented here, the planetesimal surface density may extend to distances that
are a factor of two less than what the gaseous disk occupies.  

\section{Discussion and Implications}

The results of the different case studies presented in the previous section 
demonstrate the sensitivity of the specific results to
the values of the free parameters used.  We cannot expect to know the precise 
starting conditions of the solar nebula or other disks;
rather, we hope to use the general results presented here to understand different features of
protoplanetary disks as well as the different stages through which our
own solar system might have evolved.  Below we discuss the implications our 
models have for such studies.

\subsection{Inner Nebula and Chondritic Parent Bodies}

The results of the various cases discussed show that the region of the disk
interior to the snow line would have experienced fluctuations in the 
concentration of water vapor relative to hydrogen.  The concentration may
have been enhanced by as much as an order of magnitude and the overall enhancement
may have lasted for a million years or more.  As the influx of material from the outer
nebula decreased, the inner nebula became depleted in water
vapor, reaching concentrations that were less than expected under
canonical nebular conditions.  Because of its importance in determining
the oxidation state (the so-called oxygen ``fugacity'') 
of the gas, the water vapor concentration
affected the chemistry and mineralogy of primitive materials 
in the inner disk.  Thus, we can look
at the properties of the chondritic meteorites to determine how much the
oxidation state of the inner solar nebula changed and estimate the timescales
on which these fluctuations took place.  By doing so, we may be able to constrain the
various parameters used in this model for our solar nebula.

Enhanced nebular oxygen fugacities have been invoked to explain a number
of chemical and mineralogical features in chondritic meteorites.  Among
these  \citep[as reviewed by][]{krot00}
are the valence state of titanium in Wark-Lovering rims around
CAIs which requires a 5 order of magnitude increase in oxygen fugacity over
the canonical solar value \citep{dyl05}, 
observed Mo and W depletions in CAIs \citep[3-4 orders of magnitude;][]{fegpalme85},
fayalite (FeO) grains in carbonaceous chondrites \citep[3-4 orders of magnitude;][]{huabus95},
fayalitic olivine in carbonaceous and ordinary chondrite meteorites \citep[2-4 orders
of magnitude;][]{palmefeg90},
FeO content in chondrules \citep[4-6 orders of magnitude;][]{huang96,hewins97}, and
retention of volatiles during formation of FeO rich chondrules 
\citep[2-5 orders of magnitude;][]{yu98}.  In the
model cases presented here, the maximum enhancement produced was 
no more than one order of magnitude, suggesting
that enhanced oxygen fugacities of the type needed to produce
the observed signatures in meteorites cannot be produced by the radial transport
of water alone.

While this is true in the specific cases presented here, we can also calculate 
whether this limited enhancement is true for all disks.  The enhanced water vapor
concentration of the inner disk will result from water ice from the outer disk 
being brought inside of the snow line where it would vaporize.  Thus the maximum possible
enhancement will be achieved when all of the water in the disk is concentrated
inside of the snow line.  If the concentration of water relative to hydrogen
is initially some constant fraction throughout the disk,
then the maximum enhancement possible will go roughly as the mass of the
disk divided by the mass of the disk interior to the snow line.
The amount of mass in a disk inside
a given radius, $R$, whose
surface density is described as a power law (as is assumed here) is:
\begin{equation}
M_{D}(r<R) = \frac{2 \pi r_{0}^{p} \Sigma_{0}}{2-p} R^{2-p},
\end{equation}
where $r_{0}$ is 1.5$\times$10$^{13}$ cm (1 AU, where $\Sigma_{0}$ is defined).  
Thus, initially the ratio of the 
amount of mass in the disk to the amount inside the snow line will be 
($R_{0}$/$R_{sl}$)$^{2-p}$.  As an example, this formula predicts that the 
maximum enhancement possible in case 2 would be 40 AU/5 AU $\sim$8 for $p=1$, which
is close to the value found in the model runs with $t_{coag} \sim 10^{4}$ years.  
In order to enhance the inner
solar nebula with water vapor by just 3 orders of magnitude (roughly the lower
level needed to produce some of the oxidized features of meteorites identified above) 
requires that the disk
extended outwards to $\sim$5000 AU (for $p=1$) if the snow line were located at 5 AU. 
Smaller disks would be needed if the surface density distribution
had a slope of $p < 1$, so that most of the disk mass was concentrated at
larger radii. In the unlikely case of $p=0$, 
which would represent a disk with a constant surface density at all radii,
the disk would have to extend to 160 AU to enrich the inner disk by 3 
orders of magnitude.  More plausible disk structures with $R_{0} \sim 50-200$ AU,
$p \sim 1$ and $R_{sl} \sim 2-10$ AU predict enhancements of 5-100.  However,
it must be remembered that these are \emph{maximum} values as these calculations
assume that no water would be left beyond the snow line.  
Transporting all water in the disk
to inside the snow line is likely not feasible, however, and thus in real disks, 
it is unlikely that the maximum enhancements would be reached.  Our findings 
suggest that the maximum enhancement of water vapor a real disk would experience
is roughly an order of magnitude.

The large oxygen fugacities recorded by some chondritic meteorites thus are likely
not
due to the inward migration of icy bodies alone.  In reviewing the different
features, \citet{krot00} argued that many could have resulted from asteroidal
processing as primitive materials reacted with oxidizing fluids (liquid water)
after they were accreted by their parent bodies.  However, nebular processes
may have still played a role as the enhanced water vapor
concentration would lead to a more oxidizing nebular gas that could combine with
other processes that would concentrate silicate dust 
to enhance the oxygen fugacity in the inner disk, provided that there was a way
to vaporize the dust, perhaps in a chondrule forming event.  

The enhancement of water vapor in a chondrule formation event was considered by
\citet{ciesla03} who demonstrated that if
shock waves, such as the type thought to be responsible for the formation of
chondrules, occurred in an icy region of the solar nebula, large water vapor
pressures could result.  The shock wave alone would increase the vapor pressure
by almost two orders of magnitude, though hydrogen would increase by an equal amount,
maintaining a constant \emph{concentration}, and thus keeping the oxygen
fugacity roughly the same as it was before the shock occurred.  However, the
higher partial pressures of all vapor species would
create a new, temporary environment which may then allow chemistry to proceed
along different paths or rates than it would under the pre-shock conditions.
If the shock waves were to occur in regions of the disk where solids were
concentrated above the solar value, then those solids may be vaporized, releasing the oxygen
they contained to the gas.  Regions of a disk that are enhanced in solids can
be caused by gravitational settling to the disk midplane \citep{weidenschilling80} or
by turbulent concentration \citep{cuzzi01}, with enhancements exceeding 100
times solar predicted to be common.  If these solids are vaporized,
vapor pressures 
10$^{4}$-10$^{5}$ larger than found in the canonical nebula would be possible,
if combined with global enhancements due to the inward migration of water
as described above.   In fact, \citet{dyl05} argue that the Ti
oxidation states that they observe in the Wark-Lovering rims of CAIs are likely
due to shock waves in a dusty gas.  Further work is needed to confirm that
such temporary environments could imprint their signatures on the various
primitive materials described above.

While some chondritic meteorites appear to have formed in regions of the nebula
with higher-than-solar oxygen fugacities, others
appear to have formed in environments
that were more reducing than canonical conditions.  Models for the
formation of the enstatite chondrites suggest that removal
of more than 50\% of the water in the nebula had to have taken place in
order for the observed minerals to become stable \citep{hutruz00,pasek05}.
Such a situation is realized in the model runs presented here, but
after millions of years of evolution.  In order for water to be depleted so that 
its concentration is low enough to allow the enstatite mineralogy to be 
reproduced, either the disk has to have a large enough viscosity 
($\alpha > 10^{-4}$) to ensure that vapor can diffuse outwards rapidly enough,
or meter-sized rubble must be prevented from reaching the inner disk. Neither
of these situations would allow for the inner disk to become substantially
\emph{enhanced} in vapor as well.  Thus it may be that large water vapor
enhancements throughout the region interior to the snow line followed by large depletions 
are difficult to get in the same protoplanetary disk.  

In addition to the different oxygen fugacities recorded by primitive 
meteorites and their components, different oxygen isotope abundances are also 
observed.  On an oxygen three-isotope plot, CAIs and chondrules fall on a line
with a slope of $\sim$1, which has been interpreted to arise due to the mixing of
an $^{16}$O rich gas  with a reservoir rich in $^{17}$O and $^{18}$O 
\citep{clayton93}.  It has been suggested that this mixing may have arisen
as the inner nebula gas, which was $^{16}$O rich, incorporated the water vapor
that was introduced by inwardly migrating water-ice boulders which were
$^{16}$O-poor.  The
excess heavy oxygen isotopes in these boulders could be due to isotopic 
self-shielding during CO dissociation in the parent molecular
cloud of the solar nebula \citep{yuri04} or the outer disk \citep{lyonsyoung}.  
Either mechanism requires
the inward drift and evaporation of outer solar system rubble to alter
the oxygen isotope ratio of inner disk solids \citep{yuri04,lyonsyoung,krot05}. 
\citet{krot05} reported on one CAI which formed in an $^{16}$O-poor environment 
with an age of less than 0.8 Myr, suggesting that icy material would have to be brought
inwards rapidly in order to alter the gaseous isotopic abundances in the nebula.
(This age was determined assuming that $^{26}$Al was uniformly distributed in the
disk and was not altered by later inection or local production of the nuclide.)
In the model developed here, enough ice was introduced to the inner nebula in
less than 0.1 Myr that the region immediately inside the snow line would reach
its maximum enhancement.  Over the next few hundred thousand years, the rest
of the inner nebula would be equally enhanced and its isotope abundances would
be modified.  Even small enhancements of the inner nebula can lead to
large changes in the oxygen isotope abundances \citep{yuri04}, and thus 
determining how much outer-nebula water ice must be brought inwards to
produce the observed isotopic trends could provide a constraint on the level of 
mixing that took place in the solar nebula.

\subsection{Implications for Giant Planet Cores}

One of the motivations for previous studies of the distribution of water in
the solar nebula was to evaluate if water would be concentrated at a location
that would allow rapid growth of large bodies which would become the
cores of the giant planets (assuming they grew through the core accretion
mechanism).  \citet{stevenson88} showed that significant
concentrations were possible immediately beyond the snow line due to the
outward diffusion of water vapor
alone, but they neglected the migration of bodies from the 
outer nebula.
\citet{stepval97} developed a more detailed model which considered the
evolving distribution of water throughout a protoplanetary disk and also concluded
that the surface density of icy planetesimals would peak immediately 
outside of the snow line.  From this, they argued that the largest icy 
body (or planet with the largest icy core) would form at this location.
\citet{cuzzahn04} noted that evaporation front effects produced a larger
enhancement than inward migration of solids or outward diffusion of
vapor alone, however the enhancements found there were larger than those
found here because of the assumption of a steady-state.

In some of the cases presented here, a similar result was observed, though, 
the largest surface density of icy planetesimals
was not always located immediately outside of the contemporary 
snow line.  This is due to 
the fact that  the location of the snow line migrates inwards over the lifetime
of the disk as a result of the decrease in internal heating and optical
thickness of the disk.  During the early 
stages of disk evolution, icy planetesimal formation would occur only in
the outer regions of the nebula. As the snow line migrates inwards, 
icy planetesimals could grow at smaller heliocentric distances.  If
planetesimal formation occurred rapidly enough, a massive planetesimal swarm
could develop near the initial location of the snow line.  As migrators move inwards
from the outer nebula, they continue to feed \emph{this} planetesimal swarm
because the planetesimals have a large accretional cross-section.  
Very few of the migrators would
survive passing through this region, so the raw material that would be
needed to build planetesimals at later times closer
to the sun would be limited to what
was already there or what was brought outwards from the vapor in the
inner disk.  Thus, the peak in the icy planetesimal surface density
may ``remember'' the snow line of earlier stages of
nebular evolution \citep[c.f.][]{kornet04}.  Our results indeed show
a relatively \emph{broad} distribution of planetesimals.  This may have 
implications for the formation of multiple gas giant cores in close
proximity \citep{thommes99}.

Another effect of solid transport is the feedback onto the viscosity of the disk.
Previous studies neglected how the opacity throughout the disk would change
as the dust concentration evolved.  In some cases, the pressure gradient of the
disk may be altered to slow the rate at which solids would drift inward due
to gas drag.  This
would result in a traffic jam of particles as they were forced to suddenly slow
down, leading to a localized region of high solid mass density.  This could be a
preferred location for the formation of planetesimals or larger bodies.  This
is slightly different from the results of \citet{haghi03} who found that
solids would drift to and concentrate at local pressure maxima in a protoplanetary
disk.  In this work, no pressure maximum exists; instead, it is a
shallow pressure  gradient that encourages the spatial concentration of solids.

Finally, as dust coagulated to form larger bodies in the outer disk, or
were removed from the region entirely, temperatures
would decrease due to the lower opacity.  This would produce much cooler 
temperatures in the outer disk than would be predicted in models where the
opacity is determined as if solids were at their solar concentration relative
to hydrogen.  As a result, this may play a role in determining how far inward
noble gases could be incorporated into solids.  As reviewed by \citet{owen99},
the noble gas content of Jupiter suggests that the planet incorporated a number
of planetesimals that formed at very low temperatures ($\sim$30 K).  
If the opacity of the outer disk dropped rapidly enough, this may allow such 
temperatures to form much closer to Jupiter than the 30 AU predicted 
by \citet{owen99}.  This is speculative and requires a better treatment of 
disk opacity than used here.

\subsection{Implications for the Structure of Protoplanetary Disks}

One striking result in all cases studied is how quickly the concentration of
water in the nebula deviates from the canonical value at all locations.  
Typically the distribution of solids (or of any chemical species 
in the gas phase) in models of the solar nebula or a protoplanetary
disk is assumed to be a constant.  Here we find
that because solids are transported through a disk at different rates
depending on their sizes,  the abundance of a species--in
both its vapor and solid phases--will vary with both
time and location in the disk.  This must be considered when
interpreting observations of disks around young stars.  For example,
to infer the mass of a disk,  observations will measure the solids with the
greatest opacity (the dust).  As can be seen in the various panel B's shown
in these simulations, the shape of the dust distribution and its
abundance relative to the gas can vary greatly.
In addition, because solid-forming species 
are preferentially lost from the outer nebulae due to the inward migration 
of meter-sized rubble, disks at 10-100 AU likely contain more hydrogen 
and helium than are inferred by assuming
a constant ratio of solids to gas.  Therefore the masses of protoplanetary
disks that have been inferred based on observations could be underestimates of
the actual masses, and perhaps the sizes as well.

Decoupling of solids and gas should also be considered when trying to 
reconstruct the
structure of the solar nebula from the solids currently present in the
solar nebula. \citet{weid77neb} and \citet{hayashi81} calculated the
structure of a ``Minimum Mass Solar Nebula,'' where the planetary material 
currently in orbit around the sun was distributed into annuli around the
sun to represent where the material was accreted from.  The amount of hydrogen
and helium needed in order to reproduce the canonical solar 
abundance ratio of this material at each radius was then calculated, and from 
that, the
surface density of the nebula was found.  In the calculations presented here,
the planetesimal surface densities did not follow the same structure as
the gas (compare panels A to D in all cases shown).  
If enough hydrogen and helium were added at every location to reproduce the
canonical water-to-hydrogen mass ratio based on the planetesimal surface densities,
the resulting nebula would have a very different radial structure and overall
mass than those actually calculated in the model.  Because solid-building
species are lost from the disk faster than the hydrogen and helium gas, the
solar nebula was likely more massive and extended to larger radii 
than the minimum mass nebula estimates.

Another result of this model is that the material which is accreted onto the
central star will vary in its chemical composition over time.  In the models
presented here, the amount of water at the very inner edge of the disk was initially
equal to the canonical amount expected.  As the inner disk became enhanced in
water vapor, the central
star would accrete more water than under canonical conditions. 
Later, as the concentration of water decreased in the inner nebula, the amount
of water accreted by the central star would be below the canonical ratio.  
\citet{carr04} reported that observations of hot molecular H$_{2}$O and CO
vapor inside of 0.3 AU around the young stellar object CSVS 13 indicate that
the abundance of water is 10 times lower than chemical equilibrium models predict.  This could
result from water being depleted in the inner nebula as it is locked up in
immobile objects further away as illustrated in the later stages of the models
presented here, or that excess CO is being introduced into the vapor as refractory
C-containing solids migrate inwards \citep{bitw2}.  The varying chemical 
abundances of accreting material may also have implications for determining
the metallicity of stars, if the accreted material is concentrated in the
surface layers of the star and not well mixed.

\section{Summary and Discussion}

In this paper we have developed a model to track the evolution of water in
a turbulently evolving protoplanetary disk.  Such a disk is capable of explaining
the mass transport observed to take place in protoplanetary disks as well as
expected to have occurred
in our own solar nebula.  We find that the transport of material in an 
evolving disk can lead to 
large enhancements and depletions of a chemical species that vary with time
and location.  This is a natural, unavoidable, result of disk evolution, and
these abundance variations may be able to explain some of the observed 
properties of primitive meteorites,
giant planets, and protoplanetary disks.  In particular, transport would lead
to fluctuating oxidation and isotopic conditions inside the snow line which are thought
to be reflected in various components of chondritic meteorites.  In addition,
the changes in the distribution of solids in the disk will cause opacity variations
that will change the thermal and pressure structure of the disk.  This could
help lead to the spatial concentration of particles in the outer disk,
possibly aiding in the rapid formation of giant planet cores.  

The results of the model developed here qualitatively agree with those of the
previous studies reviewed earlier.  The concentration of water vapor in the inner
disk is expected to change throughout the lifetime of the disk, as it is
enhanced during the early stages of evolution and then declines to sub-solar values
at later stages.  Solids preferentially concentrate immediately outside
the snow line of the disk as growth processes operate on shorter timescales there
than at larger heliocentric distances
and transport processes continue to feed that area with raw materials to grow
larger objects.  Because solids decouple from the gas as they grow larger, their
distribution in a protoplanetary disk can vary greatly from the distribution of
hydrogen and helium which dominate the mass of the disk.
Finally, the mass of solids retained in a disk is higher for more 
extended disks as the greater distances overwhich they would be transported 
provide more opportunity to be accreted into immobile objects.  For instance,
future studies should explore larger disks, as large as perhaps 200-300 AU.

While these general results agree with previous work, there are differences in
our results due to the different assumptions and treatments used in this model.
While the water vapor concentration will be enhanced in the inner disk, the finite
supply and rate of transport of ice to the inner disk limits this enhancement to be
no more than a factor of $\sim$10, though there is no limit to the level of 
depletion for the inner disk.  Also, the transport and coagulation of the dust
particles, which determine the opacity of the disk, has a feedback effect on
the evolution of the overall disk surface density.  The temperature structure
of the disk would also be affected, leading to sharp radial variations in the
rate of both gas and solid transport.  Finally, because the different dynamic
categories of solids are tracked simultaneously rather than assuming a single size
of objects at a given radius, 
it is found that the planetesimal surface density distribution is controlled by 
both the accretion rate and rate at which the disk cools resulting in the 
snow line migrating inwards.
The peak in planetesimal surface density may exist well outside the snow line
in the later stages of disk evolution if the accretion timescale is short
compared to the cooling timescale.  This may help us explain the presence of 
water well inside the orbit of Jupiter.

Even though this paper has focused on the dynamical behavior of water, the results
can be extended to other substances in protoplanetary disks such as 
silicates or organics.  Because both of these materials vaporize at higher 
temperatures than water, their
respective evaporation fronts would be located much closer to the central star
than the snow line.  This would lead to higher levels of enhancement in the
vapor phase for each species as the vapor would be distributed over a smaller
area in the disk \citep{bitw2}.  Also, the enhanced vapor would last for a shorter period of
time as the accretional timescales would be shorter at the smaller heliocentric
distances, leading to a more rapid growth of the planetesimal swarm outside of
the evaporation front.

The above discussion assumes that the migrators are pure substances, that is they
are only
made of silicates, organics, or in the model presented here, water ice.  In
reality, solar nebula solids were likely a mixture of all solids available at the locations
that they formed and accreted.  Thus the migrators may not release vapor as
readily as found in this work, for example if a silicate crust were to form on
the surface of the bodies preventing the release of water to the gas.  
This may result in water
surviving in particles further inside the snowline than found here.  However,
collisions with other objects expose buried ice, minimizing this
effect.  More work is needed to understand how bodies of mixed composition can
survive transport into different thermal environments.

One result that stands out from this work is the fact that there was an 
intimate connection between
the chemical evolution of the inner disk and the physical processes in the
disk.  The chemical inventory of the inner disk will be determined 
in part by the addition of material from the outer disk or removal of 
material by outward diffusion.   The efficiency of these
processes will be determined
by the growth rate of solids in the outer disk and the processes in the disk
that would determine how they are transported.  This means that constraints
on outer disk processes may be recognized by studying primitive meteorites.
For example, determining the time over which meteoritic components with
different oxidation states formed could provide information on when large
bodies in the outer solar system began to form.  Likewise, by understanding
the variations in oxygen isotope ratios we may be able to constrain the level
of mixing which took place between the inner and outer solar nebula.

While this work provides insight into the dynamical behavior of water in an
evolving protoplanetary disk, it can be improved in a number of ways by
including more detailed considerations of the physical processes discussed.  The
simplifications made here were generally done in order to ease the computational
burden of the model.  One of the fundamental issues that requires
more work is understanding the details of what drives protoplanetary disk 
evolution.  The benefit of the $\alpha$-disk model used here is that it allows
us to calculate how a disk evolves over millions of years; however there is still much 
debate as to what the proper value of $\alpha$ is and whether the model is a truly
valid parametrization of what drives the disk evolution.  In this work, and in 
most other models, $\alpha$ is assumed to be spatially and temporally constant,
despite the fact that we are unsure of what determines its value.  If the
turbulence in the disk is due to the magnetorotational instability \citep{bh91},
then turbulence may be limited to those regions of the disk where the
ionization of the gas was above some critical value.  This could lead to
spatially heterogeneous turbulence, where the very inner and outer parts
of the disk were active while the region from $\sim$1-10 AU would be nearly
dead \citep[Gammie 1996, see however][]{flemstone03}.  If $\alpha \sim 0$
near the snow line, the vapor that is produced from the migrators would pile
up immediately inside of the snowline could reach enhancements well above the
order of magnitude found here, though over only a narrow radial band.  Similarly,
if the viscosity of the inner disk exceeded that of the outer disk (either due
to higher $\alpha$ or the opacity effects illustrated by Case 1), hydrogen
gas could be preferentially removed faster than the water is delivered to
the inner disk, resulting again in higher water-to-hydrogen ratios than predicted
here.  There are mechanisms other than the magnetorotational instability
that may have played a role in driving disk evolution which may have all
played a role at different times and locations in the disk \citep{stone00}.  
It is also possible that once larger bodies grow in the outer disk, they may
excite waves that will generate or increase the turbulence in the disk
\citep{paulig04,boley05}.  

In terms of the evolution of the solids,
we have only accounted for particle growth in the timescale formalisms described
in Section 4.  In reality, particle growth in a protoplanetary disk is a difficult
problem that we are still trying to understand.  We are unsure how fine-grained
dust particles coagulate together to form large, coherent objects rather than
simple dustball structures of loosely bound monomers.  These individual grains
and aggregates likely encountered each other at a range of velocities, some of
which were accretional and others that were destructive.   Models of these
processes tend to focus on looking at how these objects grow at a single location
in the disk, investigating the vertical distribution and settling of the
solids \citep[e.g.][]{weidenschilling97,dull05}.  The disk-wide radial transport
of objects, and the local solid enhancements and depletions of material that
we have shown here to occur, are neglected.  Again, the computational rigor of such
models prevent calculations of full-scale disk evolution.  However, these 
models are critical to determining how particle growth occurs in disks under
a variety of conditions.  These will prove useful in constraining not only
the timescales used in this model, but also allow investigation of possible
delays in the initiation of planetsimal formation, for instance, which may only
be able to occur once a certain physical situation (e.g. low turbulent velocities
or large concentration of solids) is realized.  An initial delay in planetesimal
formation may lead to slightly larger vapor enhancements in the inner disk, and then 
once the formation process begins, allow for the rapid depletion of vapor as
sites for the sequestering of water in the outer disk grow.  

Finally, while more numerical modeling is clearly needed to evaluate the robustness
of 
the assumptions made and to constrain
the free parameters used in this work, further constraints
and insights may be gained from studies of primitive meteorites.   Radiometric
dating may help set the accretional timescale for meteorite parent bodies or
identify the timescale over which the chemical environment in the solar nebula
changed.  Also, astronomical observations of protoplanetary disks that are able
to map different chemical species (i.e. dust versus gas or H$_{2}$O versus CO) will
show how the distribution of various species differ and how they
are separated by disk dynamics.
Indeed, measurements of these different objects will provide the needed data to which
model predictions can be compared.  

\begin{center}
\textbf{Acknowledgements}
\end{center}

FJC was supported by the National Research Council while this work was done.
JNC was supported by the Origins of Solar System program and the Planetary
Geology \& Geophysics program.  Conversations with Paul Estrada, Sanford
Davis, Sasha Krot, and Misha Petaev were immensely beneficial and appreciated.
We are also grateful for the detailed reviews by Tomasz Stepinski and 
Joseph Nuth III which led to a much improved manuscript.

\bibliographystyle{natbib}
\bibliography{nebula}

\input{variables}

\input{parameters}

\input{models2}

\newpage

\begin{figure}
\includegraphics[angle=90,width=3.4in]{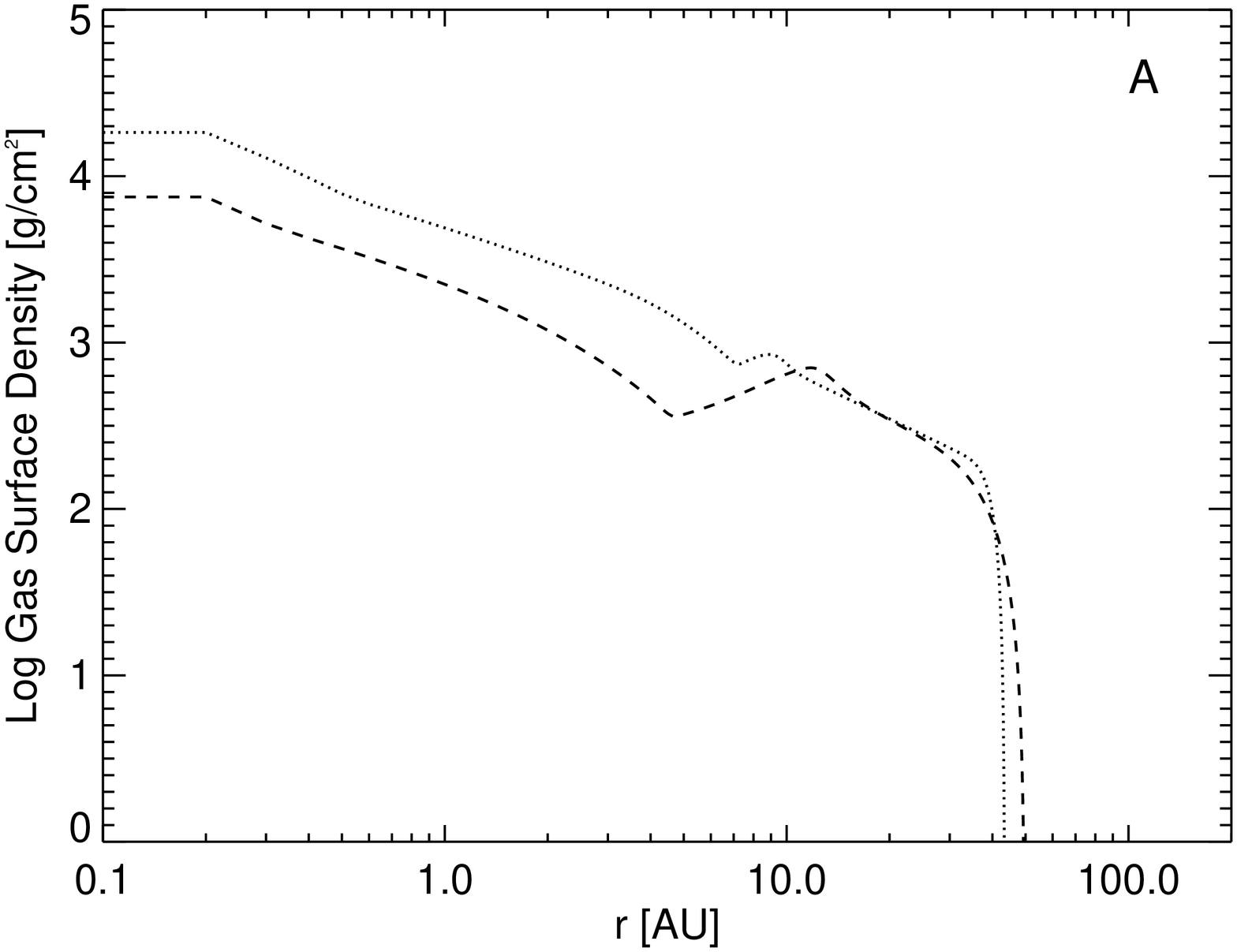}
\includegraphics[angle=90,width=3.4in]{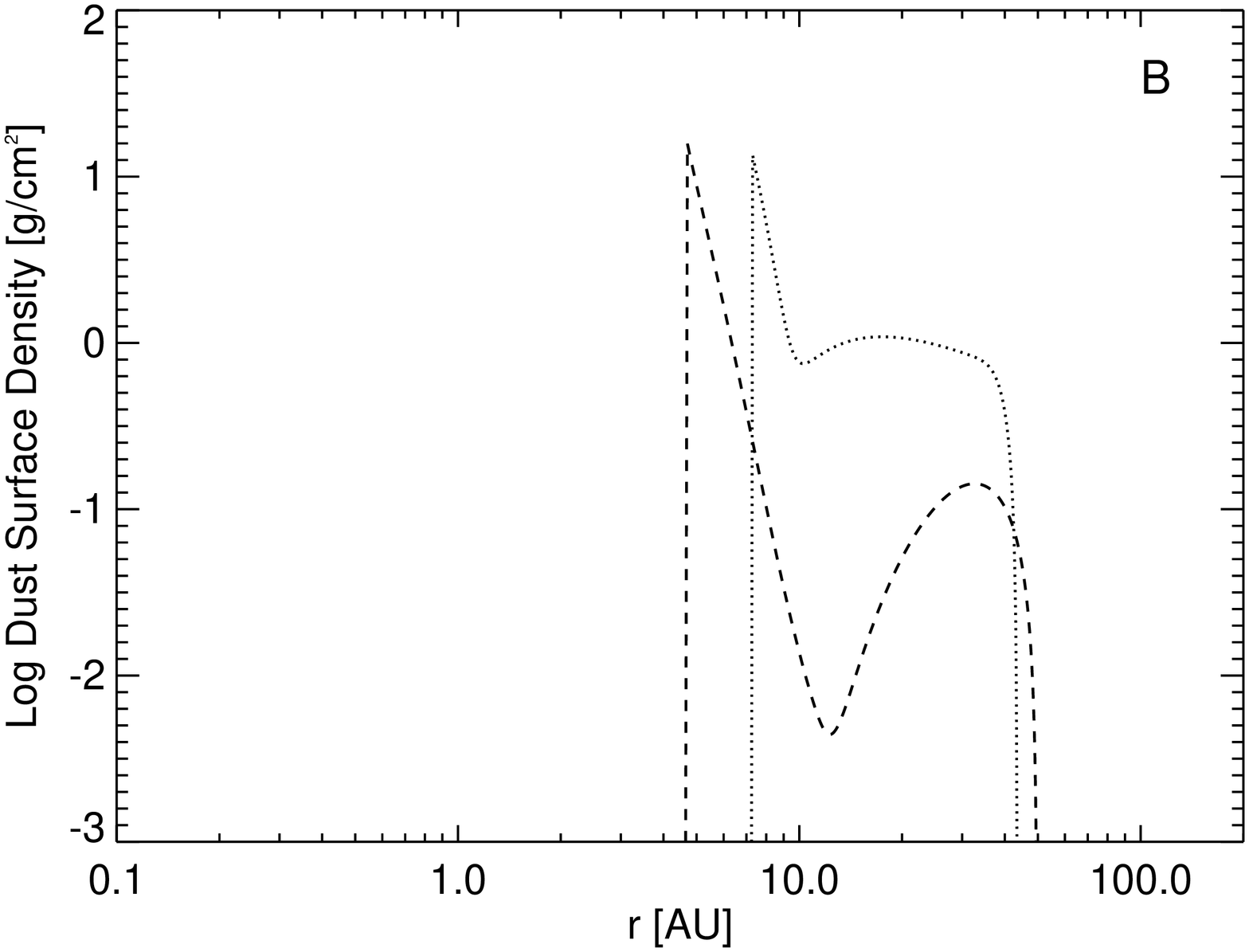}
\includegraphics[angle=90,width=3.4in]{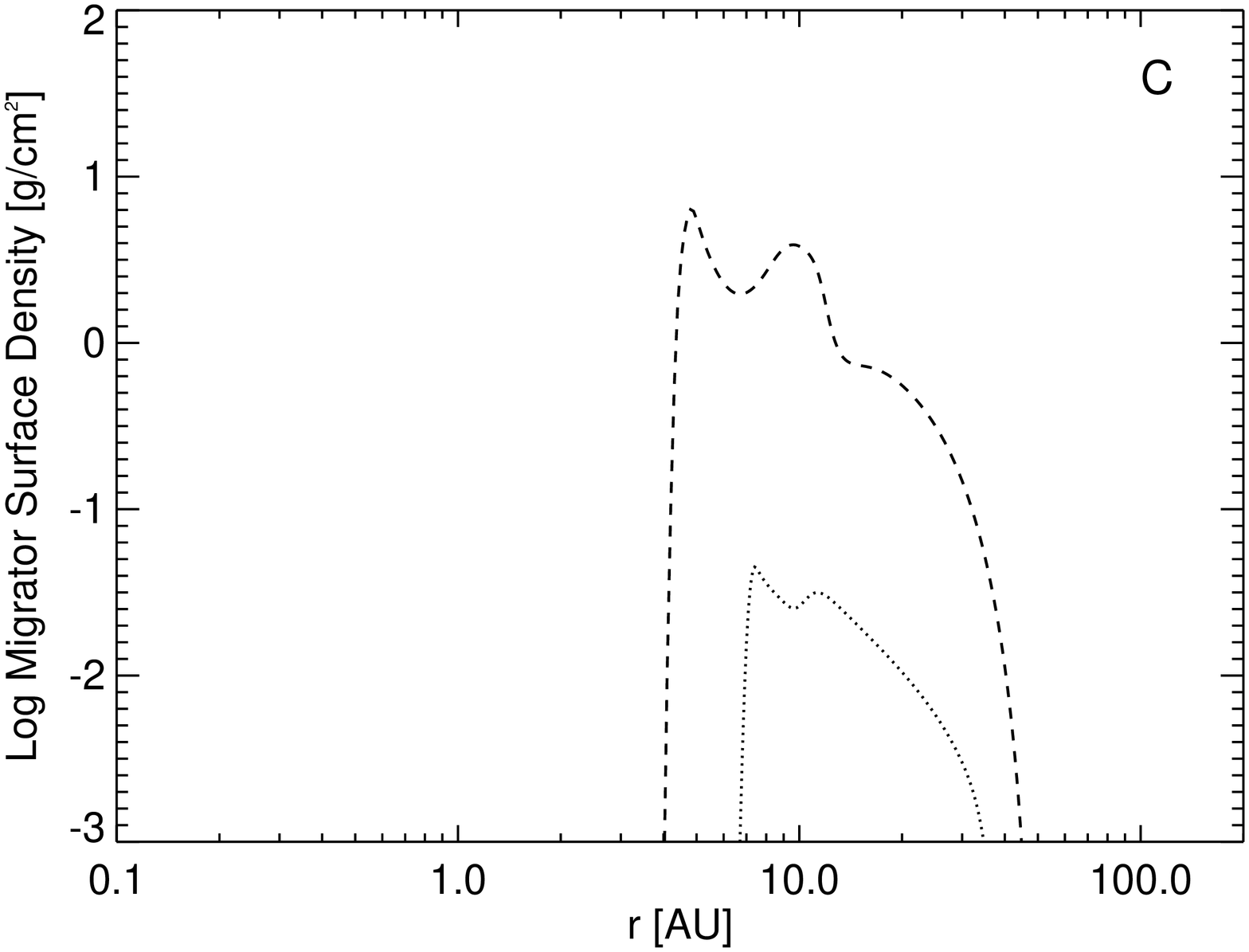}
\includegraphics[angle=90,width=3.4in]{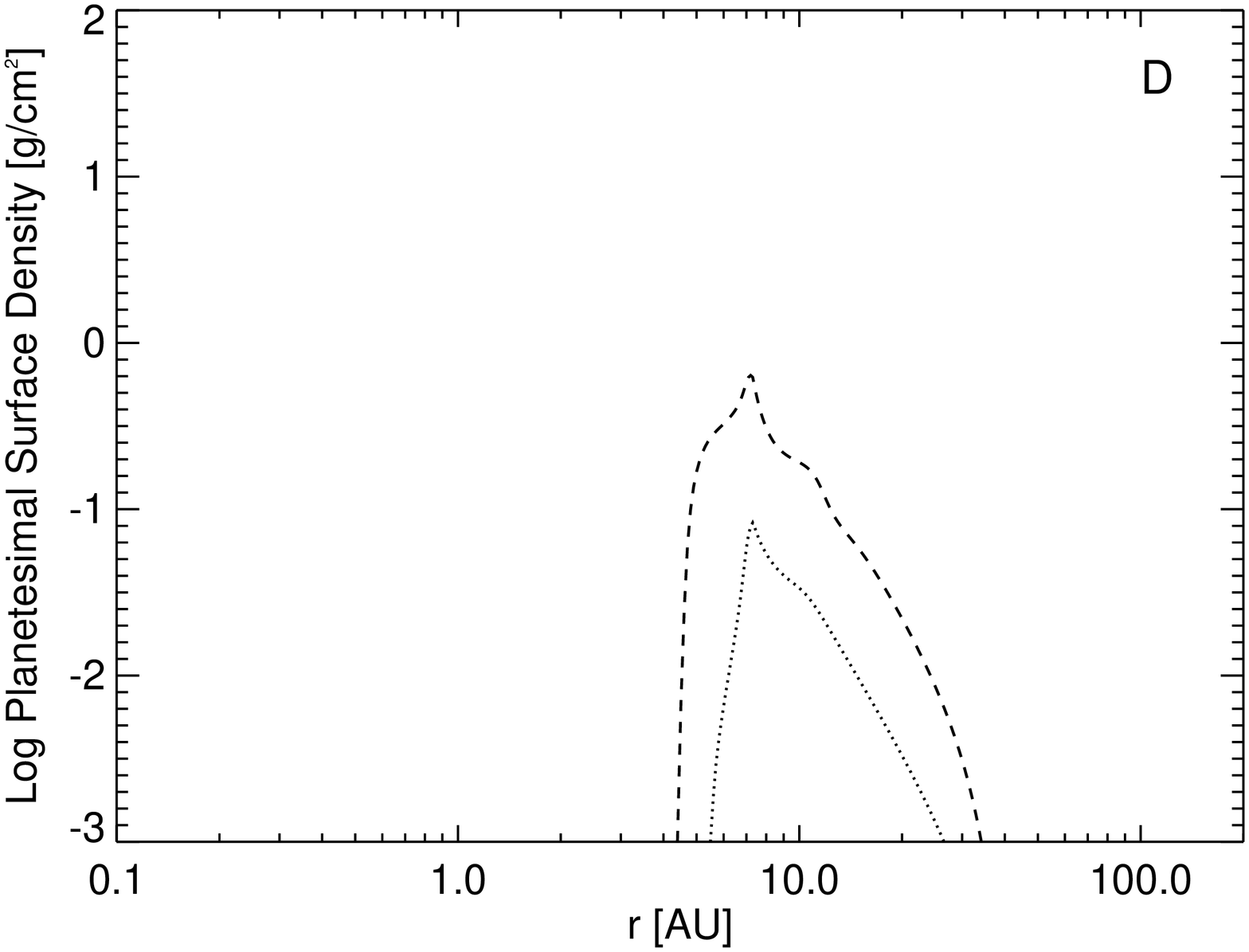}
\includegraphics[angle=90,width=3.4in]{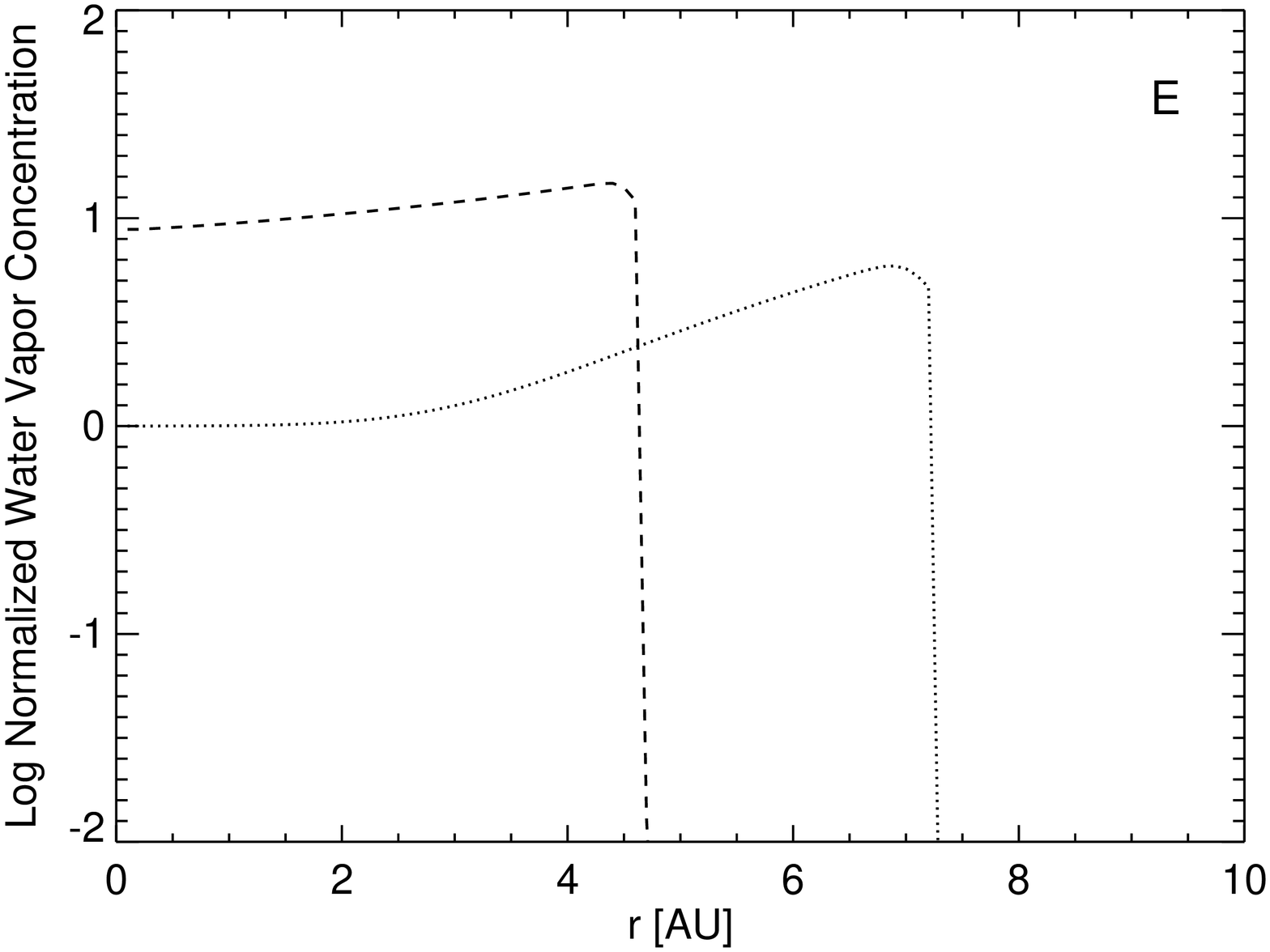}
\includegraphics[angle=90,width=3.4in]{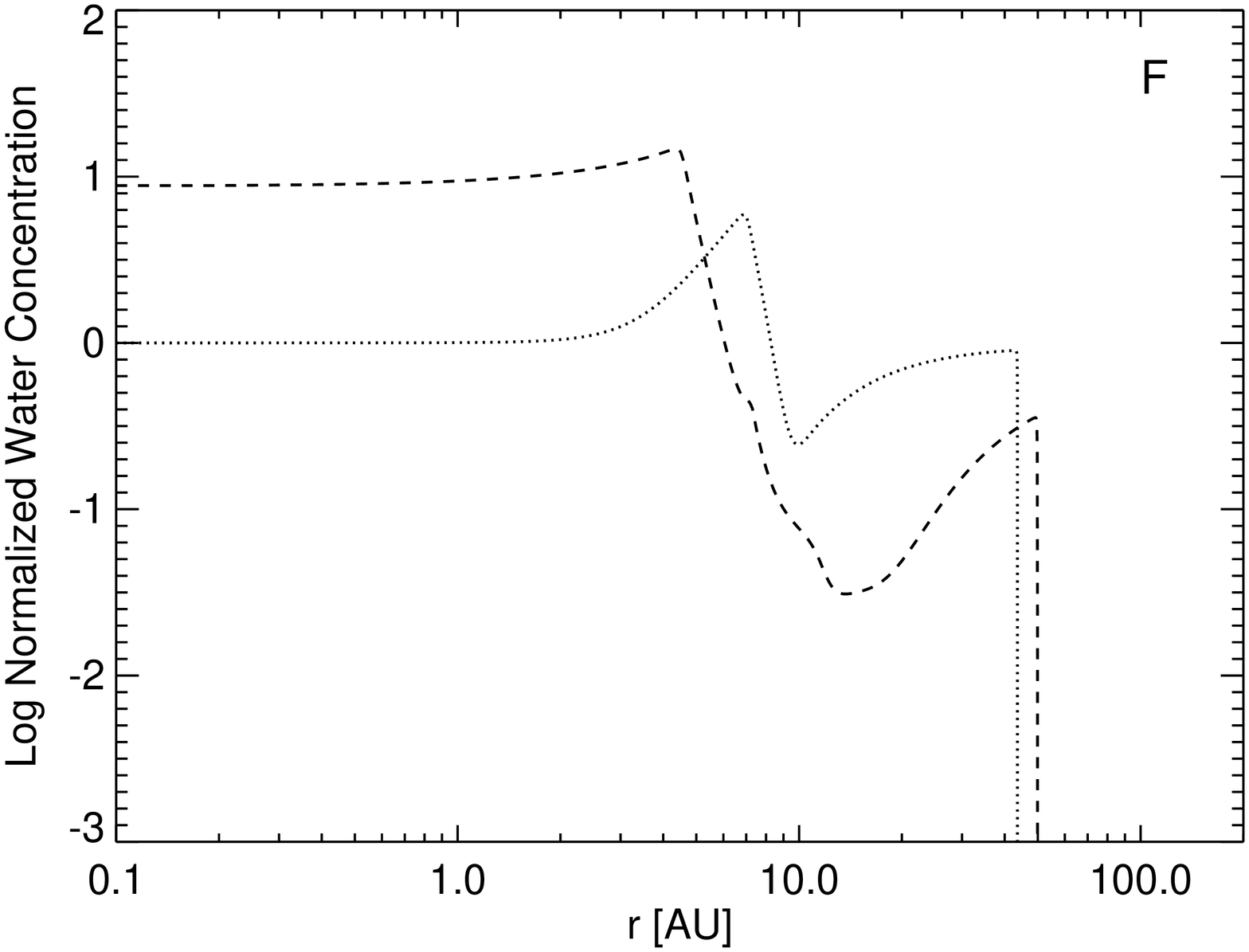}
\caption{The evolution of the disk surface density (A), dust surface density
(B), migrator surface density (C), planetesimal surface density (D), water
vapor concentration (E), and total water concentration (F) for Case 1.  Plotted
are the distributions after 10$^{5}$ (dotted) and 10$^{6}$ (dashed) years.}
\end{figure}

\begin{figure}
\includegraphics[angle=90,width=3.4in]{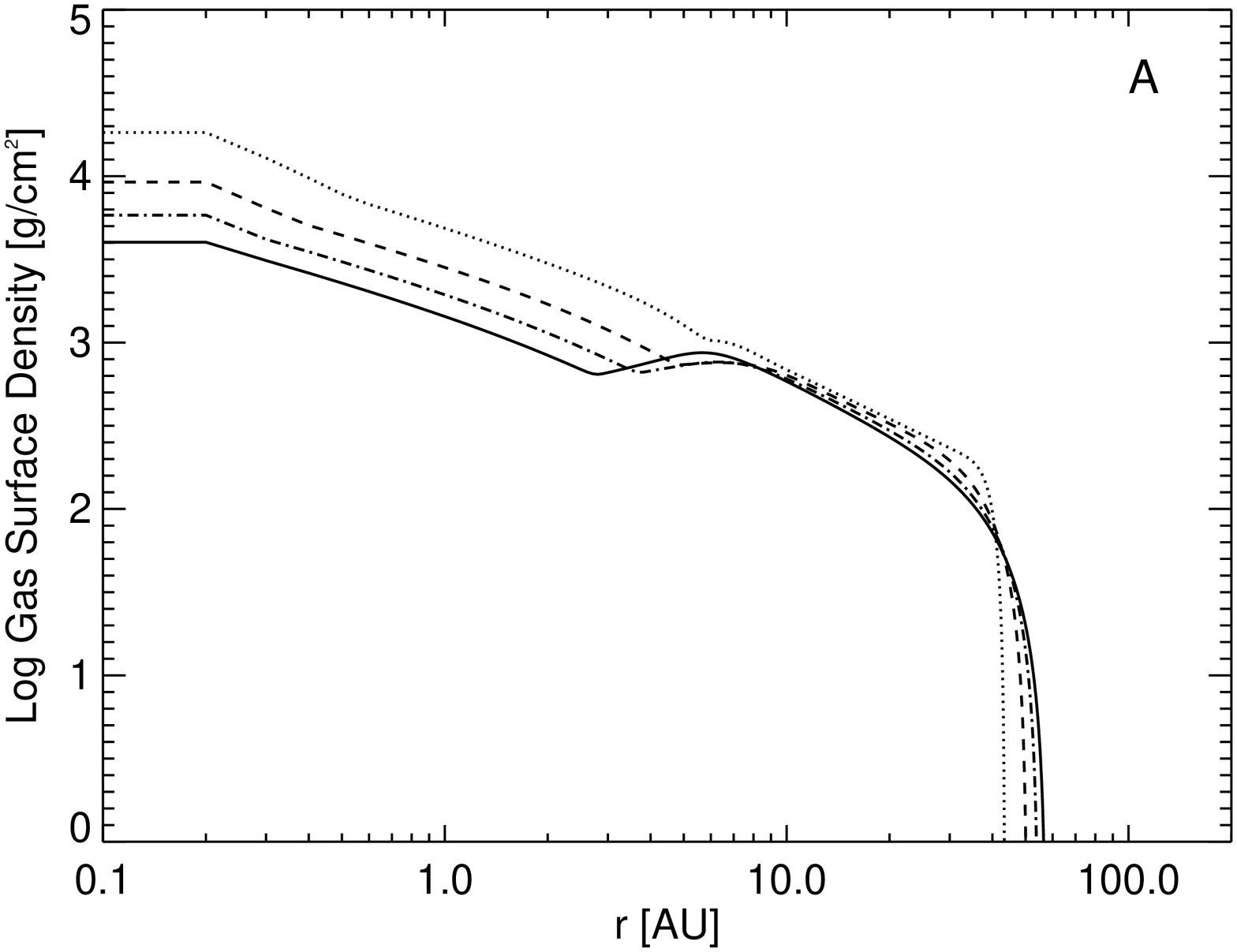}
\includegraphics[angle=90,width=3.4in]{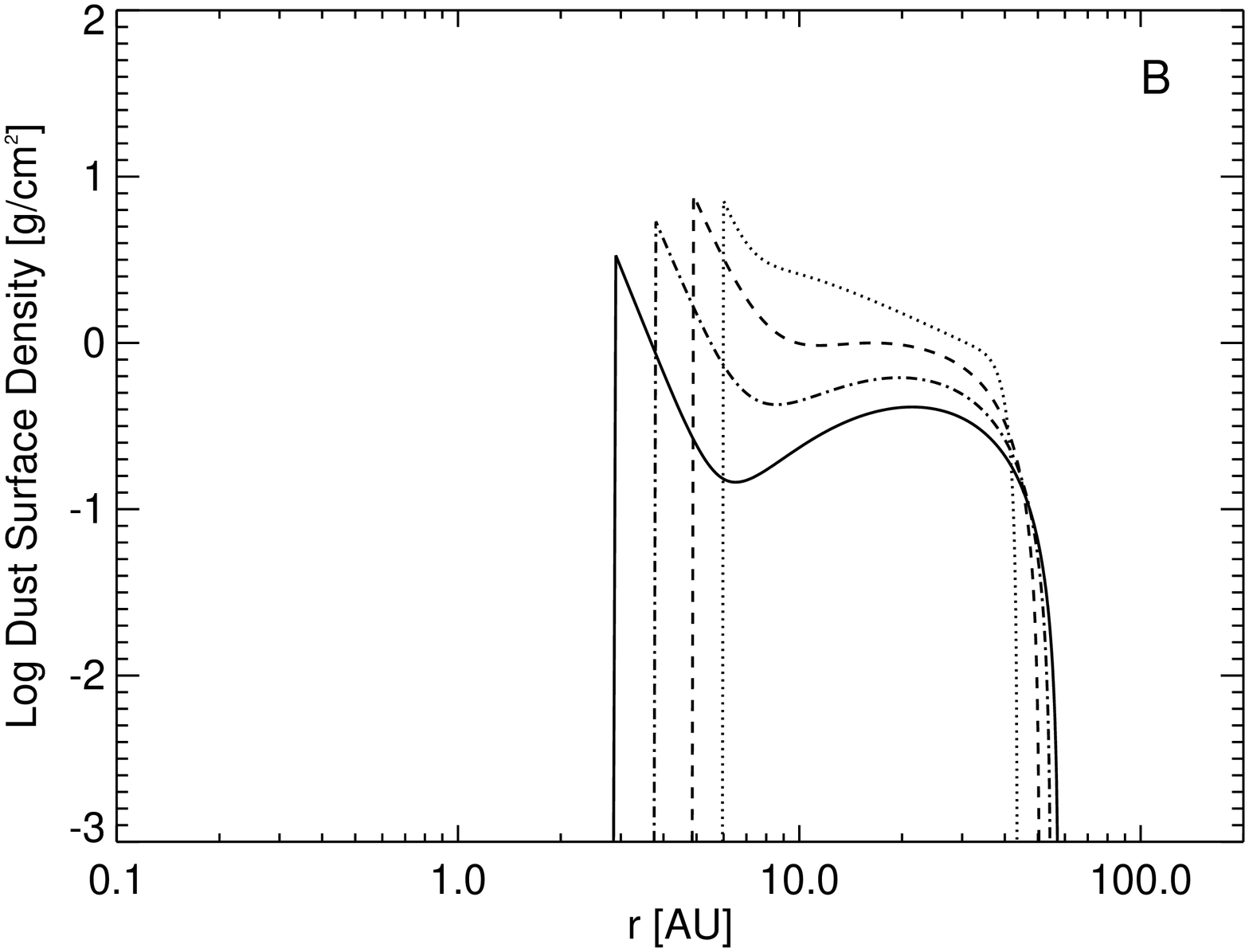}
\includegraphics[angle=90,width=3.4in]{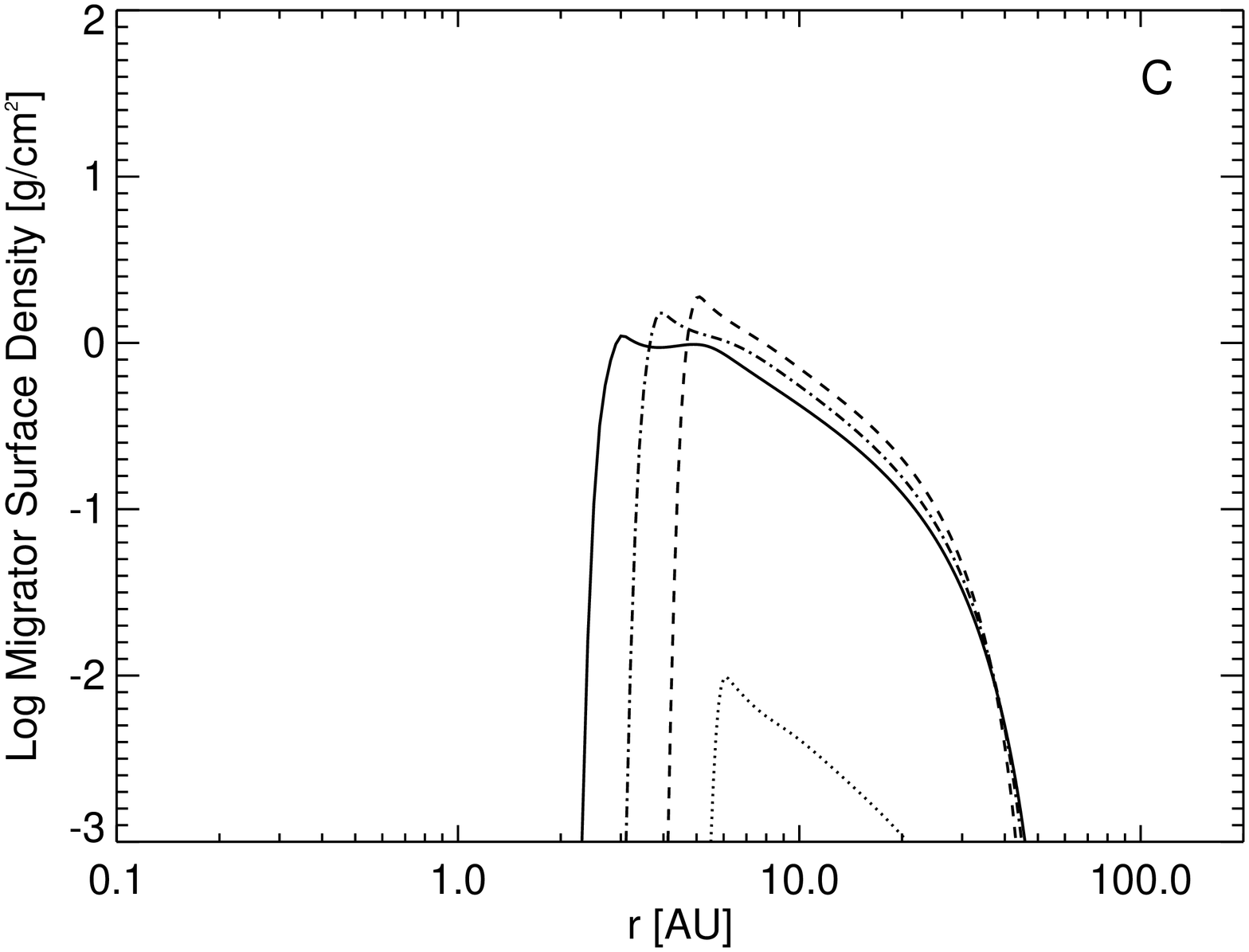}
\includegraphics[angle=90,width=3.4in]{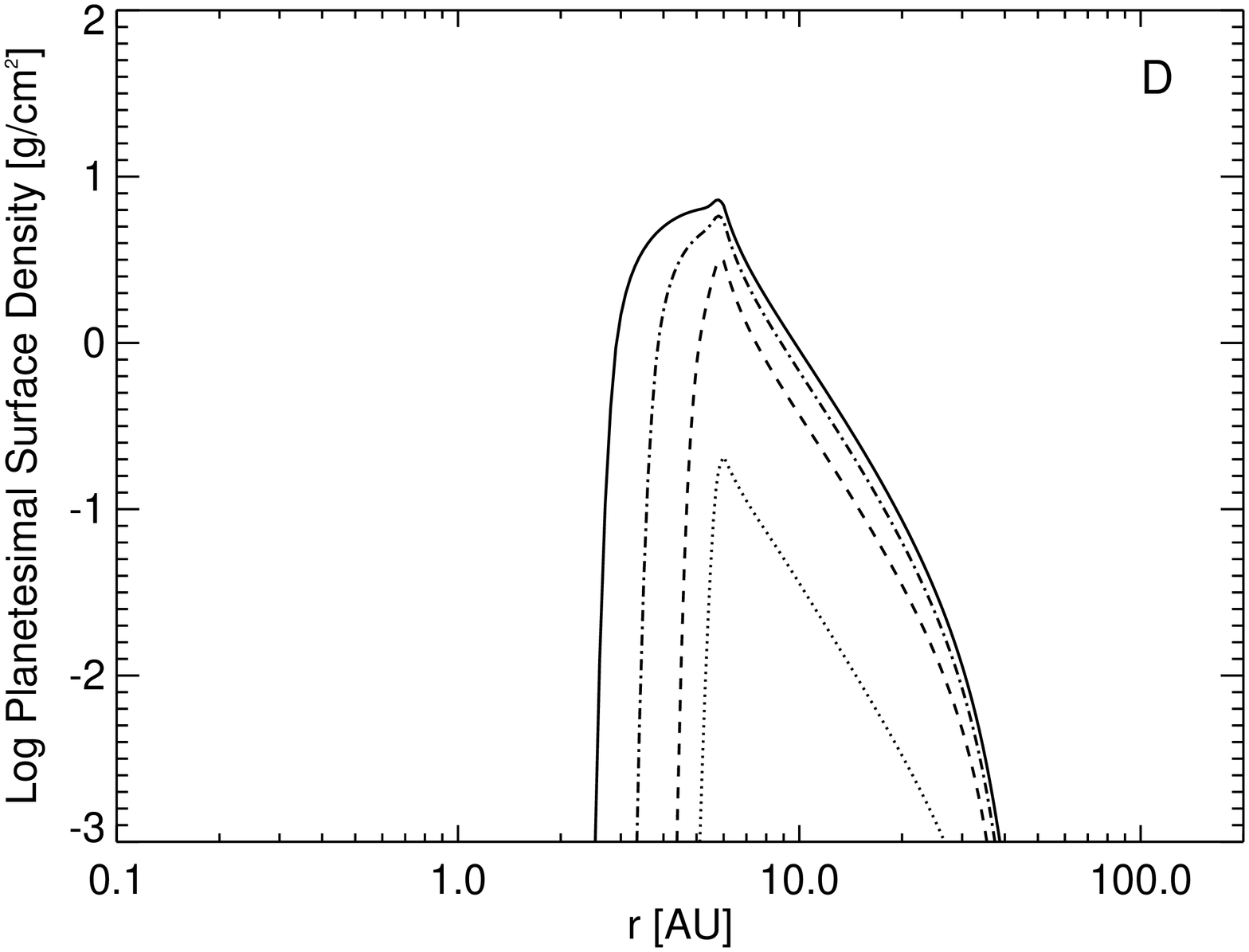}
\includegraphics[angle=90,width=3.4in]{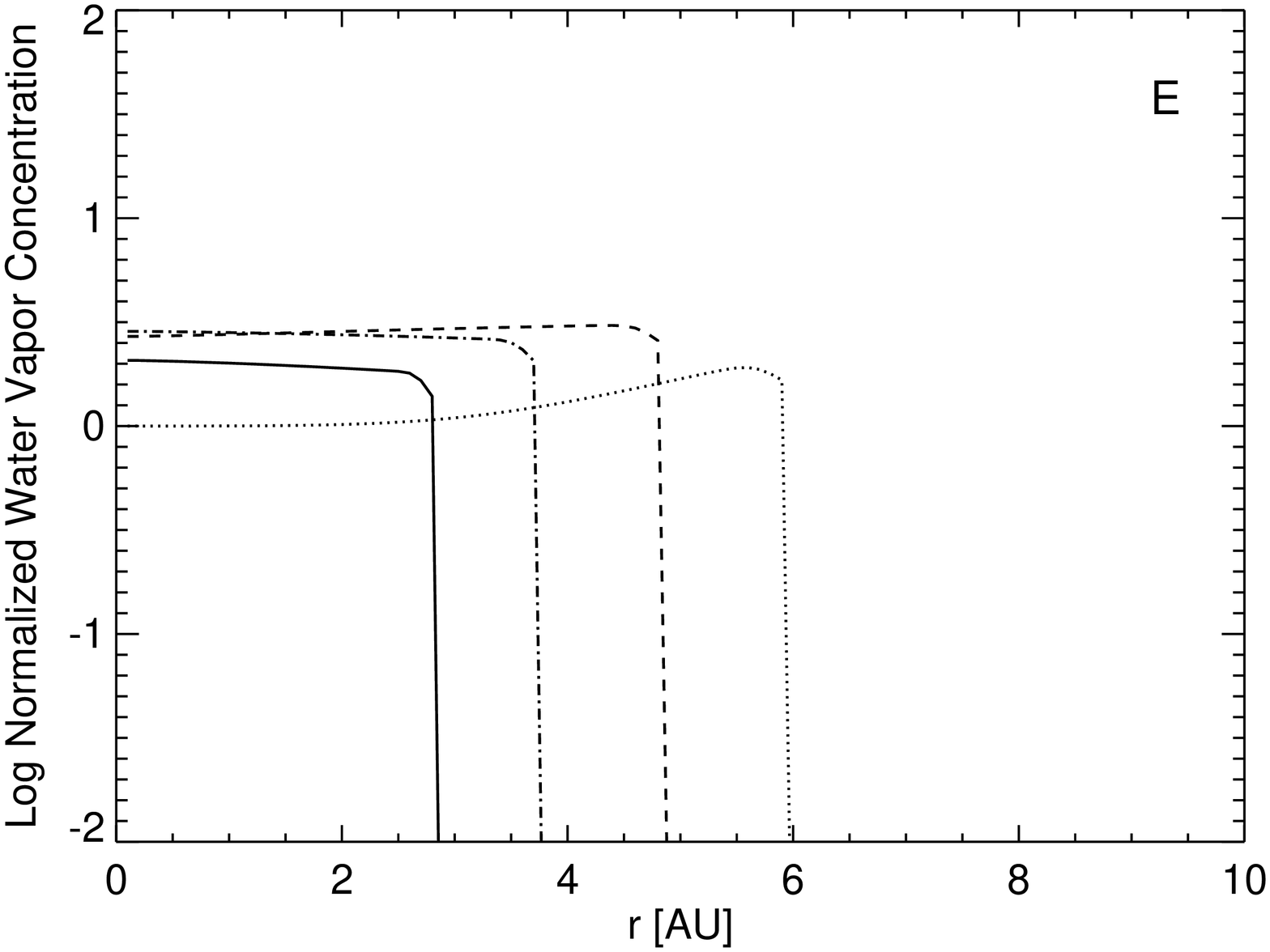}
\includegraphics[angle=90,width=3.4in]{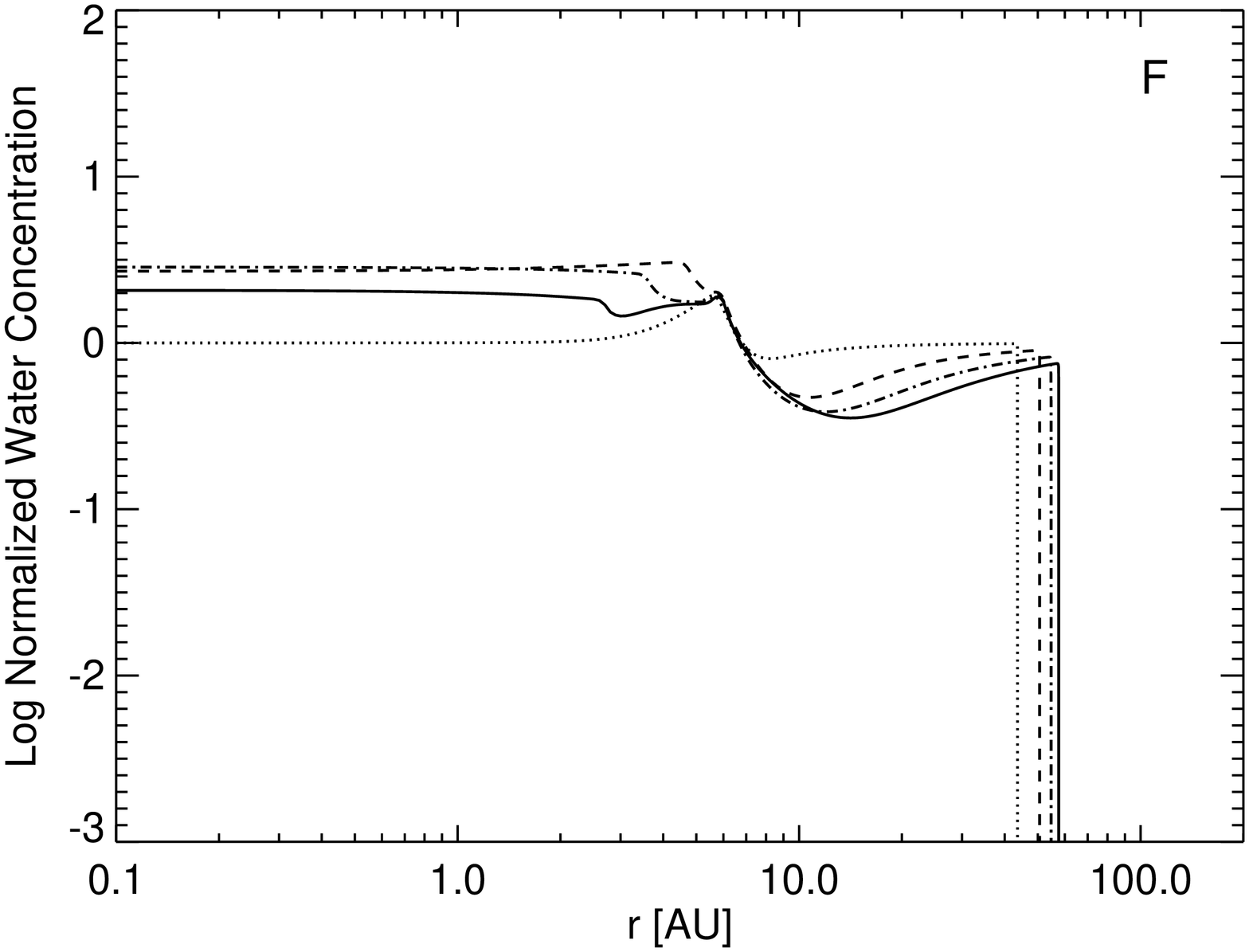}
\caption{Disk and water evolution for Case 2.  Same as Figure 1, with 
the 2$\times 10^{6}$ (dash-dot) and 3$\times 10^{6}$ (solid) years of
evolution plotted.}
\end{figure}

\begin{figure}
\includegraphics[angle=90,width=3.4in]{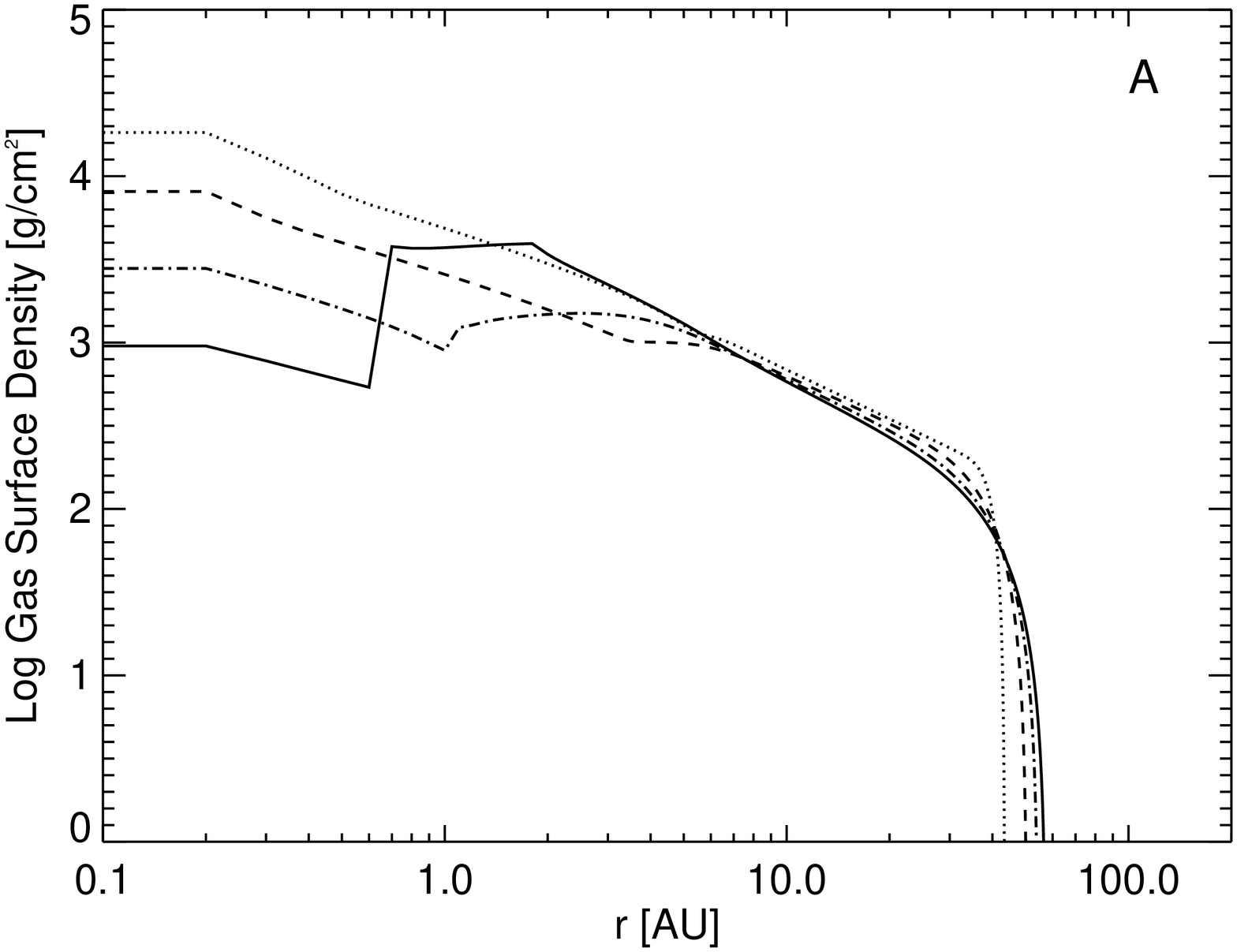}
\includegraphics[angle=90,width=3.4in]{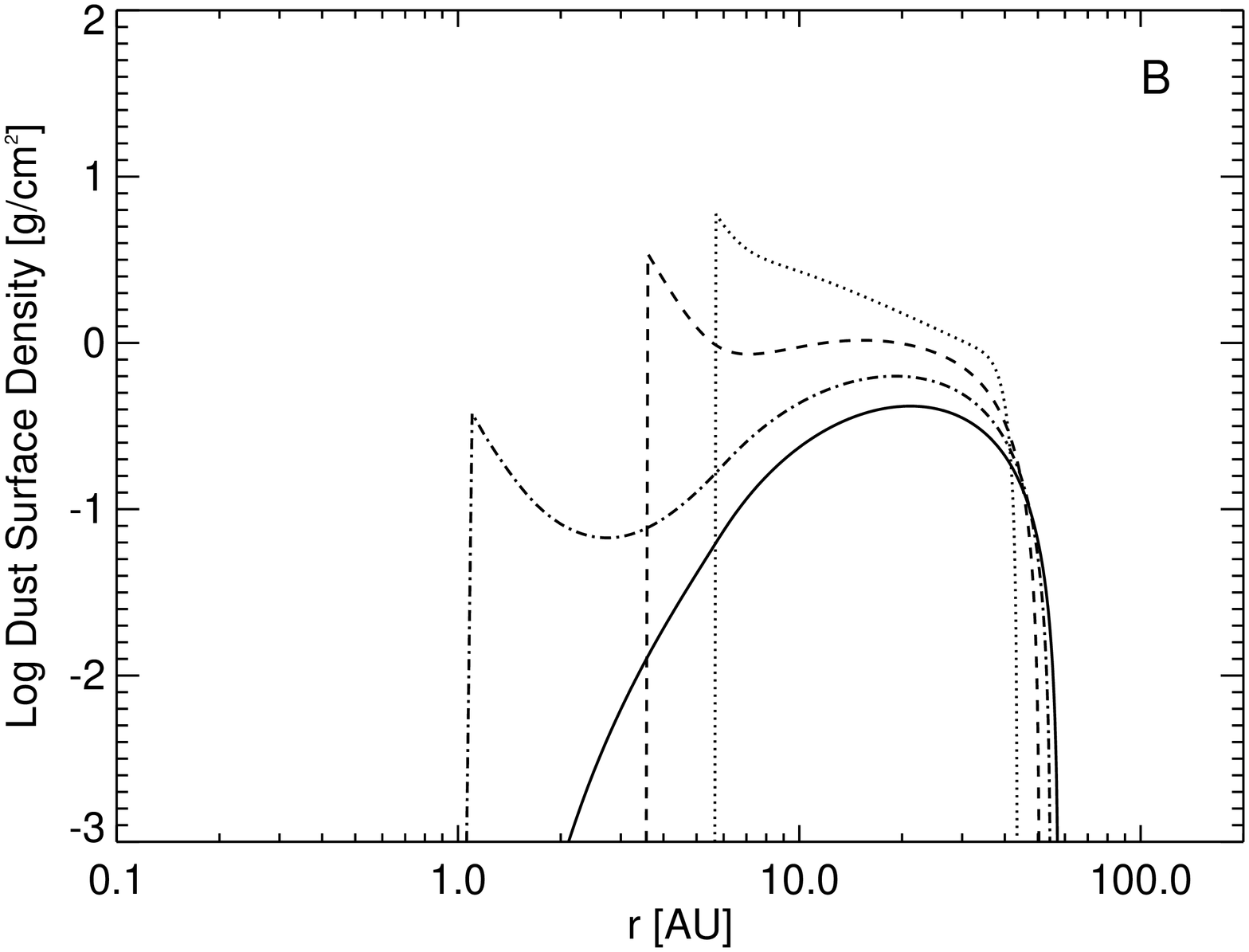}
\includegraphics[angle=90,width=3.4in]{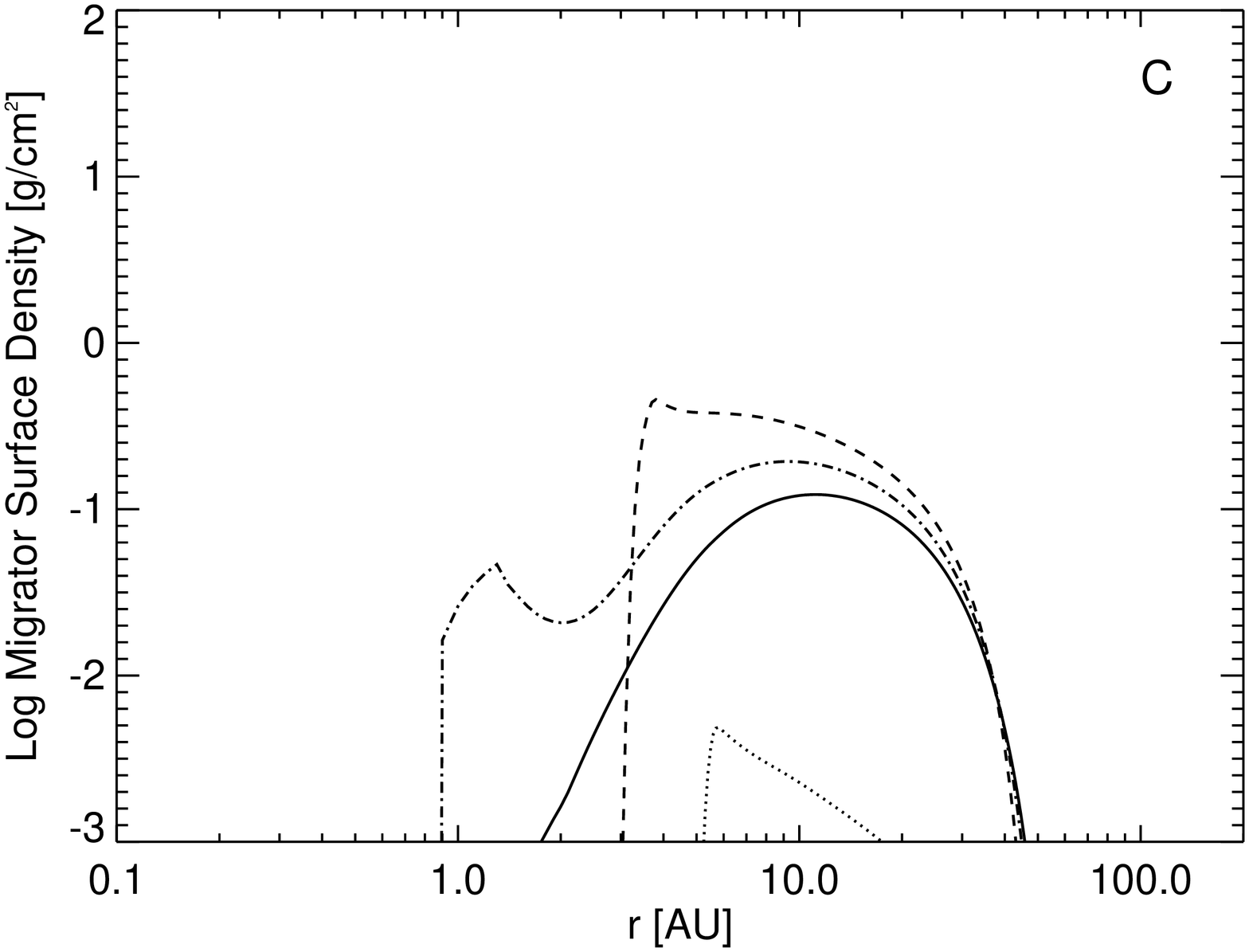}
\includegraphics[angle=90,width=3.4in]{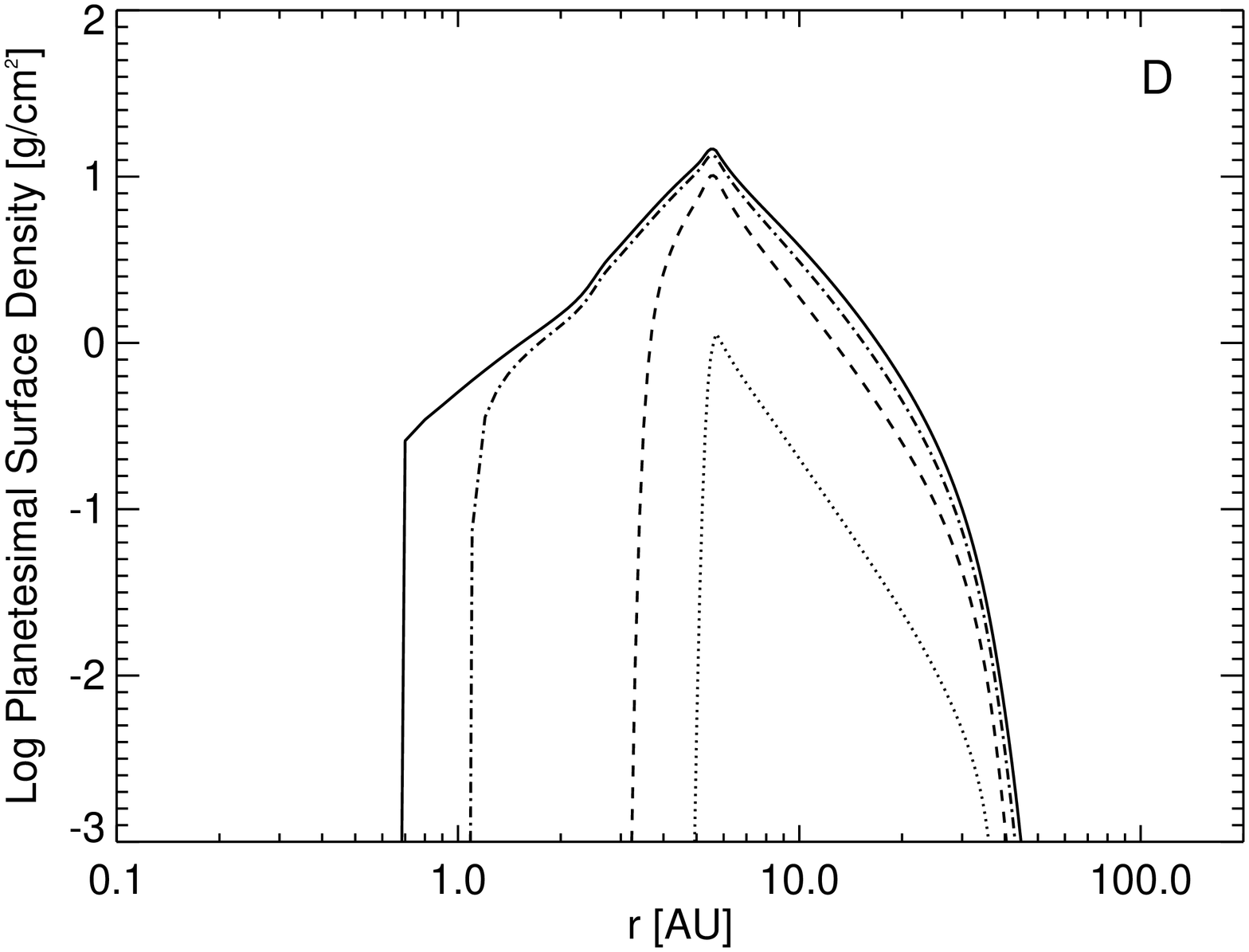}
\includegraphics[angle=90,width=3.4in]{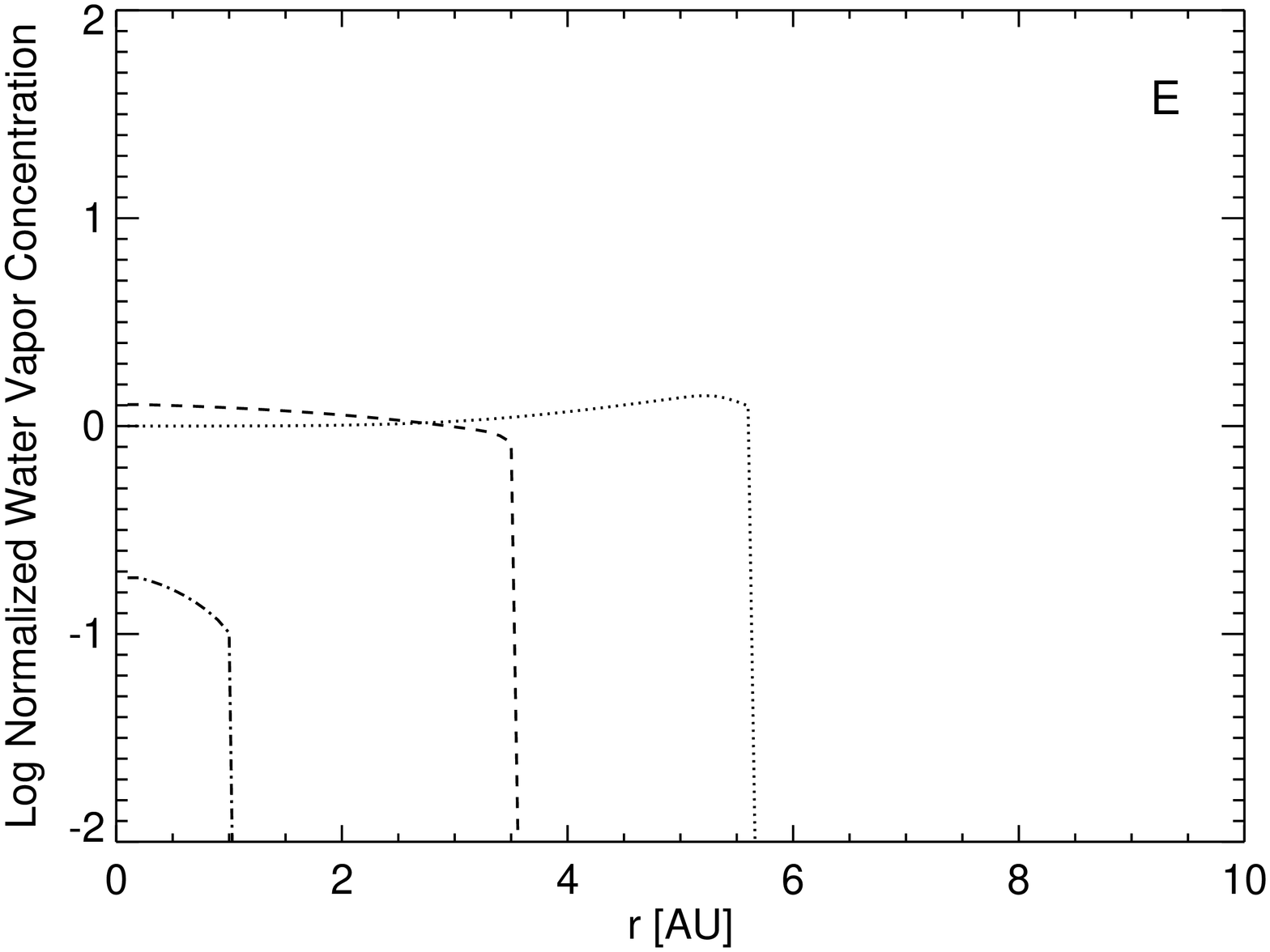}
\includegraphics[angle=90,width=3.4in]{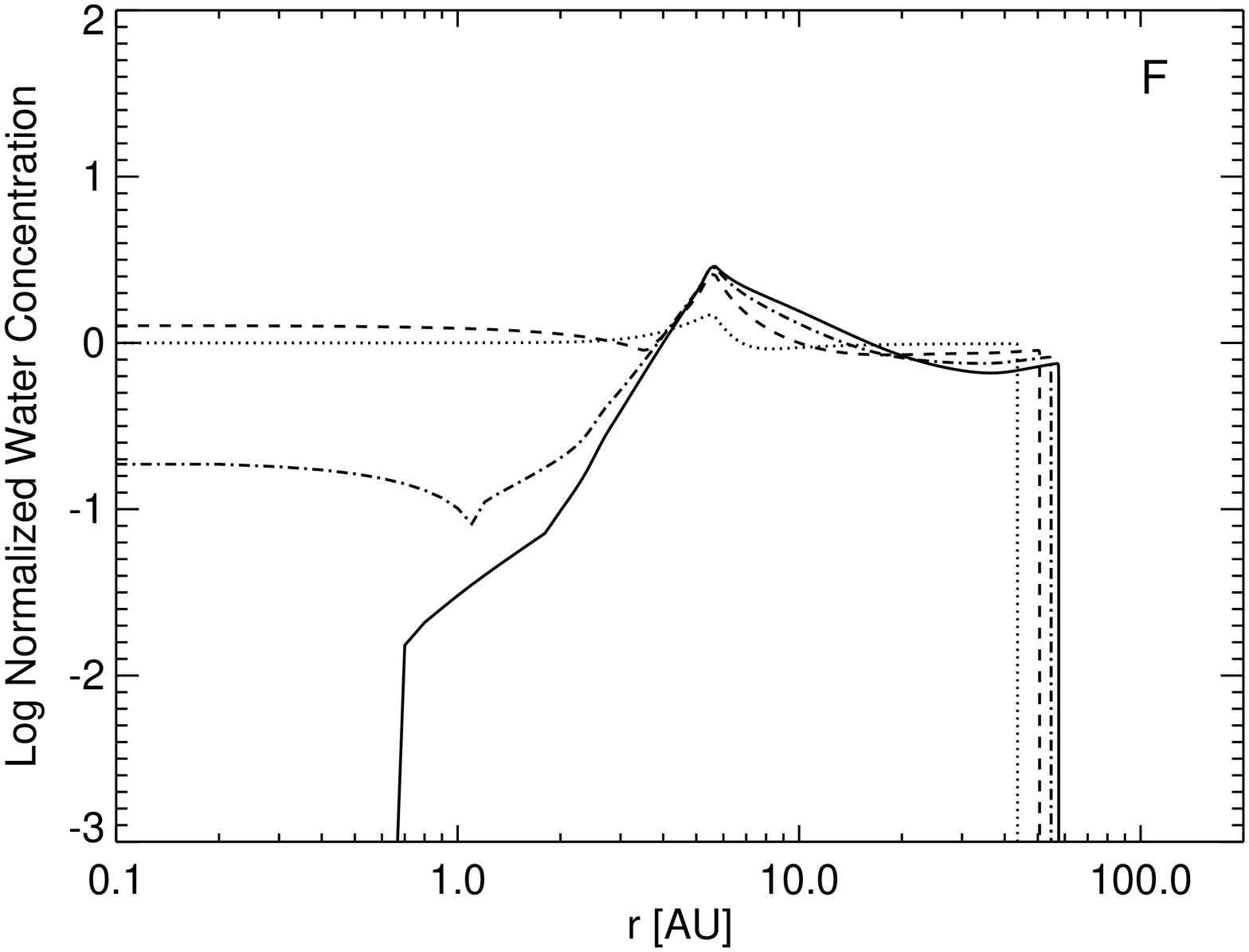}
\caption{Disk and water evolution for Case 3.  Same as Figure 2.}
\end{figure}

\begin{figure}
\includegraphics[angle=90,width=3.4in]{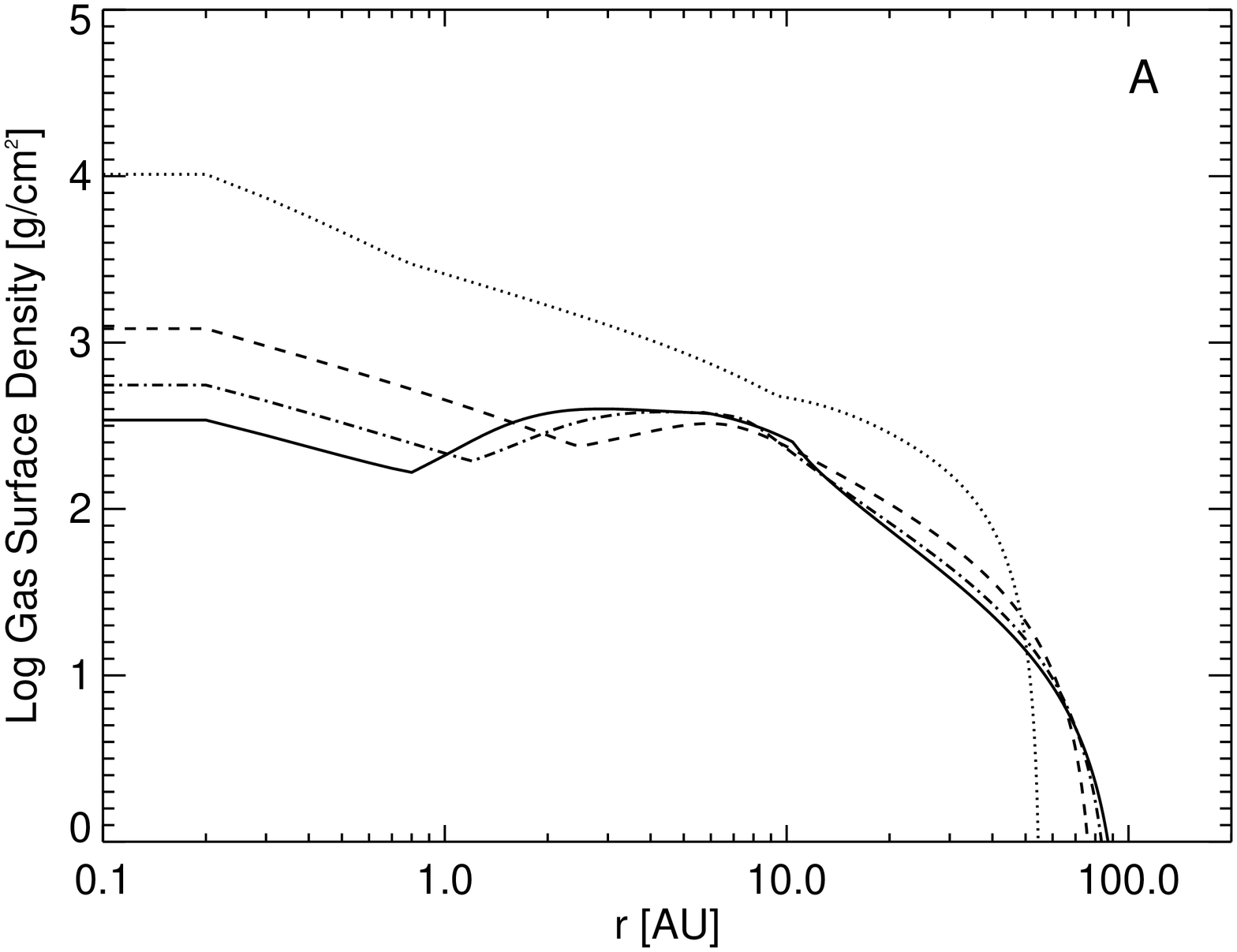}
\includegraphics[angle=90,width=3.4in]{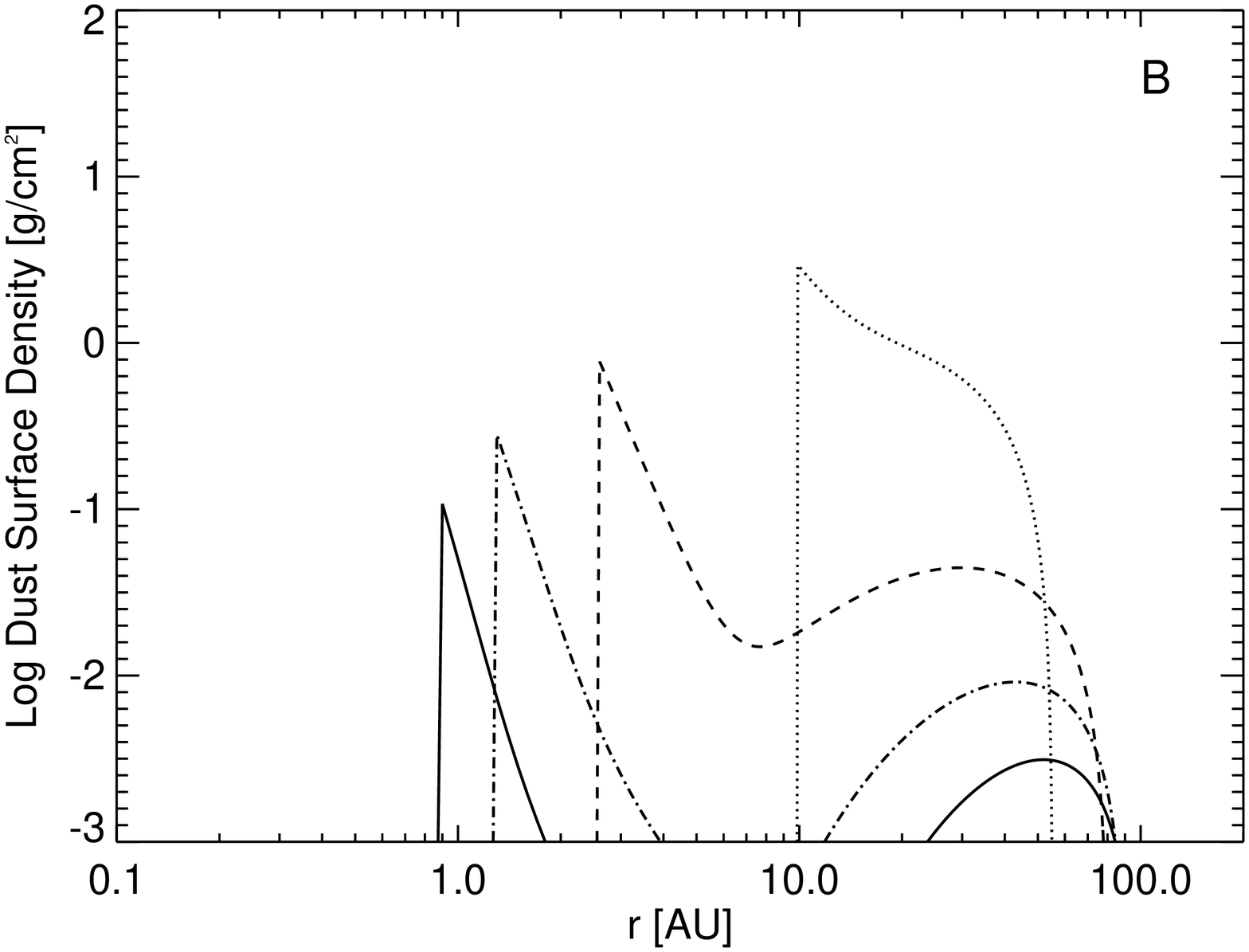}
\includegraphics[angle=90,width=3.4in]{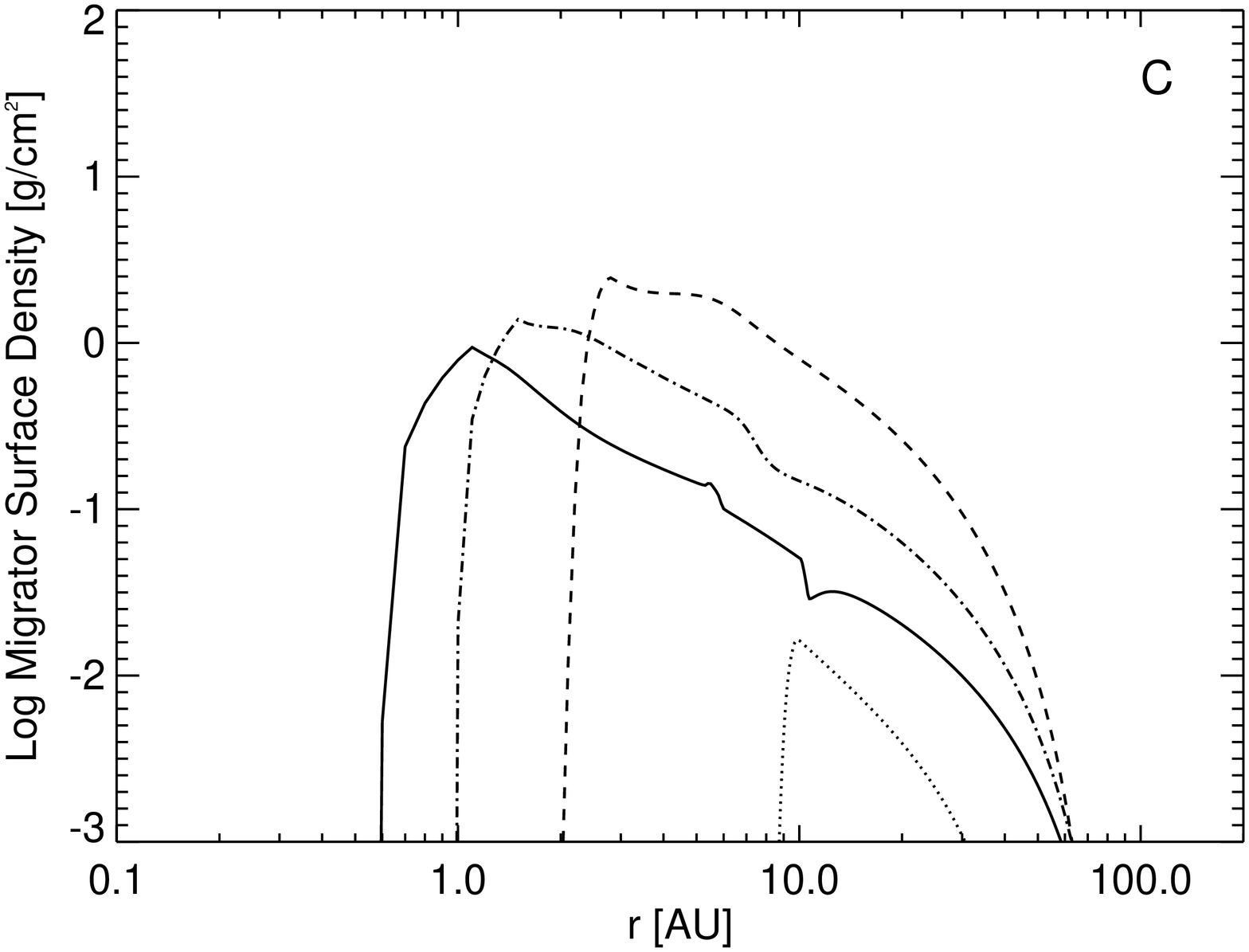}
\includegraphics[angle=90,width=3.4in]{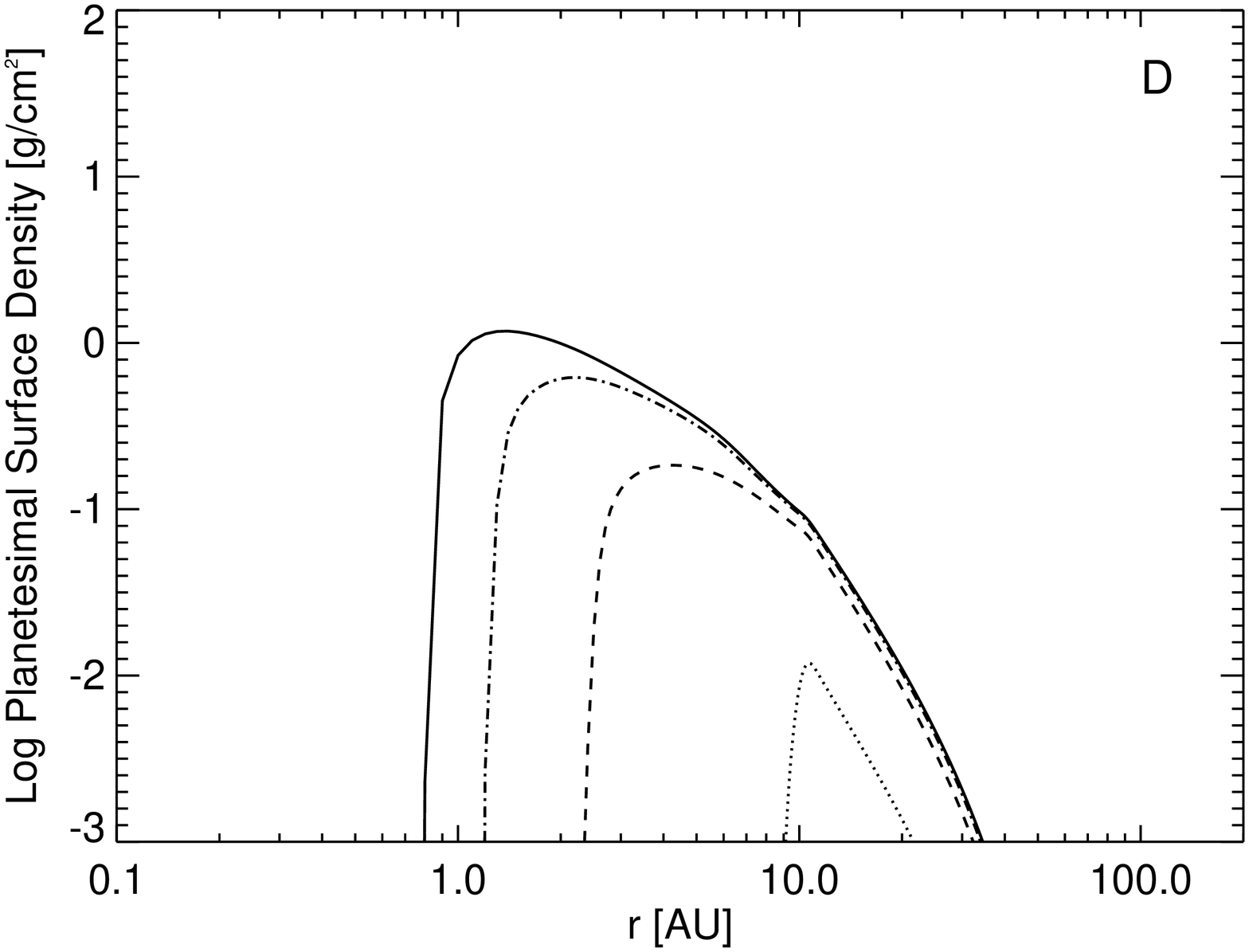}
\includegraphics[angle=90,width=3.4in]{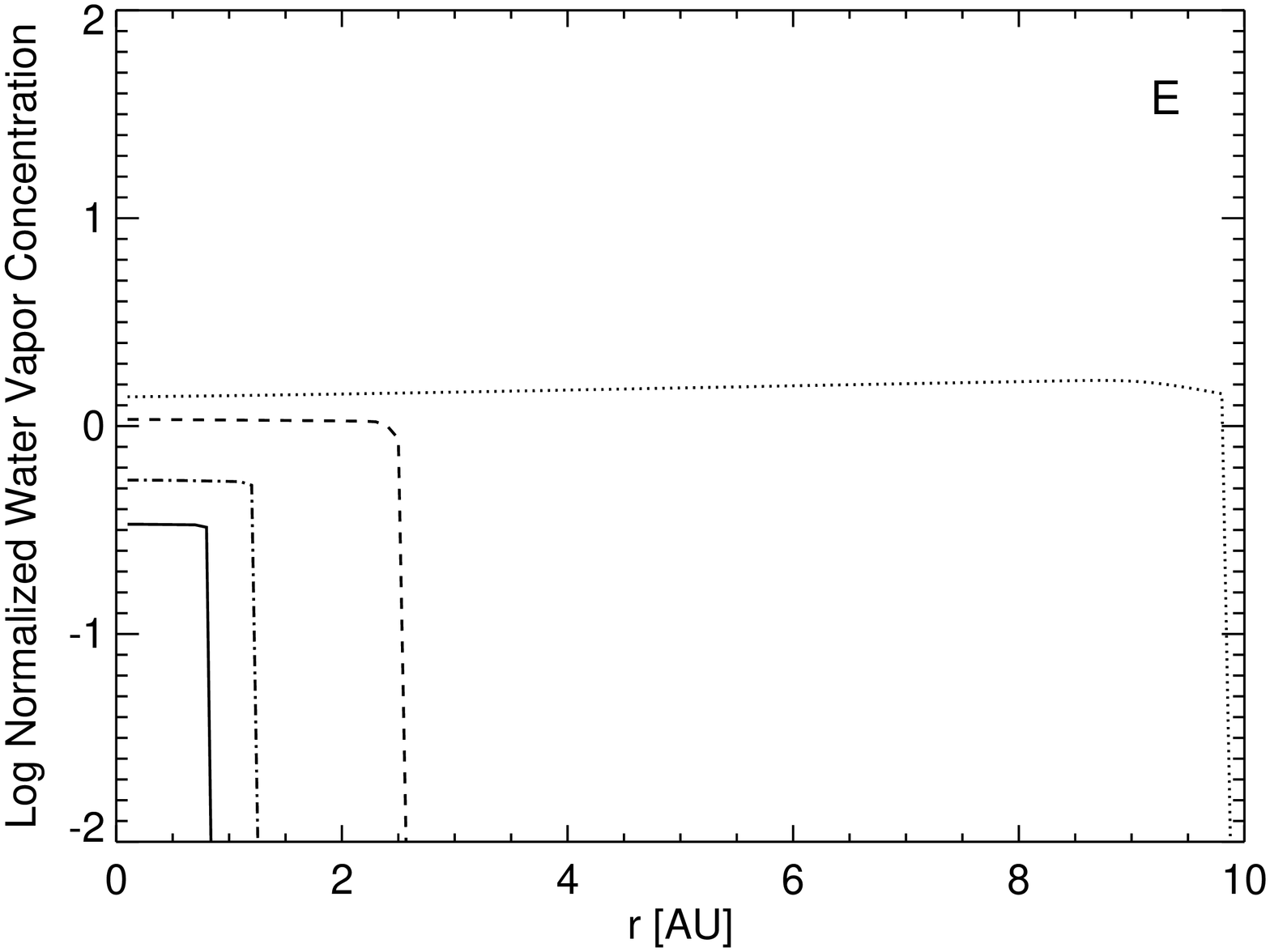}
\includegraphics[angle=90,width=3.4in]{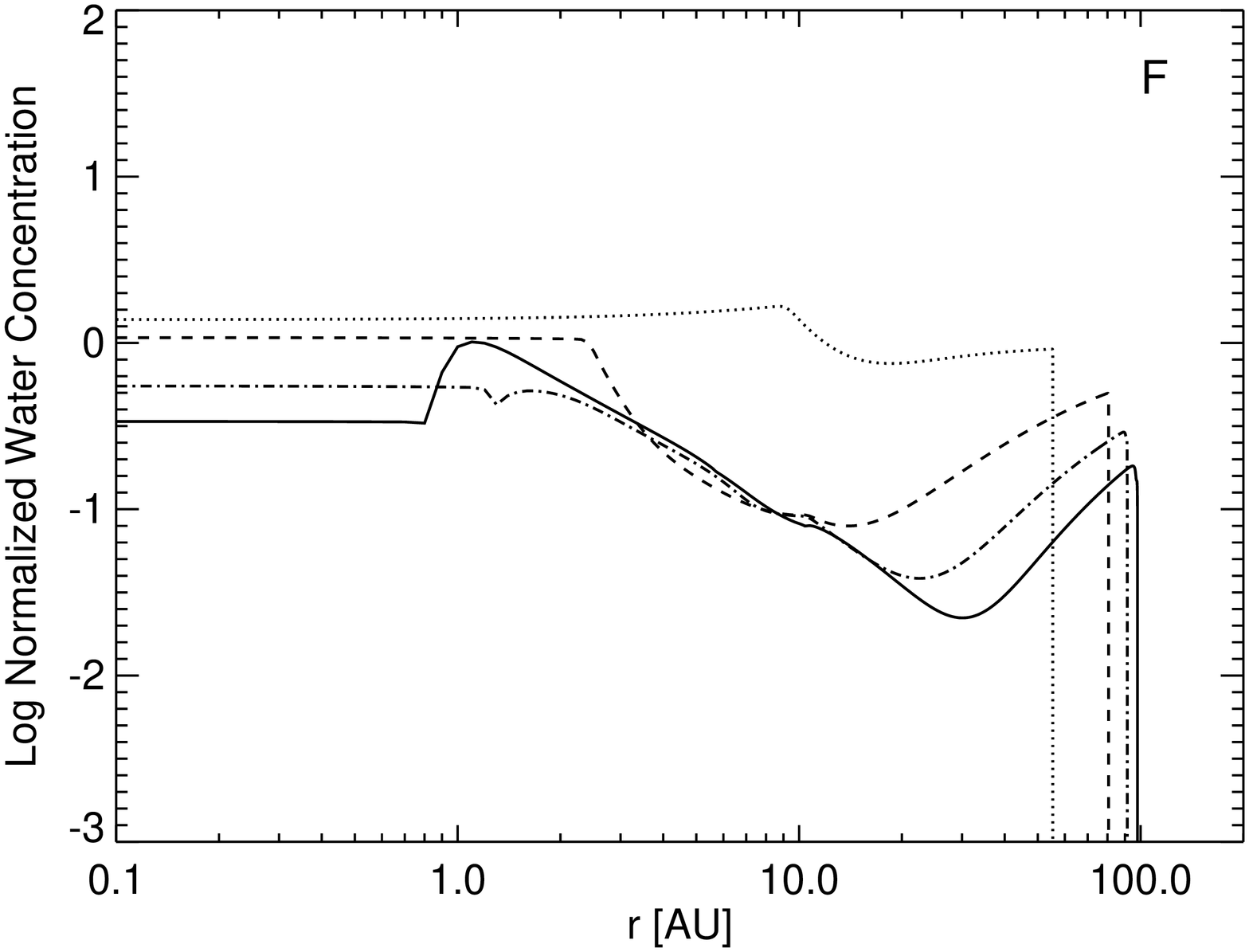}
\caption{Disk and water evolution for Case 4.  Same as Figure 2.}
\end{figure}

\begin{figure}
\includegraphics[angle=90,width=3.4in]{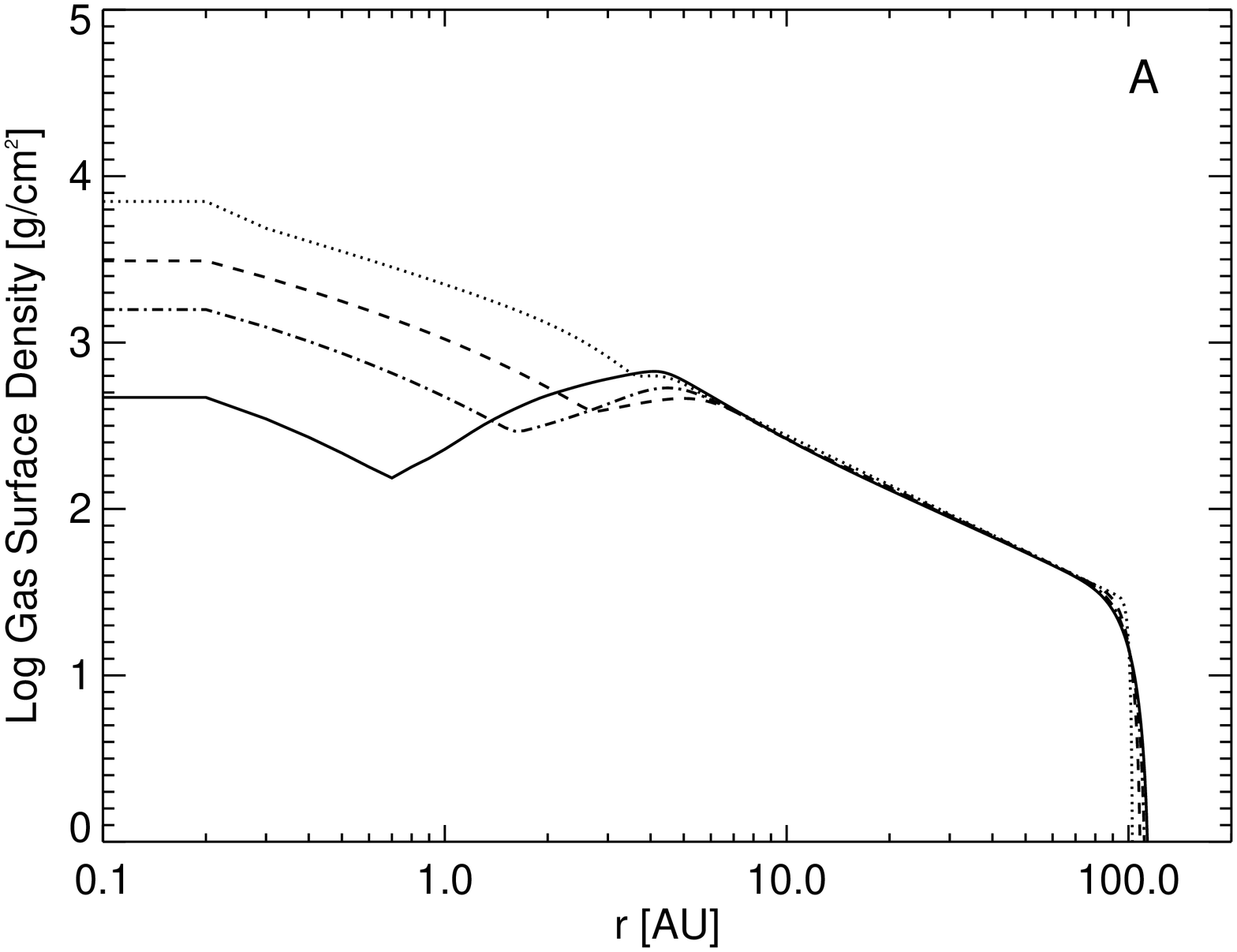}
\includegraphics[angle=90,width=3.4in]{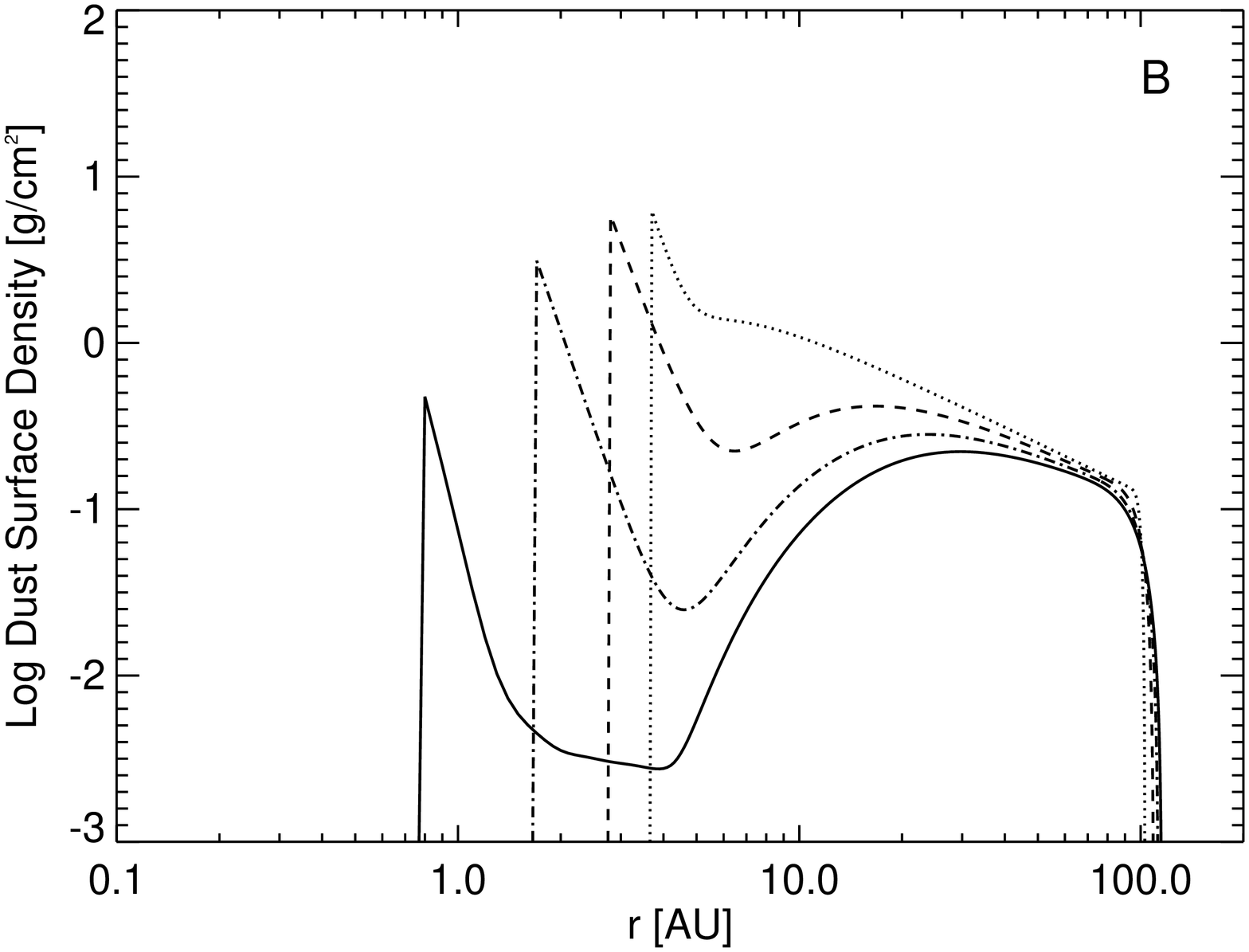}
\includegraphics[angle=90,width=3.4in]{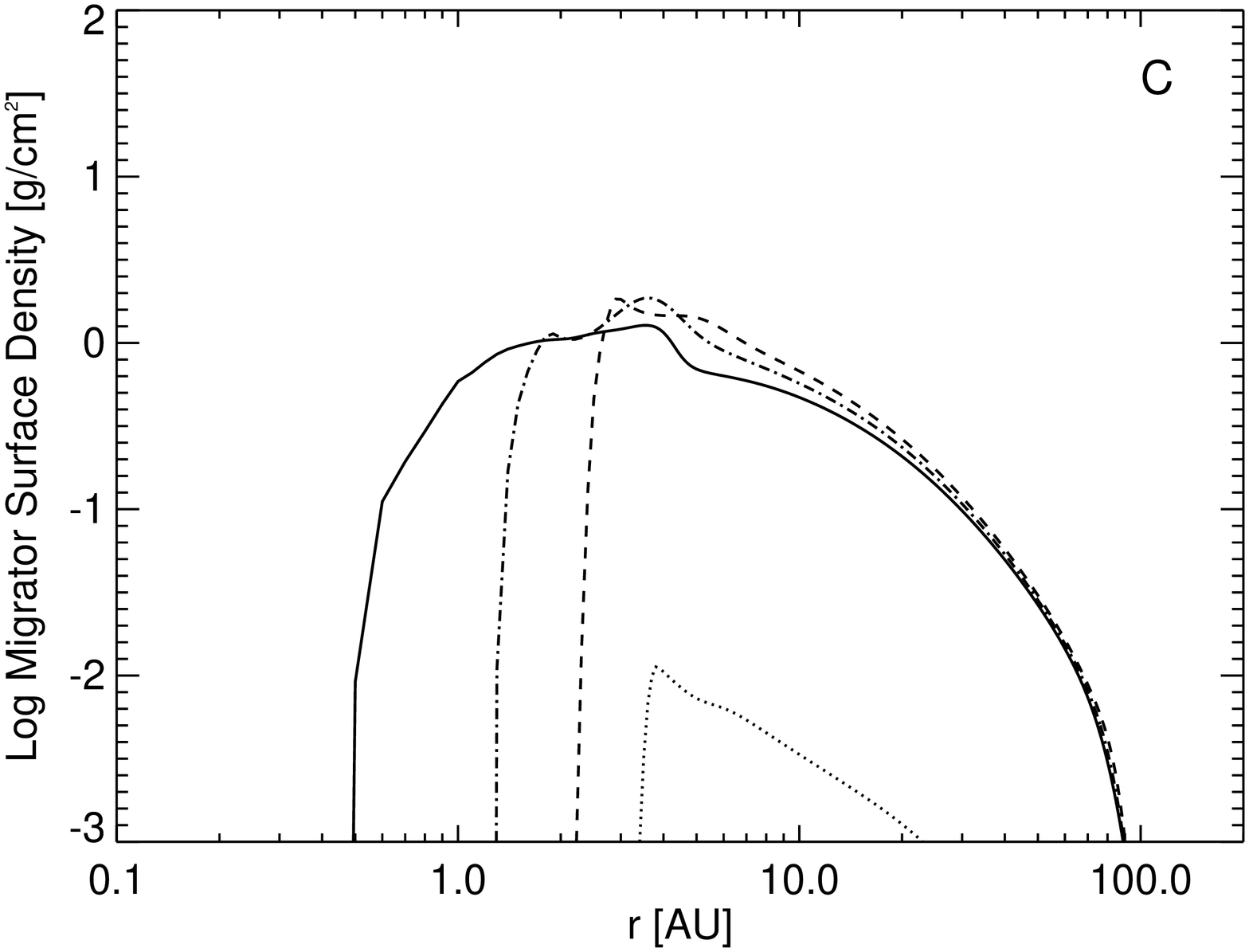}
\includegraphics[angle=90,width=3.4in]{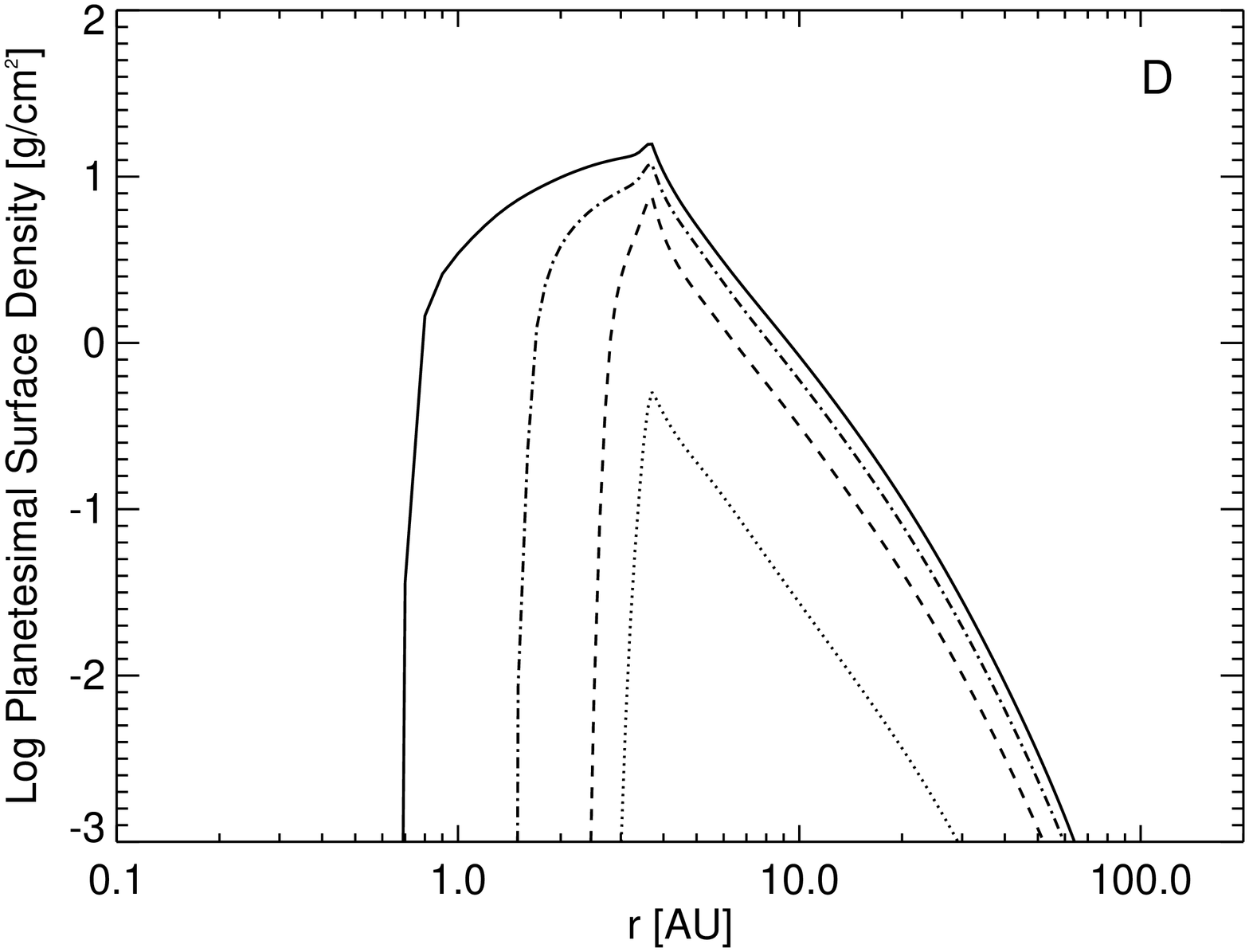}
\includegraphics[angle=90,width=3.4in]{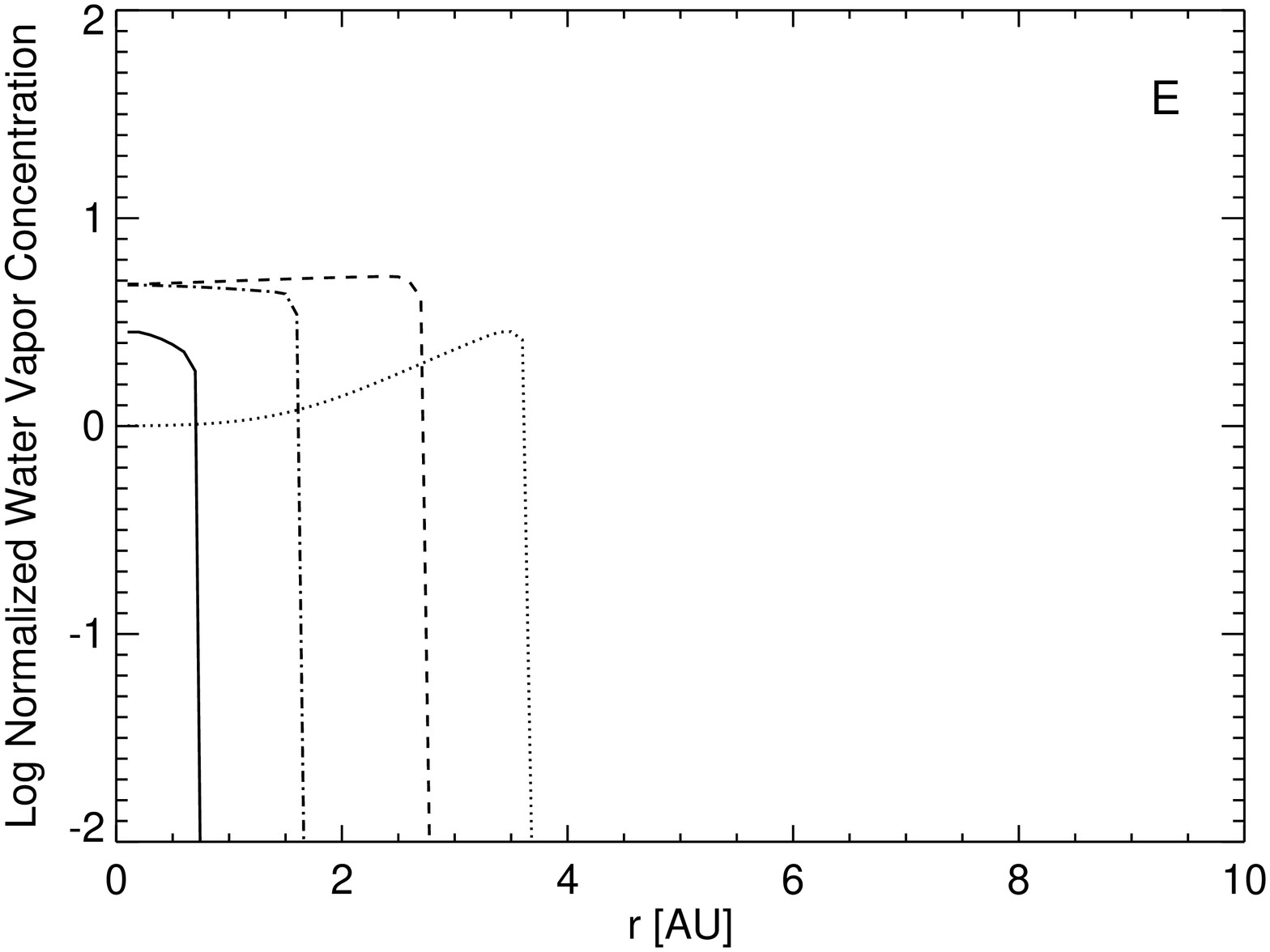}
\includegraphics[angle=90,width=3.4in]{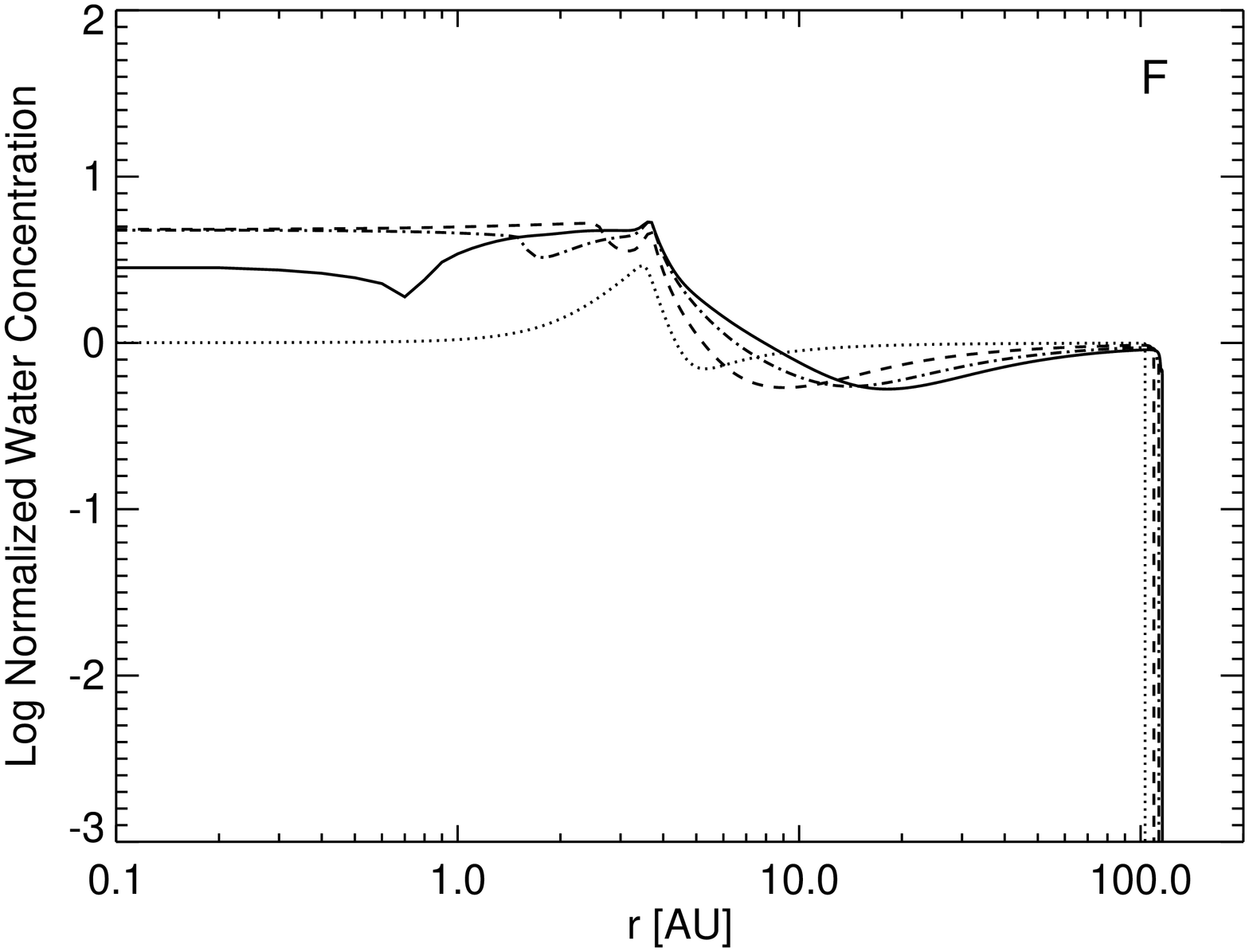}
\caption{Disk and water evolution for Case 5.  Same as Figure 2.}
\end{figure}

\begin{figure}
\includegraphics[angle=90,width=3.4in]{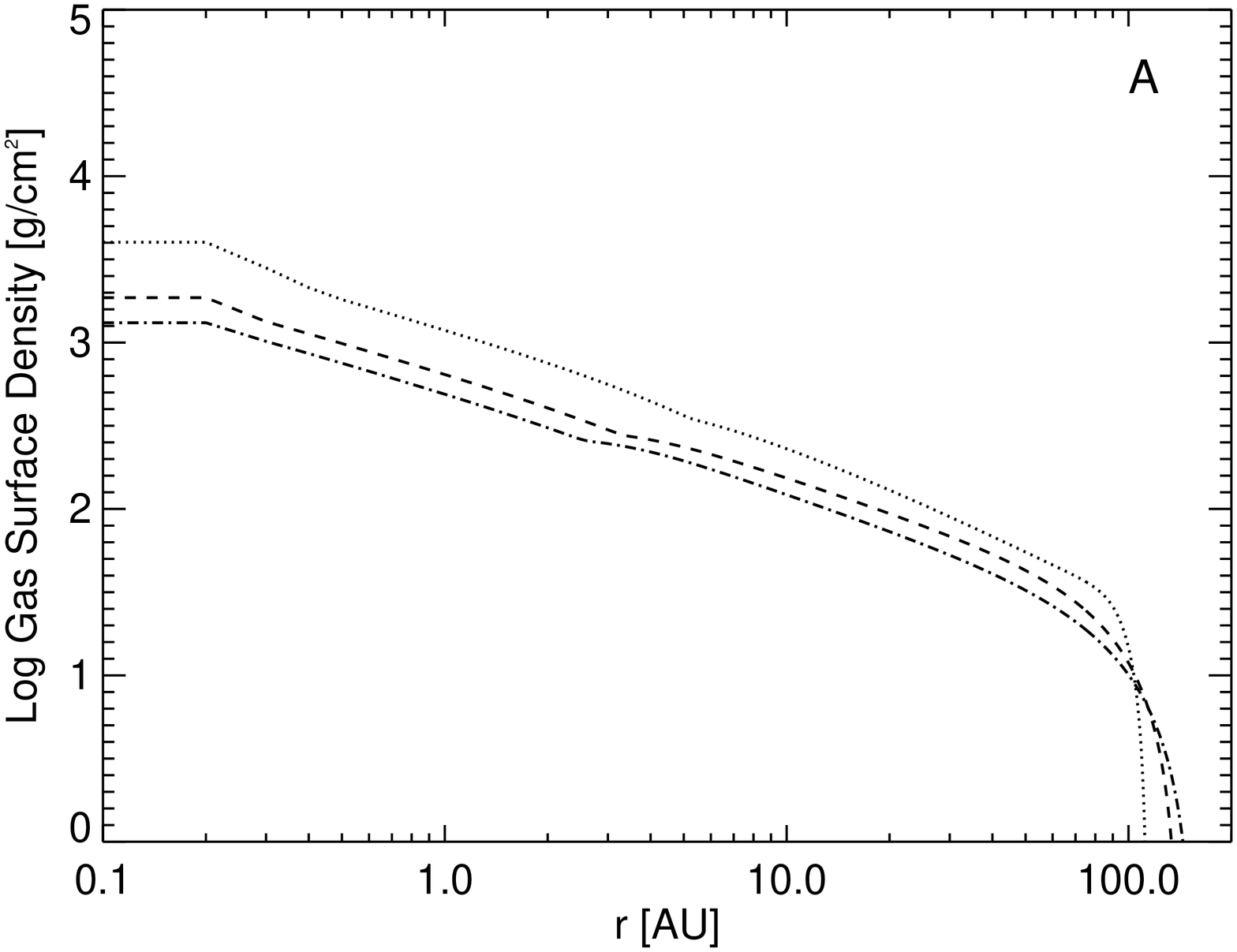}
\includegraphics[angle=90,width=3.4in]{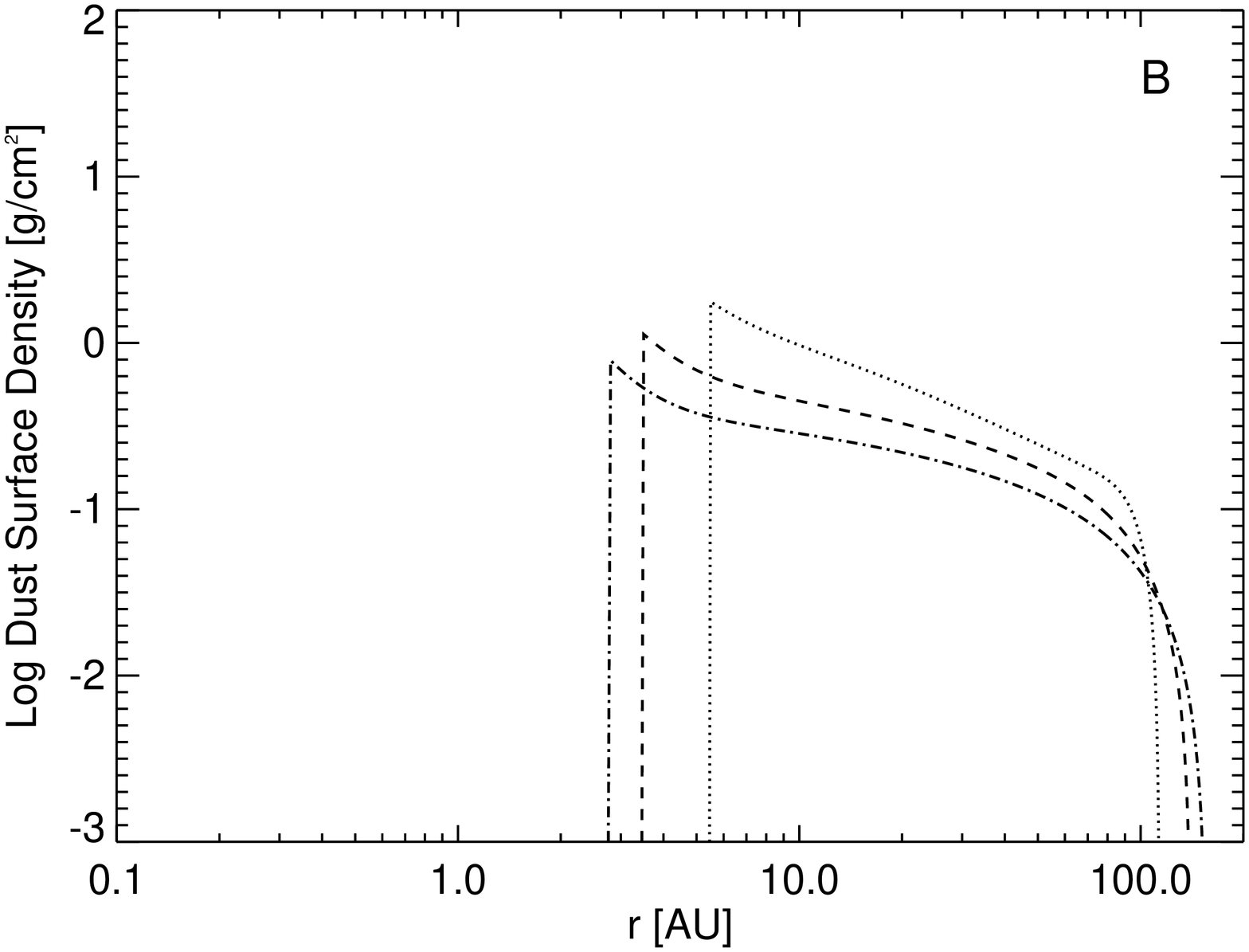}
\includegraphics[angle=90,width=3.4in]{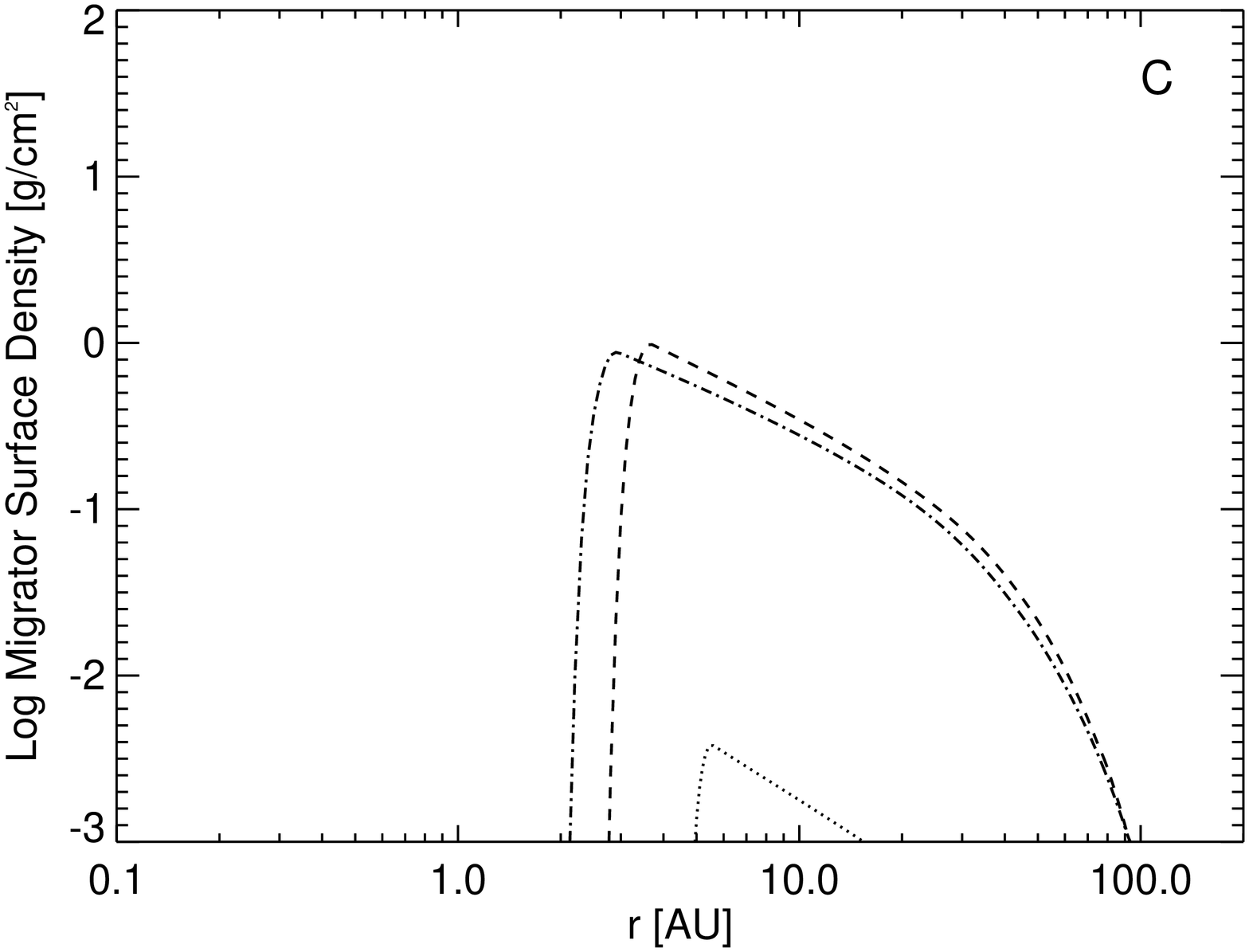}
\includegraphics[angle=90,width=3.4in]{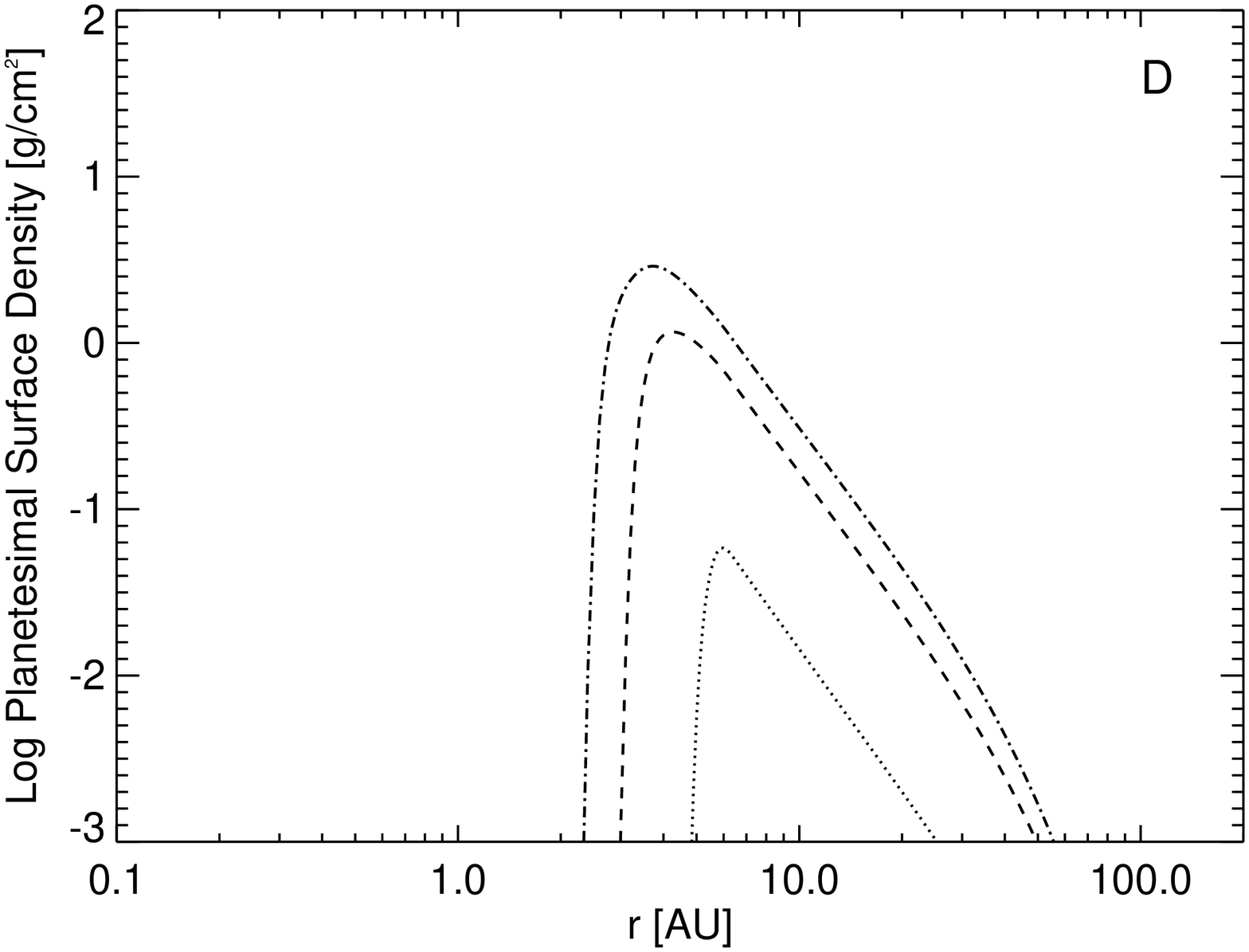}
\includegraphics[angle=90,width=3.4in]{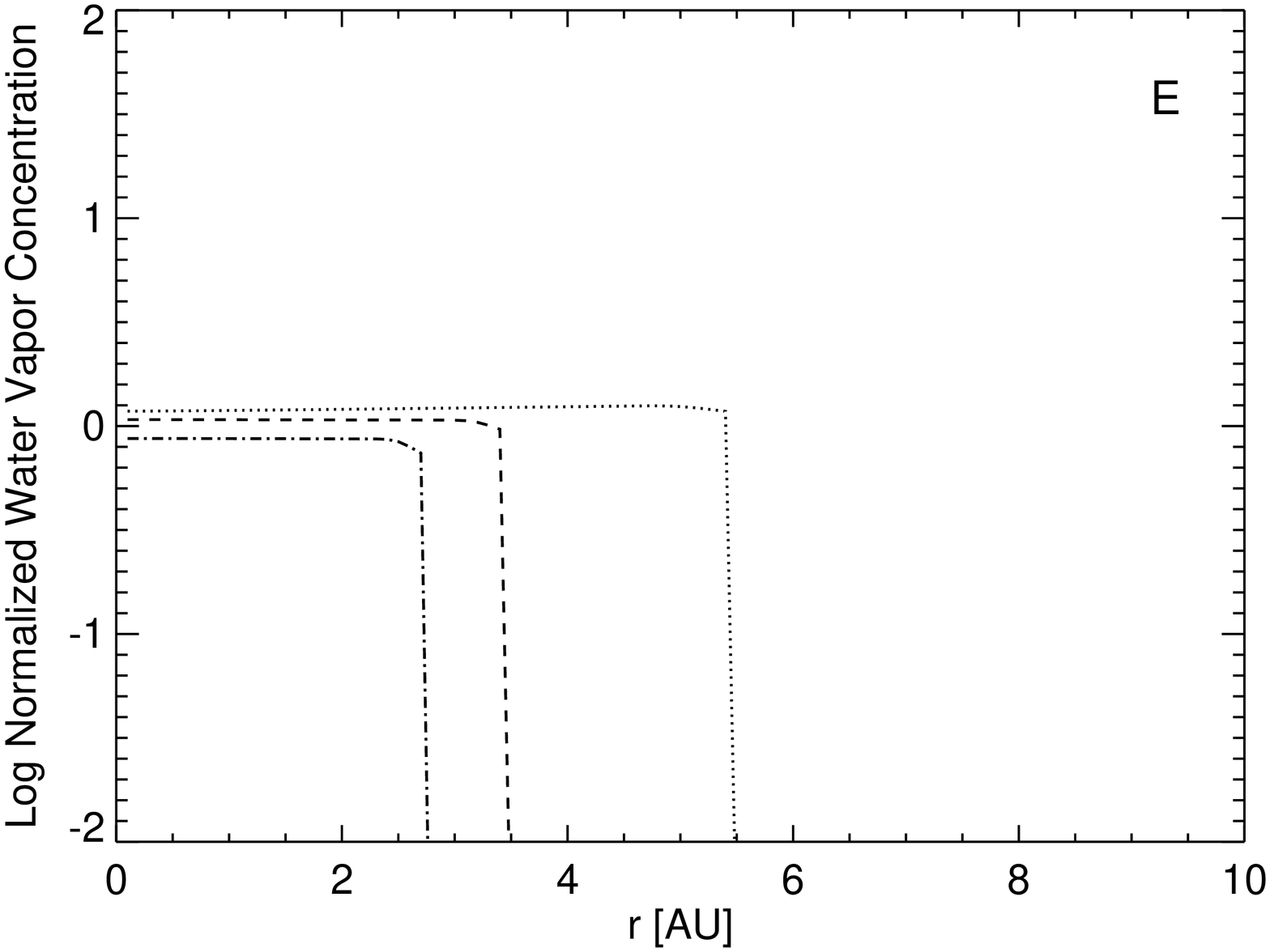}
\includegraphics[angle=90,width=3.4in]{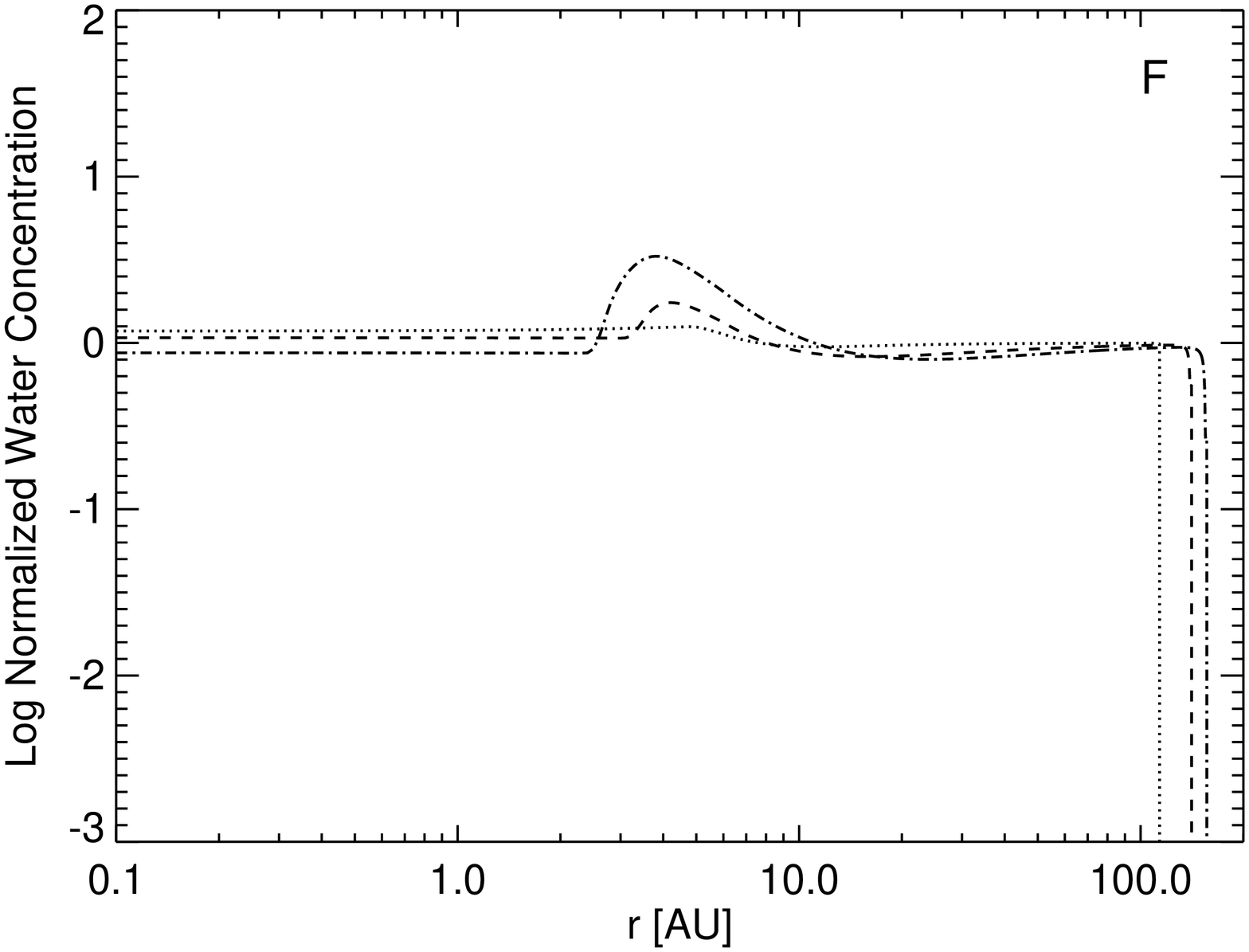}
\caption{Disk and water evolution for Case 6.  Same as Figure 2.}
\end{figure}

\begin{figure}
\includegraphics[angle=90,width=3.4in]{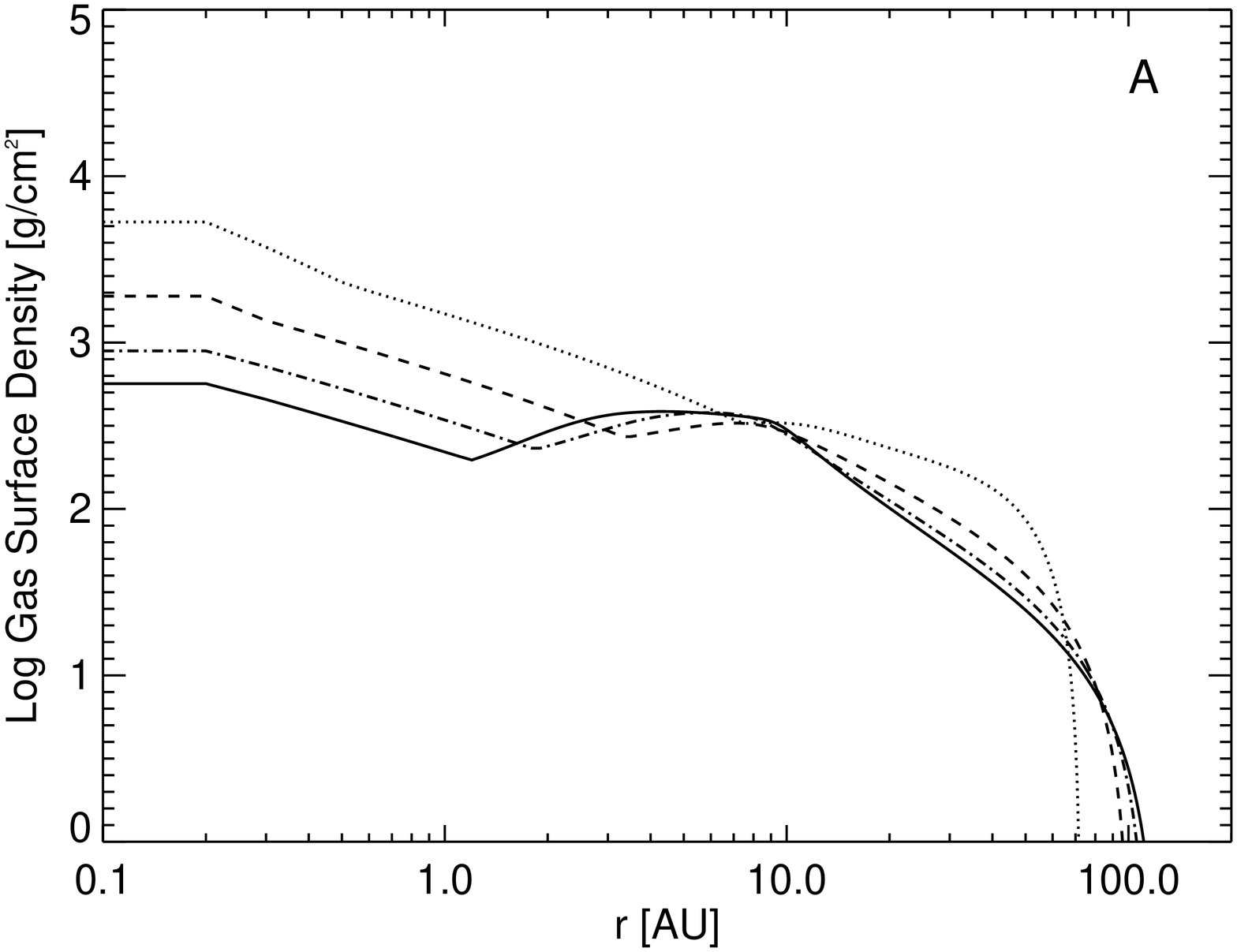}
\includegraphics[angle=90,width=3.4in]{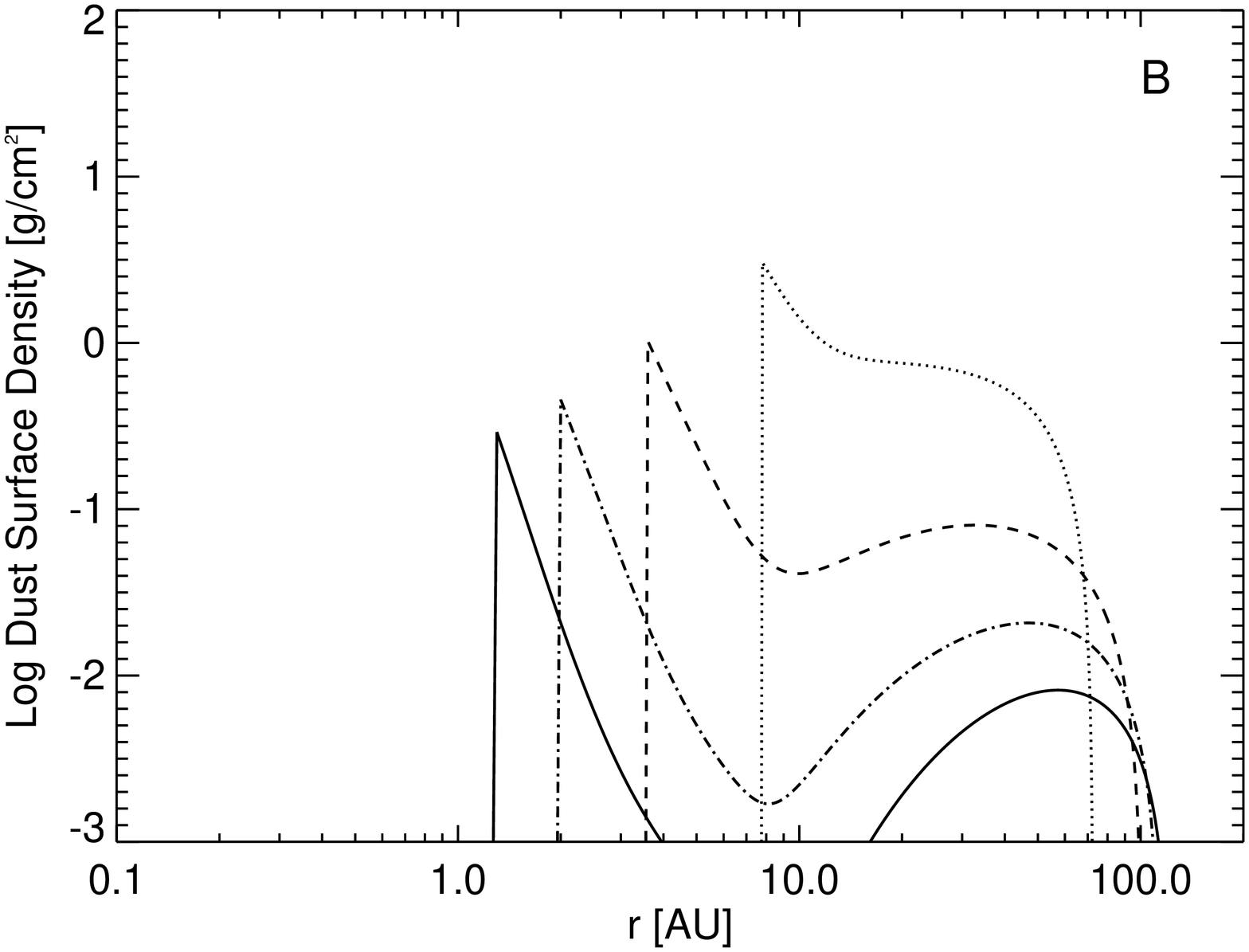}
\includegraphics[angle=90,width=3.4in]{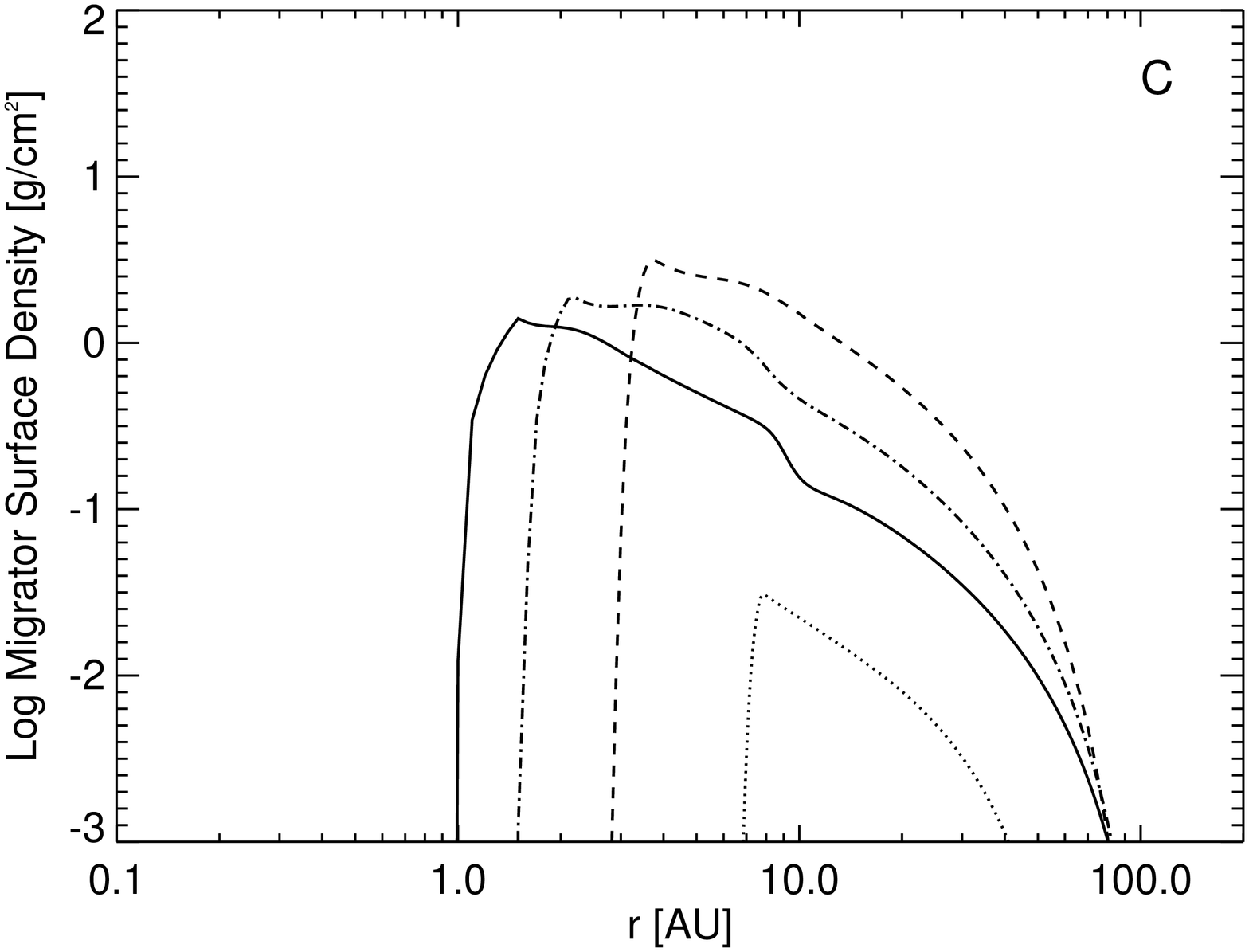}
\includegraphics[angle=90,width=3.4in]{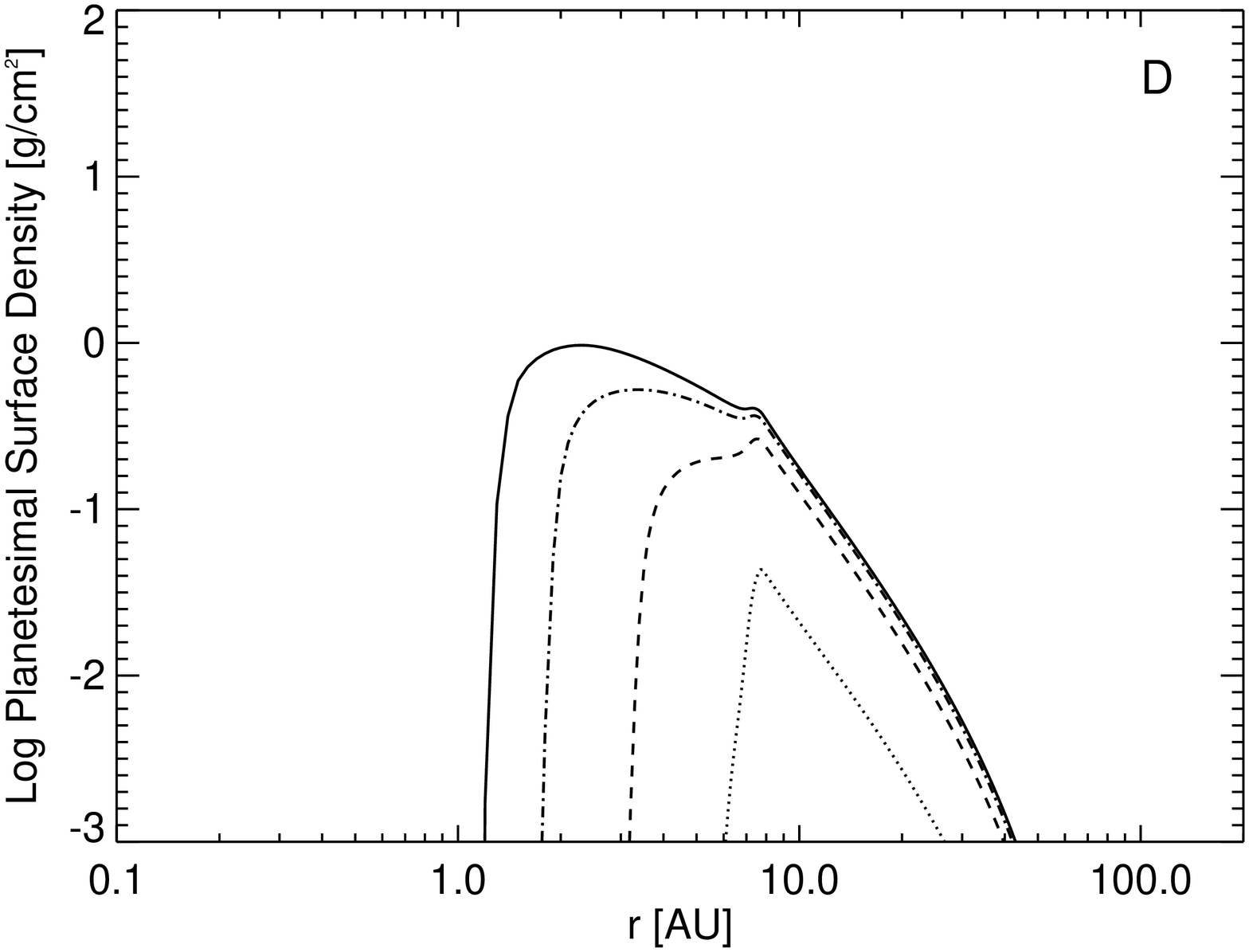}
\includegraphics[angle=90,width=3.4in]{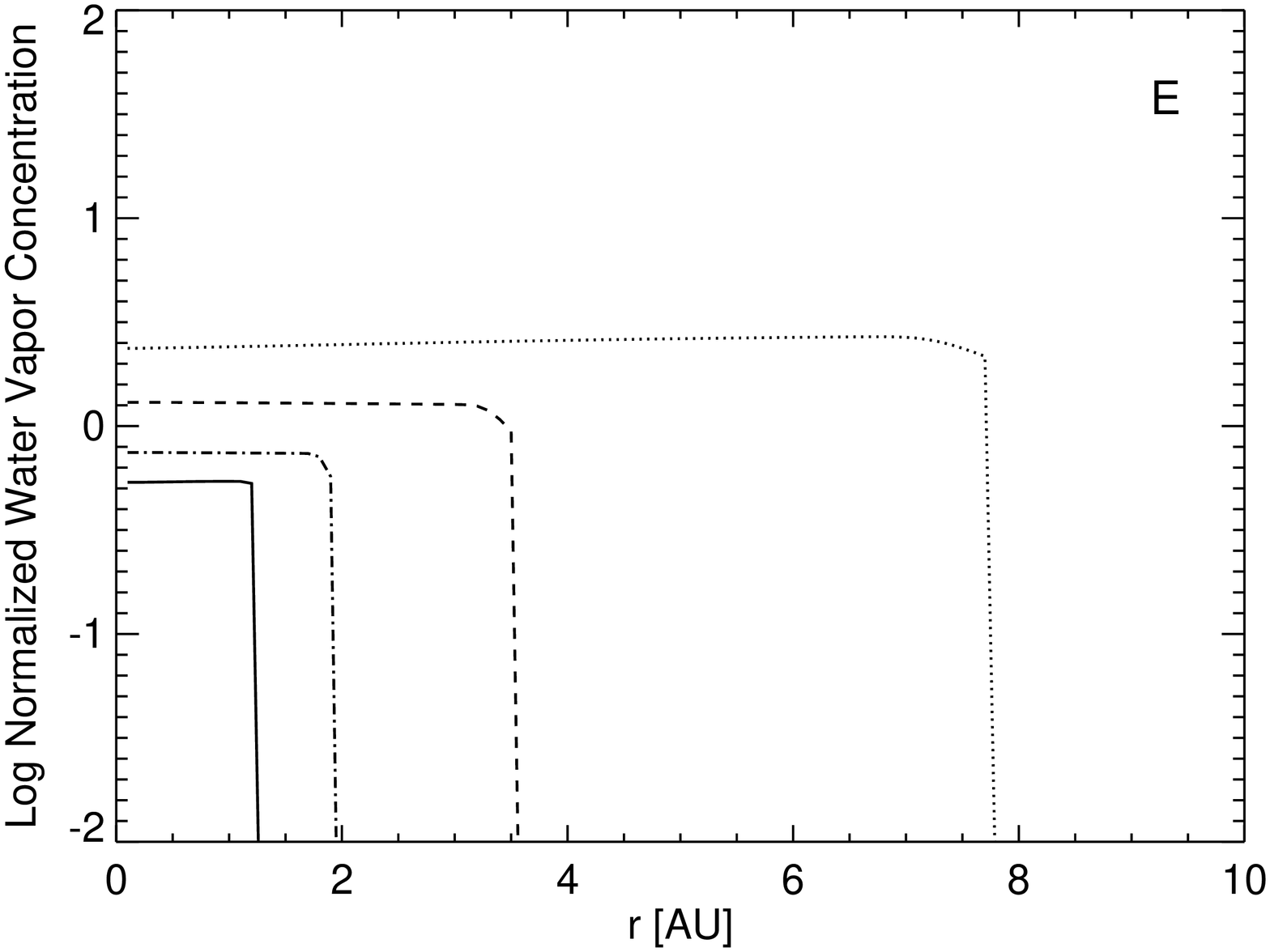}
\includegraphics[angle=90,width=3.4in]{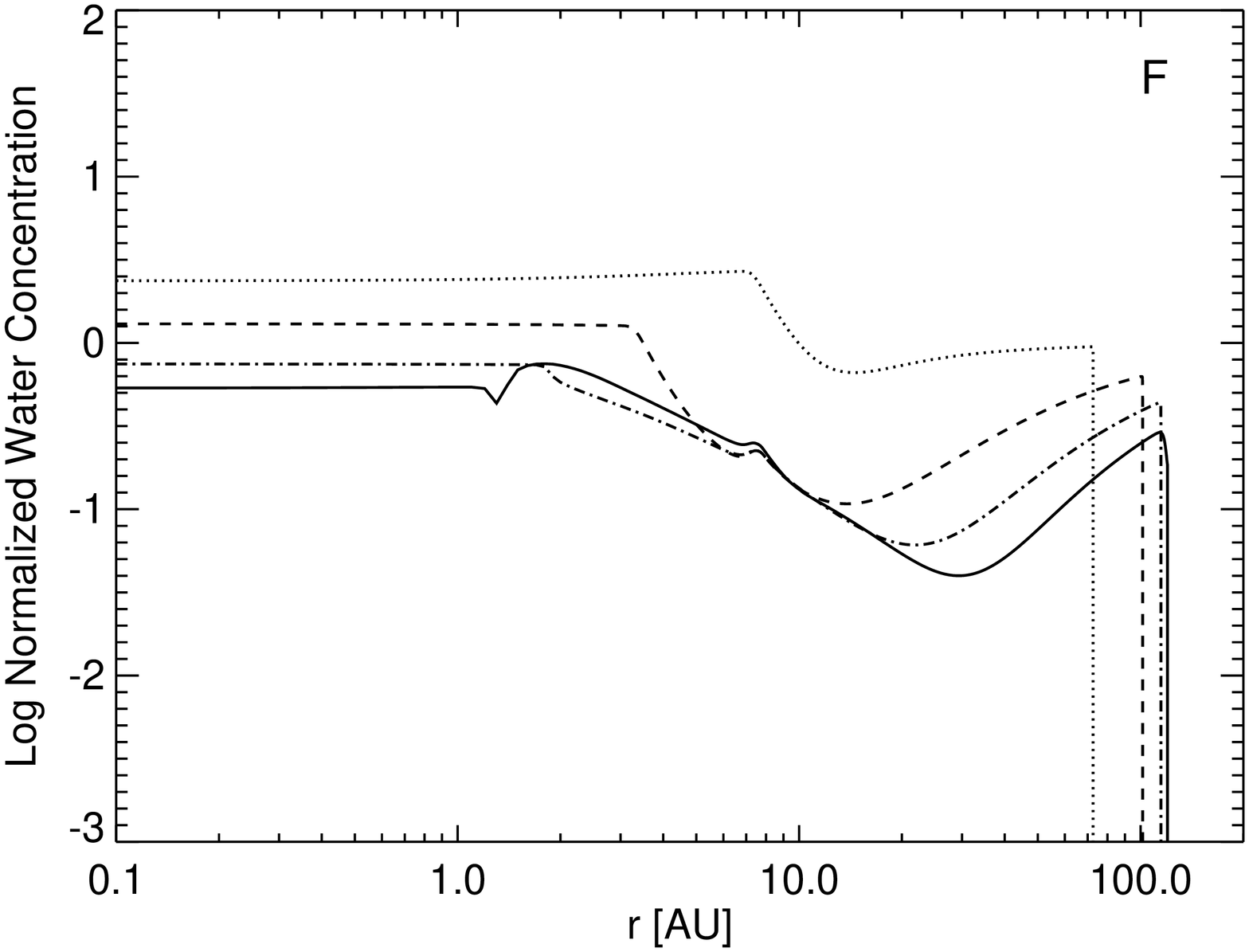}
\caption{Disk and water evolution for Case 7.  Same as Figure 2.}
\end{figure}

\end{document}

%% file: variables.tex
\begin{table}[!h]
\caption{Definition of variables used in this paper.}
\begin{center}
\begin{tabular}{ l l}
\hline
Variable & Definition \\
\hline
$a$ & Generic particle radius \\
$a_{1}, a_{2}$ & Coefficients for evaporation rate \\
$a_{d},a_{m},a_{p}$ & Radii of dust, migrators, and planetesimals \\
$c_{s}$ & Sound speed \\
$H$ & Nebular scale height \\
$H_{d},H_{m}$ & Half-thickness of the dust and migrator layers\\
$m$ & Mass of a solid particle \\
$m_{d},m_{m},m_{p}$ & Masses of individual dust, migrator, and planetesimal,
bodies \\
$\overline{m}$ & Average mass of a gas molecule \\
$m_{p}$ & Mass of a proton \\
$M_{\odot}$ & Mass of star \\
$\dot{M}$ & Mass accretion rate onto the star \\
$P_{eq},P_{vap}$ & Equilibrium and actual pressure of water in the nebula \\
$r$ & Distance from the star \\
$S_{vap},S_{d},S_{m},S_{p}$ & Source functions of vapor, dust, migrators and
planetesimals \\
$St$ & Stokes number \\
$t$ & Time \\
$T$ & Temperature of gas\\
$T_{e},T_{m}$ & Effective temperature and midplane temperature of the nebula \\
$v_{drag}$ & Inward migration velocity of migrators due to gas drag \\
$v_{r}$ & Advective velocity of the gas \\
$v_{turb}$ & Turbulent velocity of the gas \\
$Z$ & Evaporation rate of ice particles \\
$\gamma$ & Ratio of gas specific heats \\
$\nu,\nu_{m}$ & Viscosity/Diffusivity of the nebular gas and migrators\\
$\rho_{g}$ & Mass density of gas at the nebular midplane \\
$\rho_{i}$ & Mass density of ice \\
$\Sigma, \Sigma_{p}$ & Surface density of the gas and planetesimals \\
$\tau$ & Optical depth from nebular midplane to surface \\
$\Omega$ & Keplerian angular rotation velocity \\

\hline
\end{tabular}
\end{center}
\end{table}

%% file: parameters.tex
\begin{table}[!h]
\caption{Definition of parameters used in our model.}
\begin{center}
\begin{tabular}{ l c l}
\hline
Parameter & Units & Definition \\
\hline

$\Sigma_{0}$ & g/cm$^{2}$ & Nebular gas surface density at 1 AU \\
$p$ & - & Exponent of nebular surface density power law \\
$R_{0}$ & AU & Initial radius of disk \\
$\alpha$ & - & Turbulence parameter \\
$t_{coag}$(1 AU) & years & Coagulation timescale to form  migrators from dust at 1 AU \\
$t_{acc}$( 1 AU) & years & Accretion timescale to form planetesimals from migrators at 1 AU \\
$q^{*}$ & erg/g & Strength of solids \\
\hline
\end{tabular}
\end{center}
\end{table}

%% file: models2.tex
\begin{table}[!h]
\caption{Values of parameters used in the model runs presented here and mass 
remaining in disk and water at the end of the runs.}
\begin{center}
\begin{tabular}{ l c c c c c c c}
\hline
Parameter/Case & 1 & 2 & 3 & 4 & 5 & 6 & 7\\
\hline

$\Sigma_{0}$ & 7$\times$10$^{3}$ & 7$\times$10$^{3}$ & 7$\times$10$^{3}$ & 
7$\times$10$^{3}$ & 2.8$\times$10$^{3}$ & 2.8$\times$10$^{3}$ & 10$^{3}$ \\
$p$ &  -1  &   -1 &  -1 & -1 & -1 & -1 & -0.5 \\
$R_{0}$ & 40  &   40  &  40  & 40 & 100 & 100 & 56\\
$\alpha$ & 10$^{-4}$ & 10$^{-4}$  & 10$^{-4}$  &  10$^{-3}$  & 10$^{-4}$  
& 10$^{-3}$  & 10$^{-3}$ \\
$t_{coag}$(1 AU) & 10$^{4}$  &  10$^{5}$ &  10$^{5}$  &  10$^{4}$  &  10$^{5}$  
&  10$^{5}$ & 10$^{4}$  \\
$t_{acc}$( 1 AU) & 10$^{4}$  &  10$^{3}$ &  10$^{2}$  &  10$^{4}$  &  10$^{3}$
&  10$^{3}$ & 10$^{4}$  \\
$q^{*}$ & 10$^{6}$  & 10$^{6}$  & 10$^{6}$  & 10$^{6}$  & 10$^{6}$  & 10$^{6}$ 
& 10$^{6}$\\
\hline
Disk Properties \\
\hline
$M_{D}$ start & 0.2$M_{\odot}$ & 0.2$M_{\odot}$ & 0.2$M_{\odot}$ & 0.2$M_{\odot}$ &
0.2$M_{\odot}$ & 0.2$M_{\odot}$ & 0.2$M_{\odot}$\\
Length of model run & 10$^{6}$ yrs & 3x10$^{6}$ yrs & 3x10$^{6}$ yrs & 3x10$^{6}$ yrs &
3x10$^{6}$ yrs & 3x10$^{6}$ yrs & 3x10$^{6}$ yrs \\
$M_{D}$ end & 0.16$M_{\odot}$ & 0.14$M_{\odot}$ & 0.16$M_{\odot}$ & 0.06$M_{\odot}$ &
0.19$M_{\odot}$ & 0.10$M_{\odot}$ & 0.09$M_{\odot}$\\
$M_{H_{2}O}$ start & 302$M_{\oplus}$ & 302$M_{\oplus}$ & 302$M_{\oplus}$ & 302$M_{\oplus}$ &
302$M_{\oplus}$ & 302$M_{\oplus}$ & 302$M_{\oplus}$ \\
$M_{planetesimals}$ end & 5.8$M_{\oplus}$ & 47$M_{\oplus}$ & 156$M_{\oplus}$ & 4.9$M_{\oplus}$ &
62$M_{\oplus}$ & 27$M_{\oplus}$ & 8$M_{\oplus}$ \\
$M_{solids}$ end & 41$M_{\oplus}$ & 126$M_{\oplus}$ & 233$M_{\oplus}$ & 6.9$M_{\oplus}$ &
244$M_{\oplus}$ & 142$M_{\oplus}$ & 16$M_{\oplus}$ \\
\hline
\end{tabular}
\end{center}
\end{table}